\documentclass[%
reprint,
superscriptaddress,
 amsmath,amssymb,
 aps, physrev,
]{revtex4-2}

\usepackage{color}
\usepackage{xcolor}
\usepackage{graphicx}
\usepackage{dcolumn}
\usepackage{siunitx}
\usepackage{bm}
\usepackage{hyperref}
\hypersetup{colorlinks=true,linkcolor=blue,citecolor=blue,urlcolor=blue}
\usepackage{soul} 
\usepackage[version=4]{mhchem}
\usepackage[colorinlistoftodos]{todonotes}

\begin{document}

\preprint{APS/123-QED}

\title{Single crystal growth, structural and physical properties,\\ and absence of a charge density wave in \texorpdfstring{\ce{Ti_{0.85}Fe6Ge6}}{}}%


\author{Dechao Cheng}
\thanks{These authors contributed equally to this work}
\affiliation{Low Temperature Physics Laboratory, College of Physics and Center of Quantum Materials and Devices, Chongqing University, Chongqing 401331, China.
}

\author{Nour Maraytta}
\thanks{These authors contributed equally to this work}
\affiliation{Institute for Quantum Materials and Technologies, Karlsruhe Institute of Technology, Kaiserstraße 12, 76131 Karlsruhe, Germany.}

\author{Xiuhua Chen}
\affiliation{School of Emerging Technology, University of Science and Technology of China, Hefei, Anhui 230026, China.
}

\author{Xizhi Li}
\affiliation{Low Temperature Physics Laboratory, College of Physics and Center of Quantum Materials and Devices, Chongqing University, Chongqing 401331, China.
}

\author{Xueliang Wu}
\affiliation{Low Temperature Physics Laboratory, College of Physics and Center of Quantum Materials and Devices, Chongqing University, Chongqing 401331, China.
}

\author{Xiangxiang Jing}
\affiliation{College of Chemistry and Chemical Engineering,  Chongqing University, Chongqing 401331, China}

\author{Yong Hu}
\affiliation{Low Temperature Physics Laboratory, College of Physics and Center of Quantum Materials and Devices, Chongqing University, Chongqing 401331, China.
}

\author{Youpin Gong}
\affiliation{Low Temperature Physics Laboratory, College of Physics and Center of Quantum Materials and Devices, Chongqing University, Chongqing 401331, China.
}

\author{Mingquan He}
\affiliation{Low Temperature Physics Laboratory, College of Physics and Center of Quantum Materials and Devices, Chongqing University, Chongqing 401331, China.
}

\author{Yisheng Chai}
\affiliation{Low Temperature Physics Laboratory, College of Physics and Center of Quantum Materials and Devices, Chongqing University, Chongqing 401331, China.
}

\author{Xiaoyuan Zhou}
\affiliation{Low Temperature Physics Laboratory, College of Physics and Center of Quantum Materials and Devices, Chongqing University, Chongqing 401331, China.
}

\author{Pengfei Jiang}
\affiliation{College of Chemistry and Chemical Engineering, Chongqing University, Chongqing 401331, China}

\author{Yilin Wang}
\affiliation{School of Emerging Technology, University of Science and Technology of China, Hefei, Anhui 230026, China.
}

\author{Michael Merz}
\email{michael.merz@kit.edu}
\affiliation{Institute for Quantum Materials and Technologies, Karlsruhe Institute of Technology, Kaiserstraße 12, 76131 Karlsruhe, Germany.}
\affiliation{Karlsruhe Nano Micro Facility, Karlsruhe Institute of Technology, Kaiserstraße 12, 76131 Karlsruhe, Germany.}

\author{Aifeng Wang}
\email{afwang@cqu.edu.cn}
\affiliation{Low Temperature Physics Laboratory, College of Physics and Center of Quantum Materials and Devices, Chongqing University, Chongqing 401331, China.
}

\date{\today}

\begin{abstract}
Kagome materials with charge density waves (CDWs) are fascinating quantum systems, offering an ideal platform to explore intertwined orders and to uncover novel mechanisms behind CDW formation. Chemical models have been developed and applied to predict CDW in $AM_6X_6$-type kagome materials, such as the rattling chain model based on \ce{ScV6Sn6} and the magnetic energy-saving model based on FeGe. In this study, we successfully synthesized \ce{Ti_{0.85}Fe6Ge6} single crystals using the vapor transport method. As predicted by the rattling chain model, these crystals are expected to exhibit kagome CDW behavior. Magnetization measurements indicate that \ce{Ti_{0.85}Fe6Ge6} is an easy-axis antiferromagnet with $T_\mathrm{N} =$ 488 K and transport measurements reveal metallic behavior primarily driven by electron-type carriers. However, no clear signatures of a CDW were observed in \ce{Ti_{0.85}Fe6Ge6}.\@ Density functional theory calculations demonstrate a markedly distinct electronic structure compared to related compounds: instead of a carrier-doping-induced rigid shift, the density of states shifted away from the Fermi level. Consistent with our structural investigations, the absence of a CDW and the unusual band structure can be attributed to the bonding characteristic within \ce{Ti_{0.85}Fe6Ge6}. The strong covalent bonds of \ce{Ti-{Ge1b}}, along with the solid \ce{{Ge1b}-{Ge1b}} dimers, prevent the \ce{Ti-{Ge1b}-{Ge1b}-Ti} chain from rattling. The presence of \ce{Fe-Fe} antibonding state at the Fermi level enhances the spin polarization and depletes the electronic density around the Fermi level. Our results suggest that both the ionic radius and the bonding characteristics of the filler atom are crucial for the formation of CDWs in kagome materials. These factors can serve as supplementary terms to the rattling chain model, providing new insights for the discovery of novel kagome CDW materials.
\end{abstract}

\maketitle

\section{Introduction}
Kagome materials have emerged as a promising platform for exploring a range of topological quantum phases and their intricate interactions, including intertwined orders, generating considerable interest in recent research \cite{yin_topological_2022, wang_quantum_2023, wilson_av3sb5_2024, wang_topological_2024, xu_quantum_2023}.\@ This interest stems from the unique geometry of the kagome lattice (a corner-sharing triangle network), which gives rise to the characteristic kagome band structure, exhibiting three key features: Dirac crossings at the Brillouin zone (BZ) corner, Van Hove singularities (VHSs) at the BZ boundary, and a flat band crossing the BZ \cite{yin_topological_2022}.\@ These band features make kagome materials natural hosts for band topology, electronic instabilities, and strong correlations simultaneously. Therefore, when a physical order is identified in kagome materials, its interaction with the kagome bands can trigger rich quantum phenomena. A prime example is \ce{CsV3Sb5}, where the interaction between the charge density wave (CDW) and superconductivity leads to nematicity and pair density wave (PDW) formation \cite{wilson_av3sb5_2024, jiang_kagome_2023}. Consequently, extensive research has been dedicated to investigating CDWs in kagome materials.

To date, there are primarily three types of kagome CDW materials:\@ \ce{CsV3Sb5}, \ce{ScV6Sn6}, and FeGe \cite{OrtizCsVSb,arachchige_charge_2022,teng_discovery_2022}.\@ Recent investigations have found that these kagome materials exhibit significant differences in their CDW mechanisms. In \ce{CsV3Sb5} the CDW is believed to be driven by electronic instabilities linked to the VHSs \cite{wilson_av3sb5_2024,jiang_kagome_2023}  while in \ce{ScV6Sn6},\@ the CDW is driven by unstable Sn and Sc phonon modes \cite{lee_nature_2024}, with numerous competing CDW instabilities observed \cite{tan_abundant_2023,cao_competing_2023}.\@ In contrast, since no negative phonon frequencies are detected in FeGe,\@ magnetism is thought to play a crucial role in its CDW mechanism \cite{teng_magnetism_2023,shao_intertwining_2023,wu_electron_2023,ma_theory_2024,wang_enhanced_2023,zhang_electronic_2024}.\@ Yilin Wang proposed that the CDW in FeGe is driven by the balance between magnetic energy saving and structural energy cost via partially dimerizing Ge1 sites \cite{wang_enhanced_2023}.\@ Zhang \textit{et.\@ al.\@} proposed a triple-well CDW mechanism where the CDW in FeGe is driven by the downward shift of an entire Ge band \cite{zhang_electronic_2024}.\@ We found that the CDW in FeGe is highly tunable by post-growth annealing \cite{wu_annealing-tunable_2024}, and that it is the disorder in the Ge1a and Ge1b sites at high temperatures that determines the CDW phase at low temperatures, which is characterized by the $2 \times 2$ arrangement of Ge1-Ge1 dimers \cite{miao_signature_2023,chen_discovery_2024}.\@ Both the theoretical and experimental results emphasize the role of Ge1 atoms, reminiscent of the CDW in \ce{ScV6Sn6} featured by the $\sqrt{3} \times \sqrt{3}$ arrangement of Sn-Sn dimers \cite{arachchige_charge_2022}.\@ 
Recent investigations suggest that the small ionic radius of Sc is crucial for CDW formation, as it provides additional space for the motion of Sc and Sn atoms, consistent with the ``rattling chain model" \cite{pokharel_frustrated_2023,meier_tiny_2023}.\@ This model can serve as a guideline for exploring new kagome CDWs in the $AM_6X_6$ family (\ce{HfFe6Ge6} prototype) with small ions occupying the $A$ site. Indeed, following this approach, a new kagome CDW material, \ce{LuNb6Sn6}, with a $\sqrt{3} \times \sqrt{3} \times 3$ superlattice was discovered, further corroborating the validity of the rattling chain model \cite{ortiz_stability_2025}. 

We aim to explore new CDW materials within the \ce{$A$Fe6Ge6} system where the $A$ site is occupied by a small atom. This arrangement is expected to exhibit CDW ordering, as suggested by both the magnetic-energy-saving and rattling models, thereby combining features of FeGe and \ce{ScV6Sn6}. Among the known \ce{$A$Fe6Ge6} compounds, \ce{TiFe6Ge6} is chosen primarily due to the smallest ionic radius of Ti among the known filler atoms
(see Table \ref{table3}). This suggests that \ce{TiFe6Ge6} satisfies the criteria of the rattling chain model, combining a large host lattice with a small filler atom \cite{ortiz_stability_2025}.\@ So far, only magnetic property measurements of \ce{TiFe6Ge6} polycrystals have been reported, revealing simple antiferromagnetic behavior with a magnetic moment of approximately 1 $\mu_\mathrm{B}$/Fe \cite{nishihara_magnetic_1999, mazet_neutron_2000}.\@ The detailed physical properties remain elusive, likely due to the absence of high-quality single crystals.

In this paper, we successfully synthesized high-quality \ce{Ti_{0.85}Fe6Ge6} single crystals using the chemical vapor transport (CVT) technology. Note that, to the best of our knowledge, it is the first time to grow the $AM_6X_6$ crystals using the CVT method, where clean crystals with well-defined surfaces can be obtained. The crystal structure is determined by both single-crystal and powder XRD measurements. Magnetization measurements indicate an antiferromagnetic transition at 488 K with the easy axis along the crystallographic $c$ direction. Electronic transport measurements reveal a metallic behavior dominated by electron-type carriers, and negligibly small negative magnetoresistance (MR) is observed across the whole temperature range. No signature of CDW order can be observed down to 2 K. Density functional theory (DFT) calculations reveal that instead of electron doping, the density of electrons moves away from the Fermi level ($E_\mathrm{F}$), forming an indirect gap at the $k_z = 0$ plane in the BZ. The absence of CDW and the gap opening in the band structure can be well interpreted in terms of chemical bonds. Our results indicate that not only the ion radii but also the chemical bonds associated with the filler atoms are crucial for the occurrence of CDW among $AM_6X_6$ materials.

\section{Methods}

Single crystals of hexagonal HfFe$_6$Ge$_6$-type TiFe$_6$Ge$_6$ were grown by the chemical vapor transport (CVT) method. Titanium powders (99.99\%), iron powders (99.99\%), and germanium powders (99.999\%) were weighed according to the stoichiometric ratio 1 : 6 : 6, thoroughly ground, and loaded into a quartz tube with additional iodine as transport agents. The quartz tube was sealed under a high vacuum and placed into a horizontal two-zone furnace. The source and sink temperatures were maintained at \SI{800}{\degreeCelsius} and \SI{720}{\degreeCelsius}, respectively, to grow the single crystals. After being held at the single-crystal growth temperatures for 14 days, the system was cooled naturally to room temperature by shutting down the furnace. Shiny, hexagonal platelike crystals with a typical dimension of \numproduct{2 x 1.5 x 0.15} \unit{mm^3} can be obtained at the cold end of the quartz tube. We attempted to anneal the obtained crystals to investigate whether an annealing effect similar to that reported in FeGe \cite{wu_annealing-tunable_2024} could be observed. However, no discernible changes were detected. Consequently, all physical property measurements in this study were carried out on as-grown crystals.

The crystal structure is probed by both single-crystal and powder x-ray diffraction (XRD) methods. Single-crystal XRD measurements were performed on a high-flux, high-resolution, rotating anode Rigaku Synergy-DW (Mo/Ag) diffractometer using Mo $K_\mathrm{\alpha}$ radiation ($\lambda$ = 0.7107 {\AA}). The system is equipped with a background-less Hypix-Arc150$^{\circ}$ detector, which guarantees minimal reflection profile distortion and ensures uniform detection conditions for all reflections. All samples were measured to a resolution better than 0.5 {\AA} and with the beam divergence set to 5 mrad.\@ The samples exhibited no mosaic spread and no additional reflections from secondary phases, highlighting their high quality and allowing for excellent evaluation using the latest version of the CrysAlisPro software package \cite{CrysAlis}.\@ The crystal structure of the system was refined using JANA2006 \cite{Vaclav_229_2014}.\@ 
Powder XRD measurements were carried out on a PANalytical powder diffractometer in Debye-Scherrer geometry using Cu $K_\alpha$ radiation ($\lambda$ = 1.5406 {\AA}).\@ Rietveld refinements were performed on the as-collected high-resolution powder XRD data, and the background was described by a Chebyshev polynomial function with 10 parameters. The powder XRD data were measured on crushed single crystals, and preferred orientations exist for these crushed powder samples. The preferred orientations were thus corrected by using two preferred orientation directions (100 and 001). Rietveld refinements proceeded smoothly with these preferred orientation corrections, resulting in low residual $R_{wp}$ and $R_p$ factors. Laue diffraction patterns were taken on a single crystal in a backscattering geometry with the incident white light exposed along the $c$ axis. The chemical compositions of single crystals were determined by energy-dispersive x-ray spectroscopy (EDX) measurements in a Thermo Fisher Quattro S Environmental scanning electron microscope equipped with an EDX detector from Oxford Instruments. The chemical valence state is estimated using an x-ray photoelectron spectrometer (Thermo Fisher, ESCALAB250Xi).

Transport properties and heat capacity were measured in a Quantum Design DynaCool Physical Properties Measurement System (PPMS-9T). The resistivity $\rho_{xx}$ and Hall resistivity $\rho_{xy}$ were measured simultaneously on a single crystal with the six-probe configuration. The DC delta mode with a Keithley 6221 current source and a Keithley 2182A nanovoltmeter was used to record the transport data. Magnetization data were collected in a PPMS-14T using the high-temperature vibrating sample magnetometry (VSM) option. 

DFT calculations were performed using the Vienna $ab$ initio simulation package (VASP) \cite{vasp1996}.\@ The exchange-correlation potential is treated within the generalized gradient approximation (GGA) of the Perdew-Burke-Ernzerhof (PBE) variety \cite{pbe1996}.\@ We have used our experimental lattice parameters. Integration for the Brillouin zone is done using $\Gamma$-centered $16\times 16\times 10$ $K$-point grids for FeGe ($15\times 15\times 10$ for \ce{TiFe6Ge6}) and the cutoff energy for plane-wave-basis was set to 500 eV.\@ The orbital-projected band structure and the corresponding density of states (DOS) were calculated using VASPKIT \cite{VASPKIT2021}.\@ Additionally, the crystal orbital Hamilton population (COHP) analysis was performed using LOBSTER to investigate the bonding characteristics of the materials \cite{cohp1993,cohp2011,cohp2013,cohp2016_1,cohp2016_2,cohp2020}.\@ Since the spin-orbital coupling (SOC) effect is weak, SOC was not taken into account for the DFT calculation in this paper.

\section{Results and discussion}

\subsection{Crystal structure}

\begin{figure}
\centerline{\includegraphics[width=8.8cm]{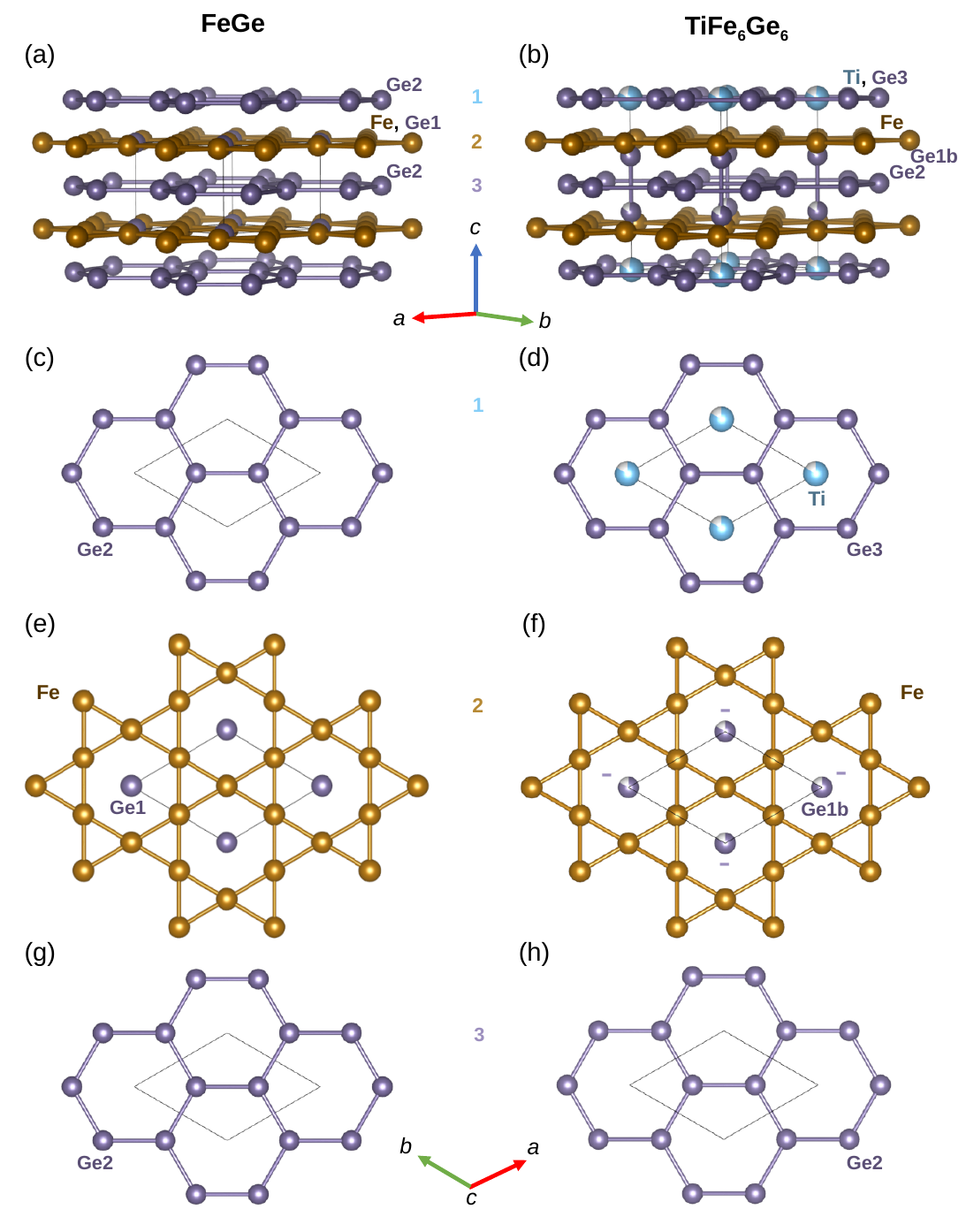}}
\caption{Perspective view of the crystal structure of (a) FeGe and (b) \ce{Ti_{0.85}Fe6Ge6}.\@ (c-h) Top views of the upper three layers corresponding to (a) and (b), respectively. The close relationship between the two systems can be easily recognized.
}
\label{fig1}
\end{figure}

The crystal structure of the obtained \ce{TiFe6Ge6} crystals was determined by both single-crystal XRD and powder XRD measurements, which are in good agreement with each other and consistent with previous reports \cite{etde_5753394}. The obtained crystal structure is displayed together with FeGe in Fig.\@ \ref{fig1} and the refined parameters are listed in Tables \ref{table1} and \ref{table2}.\@ The refined compositions were determined to be \ce{Ti_{0.85}Fe6Ge6} for single-crystal XRD measurements and \ce{Ti_{0.8}Fe6Ge6} for powder XRD measurements (Table \ref{table1}),\@ both consistent with the atomic ratio Ti : Fe : Ge = 0.80 : 6.02 : 6 obtained from EDX analysis. These results confirm a 15–20 \% deficiency of Ti.\@ EDX mapping (Fig.\@ S1, Supplemental Material \cite{SM}) further reveals that Ti, Fe, and Ge atoms are uniformly distributed within a single crystal. However, the Ti content varies across crystals from the same batch, with deficiencies ranging from 10–40 \%. Both powder and single-crystal x-ray diffraction clearly demonstrate that the materials are single-phase, showing no indications of phase separation into FeGe and \ce{TiFe6Ge6}, nor of FeGe clustering within a \ce{TiFe6Ge6} matrix. To minimize inconsistencies, physical property measurements---including single-crystal XRD, magnetization, transport, and specific heat---were performed on the same crystal. Accordingly, the single-crystal XRD composition, \ce{Ti_{0.85}Fe6Ge6}, is adopted throughout this paper as the representative chemical formula.

The crystal structure of B35-type FeGe is composed of alternatively staking layers of Fe$_3$Ge1 and Ge2$_2$ along the crystallographic $c$ axis, where Ge1 and Ge2 denote the Ge atoms located in the kagome layer [Fig.\@ \ref{fig1}(e)] and the honeycomb layer [Fig.\@ \ref{fig1}(c)],\@ respectively. The chemical formula can be rewritten in structural form as [Fe$_3$(Ge1)][(Ge2)$_2$].\@ As evidenced by its rarity (only six compounds have been reported) \cite{meier_flat_2020}, the B35 structure is relatively unstable due to the low packing density, which can be stabilized by filling the voids in the honeycomb lattice with various atoms, resulting in structurally diverse $AM_6X_6$-type compounds with more than 100 known variants \cite{venturini_filling_2006}.

In the case of FeGe,\@ Ti stuffing results in \ce{TiFe6Ge6},\@ which adopts the \ce{HfFe6Ge6}-type structure. As shown in Fig.\@ \ref{fig1}(d), Ti atoms fill the void in the Ge3 honeycomb lattice and push the adjacent Ge1 atoms [Fig.\@ \ref{fig1}(f)] out of the Fe$_3$Ge1a layer, forming Ge1b-Ge1b dimers. The structural form for the \ce{TiFe6Ge6} chemical formula results in [Ti(Ge3)$_2$][Fe$_3$(Ge1)][(Ge2)$_2$][Fe$_3$(Ge1)].\@ Interestingly, the occupancy of the Ti site is directly related to the one of the Ge1b sites, and consequently, to the formation of Ge1b-Ge1b dimers. This indicates that in all unit cells where the Ti site is occupied, Ge1b-Ge1b dimers are formed. Conversely, in unit cells where the Ti site is vacant, the Ge atom primarily stays in the kagome layer at the Ge1a site, experiencing only a minor displacement. This phenomenon, together with the 85\% Ti occupation on the $1a$ site (see Table \ref{table2}), also contributes to a certain level of statistical disorder at the Ti, Ge1a, and Ge1b sites. However, the Wyckoff positions show no occupational disorder: Ti and Ge atoms reside only on their respective $1a$ and $2e$ positions (Table \ref{table2}).\@ In the refinement process, two constraints have therefore been established: first, the occupancy of the Ti site is directly associated with the formation of Ge1b-Ge1b dimers, and second, the combined occupancy of the Ge1a and Ge1b positions is restricted to 100 \%.

\begin{table}[htp!]
  \caption{Crystallographic data of \ce{Ti_{0.85}Fe6Ge6} derived from single-crystal and powder data measured at 295 K. The goodness of fit, GOF, and the corresponding reliability factor $wR_2/R_{wp}$, $R_1/R_p$ are given as well.}
  \label{table1}
  \begin{tabular}{ccc}
    \hline\hline
    Parameters & SCXRD & PXRD \\
    \hline
    Chemical formula   & Ti$_{0.85}$Fe$_{6}$Ge$_{6}$& Ti$_{0.8}$Fe$_{6}$Ge$_{6}$   \\
    Formula weight (g/mol) & 811.6927& 809.2036  \\
    Space group & $P6/mmm$ & $P6/mmm$  \\
    $Z$ & 1 & 1\\
    $a$ ({\AA})  & 5.02377(6) &5.0219(2)\\
    $b$ ({\AA}) & 5.02377(6) & 5.0219(2)\\
    $c$ ({\AA}) & 8.03564(9) & 8.0321(2)\\
    $\alpha$ & 90$^{\circ}$ & 90$^{\circ}$ \\
    $\beta$  & 90$^{\circ}$& 90$^{\circ}$ \\
   $\gamma$ & 120$^{\circ}$ & 120$^{\circ}$\\
   $V$(\AA$^3$) & 175.636 & 175.428 \\ \hline
   GOF & 1.83 & 1.34 \\
   $wR_2/R_{wp}$ (\%) & 6.50 & 2.28 \\
   $R_1/R_p$ (\%) & 2.76 & 1.93 \\
    \hline\hline
  \end{tabular}
\end{table}

\begin{table}[htp!]
  \caption{Atomic coordinates, Wyckoff positions, occupation numbers, Occ.,\@ and equivalent atomic displacement parameters, $U_{\mathrm{eq}}$,\@ of \ce{Ti_{0.85}Fe6Ge6} derived from single-crystal XRD measured at 295 K.} 
  \centering
  \label{table2}
    \begin{tabular}{ccccccc}
    \hline\hline
    Atom  & Wyckoff & $x$ & $y$ & $z$ & Occ. & $U_{\mathrm{eq}}$ ({\AA}$^{2}$)  \\
    \hline
    Ti & 1$a$ & 0 & 0 & 0 & 0.85(1) & 0.01192(13)\\
    Fe & 6$i$ & 1/2 & 0 & 0.24869 & 1 & 0.00511(4)   \\
    Ge(1a) & 2$e$ & 0 & 0 & 0.25063 & 0.15(1) & 0.00667(5)   \\
    Ge(1b) & 2$e$ & 0 & 0 & 0.33704 & 0.85(1) & 0.00667(5)  \\
    Ge(2) & 2$d$ & 1/3 & 2/3 & 1/2 & 1 & 0.00625(4)   \\
    Ge(3) & 2$c$ & 1/3 & 2/3 & 0 & 1 & 0.00606(4)   \\
    \hline\hline
  \end{tabular}
\end{table}

The bond distance for the \ce{{Ge1b}-{Ge1b}} dimer is $d_\mathrm{Ge1b-Ge1b} =$ 2.619 {\AA}, slightly smaller than the 2.669 {\AA} in the low‑temperature CDW phase of FeGe, indicative of a similar bonding nature of the \ce{{Ge1b}-{Ge1b}} dimers in \ce{Ti_{0.85}Fe6Ge6} and FeGe \cite{wu_annealing-tunable_2024}.\@ Similarly, in \ce{ScV6Sn6}, the CDW induces an average shrinkage of 4.5 \% in the Sn–Sn dimer distance, reducing $d_\mathrm{Sn-Sn}$ from 3.218 to 3.0735 {\AA} \cite{arachchige_charge_2022}. Hence, \ce{TiFe6Ge6} can be regarded as the fully (100 \%) dimerized analogue of FeGe (85\% in the case of \ce{Ti_{0.85}Fe6Ge6}), whereas the low-temperature CDW phase of FeGe corresponds (neglecting a small but certain degree of disorder between dimer and pristine Ge site \cite{wu_annealing-tunable_2024}) to a 1/4 dimerized state \cite{chen_discovery_2024}.\@

The Ti stuffing and the formation of \ce{{Ge1b}-{Ge1b}} dimers result in the doubling of the unit cell along the $c$ axis. The obtained lattice parameters for \ce{Ti_{0.85}Fe6Ge6} are $a =$ 5.02377(6) {\AA}, $c =$ 8.03564(9) {\AA}. Compared with the lattice parameters of FeGe, where $a =$ 4.9960 {\AA} and 2$c =$ 8.1048 {\AA} obtained by single-crystal XRD measurements at 300 K on a \SI{320}{\degreeCelsius} annealed crystal  \cite{wu_annealing-tunable_2024}, Ti stuffing leads to a 0.56 \% expansion of intralayer distance and a 0.85 \% shrinkage of interlayer distance. This result is consistent with the \ce{{Ge1b}-{Ge1b}} dimer as the main feature of structural distortion.
The incorporation of the small \ce{Ti^{4+}} ions into the FeGe lattice results in reduced lattice parameters relative to other \ce{$A$Fe6Ge6} compounds: the $a$ parameter is the smallest within the series, whereas the $c$ parameter is slightly larger than that of \ce{LiFe6Ge6} but remains smaller than in all other \ce{$A$Fe6Ge6} materials (see Table \ref{table3}). When we place the lattice parameters of \ce{Ti_{0.85}Fe6Ge6} on the phase diagram of cell volume vs $A$-site ion radius for known $AM_6X_6$ materials [see Fig. 1(c) in ref.\cite{ortiz_stability_2025} or Fig. S2 in Supplemental Material \cite{SM}], it can be seen that \ce{Ti_{0.85}Fe6Ge6} locates at the bottom left corner of the phase diagram, i.e., Ti has the smallest ionic radius, but the cell volume of \ce{Ti_{0.85}Fe6Ge6} is also small among the $AM_6X_6$ materials \cite{ortiz_stability_2025}. From the rattling chain model, the occurrence of a CDW requires a combination of a large host lattice and a small filler atom. Thus, the occurrence of a CDW in \ce{Ti_{0.85}Fe6Ge6} is unclear according to the rattling chain model, requiring further experimental validation.

\subsection{Sample characterization}

\begin{figure}
\centerline{\includegraphics[width=9cm]{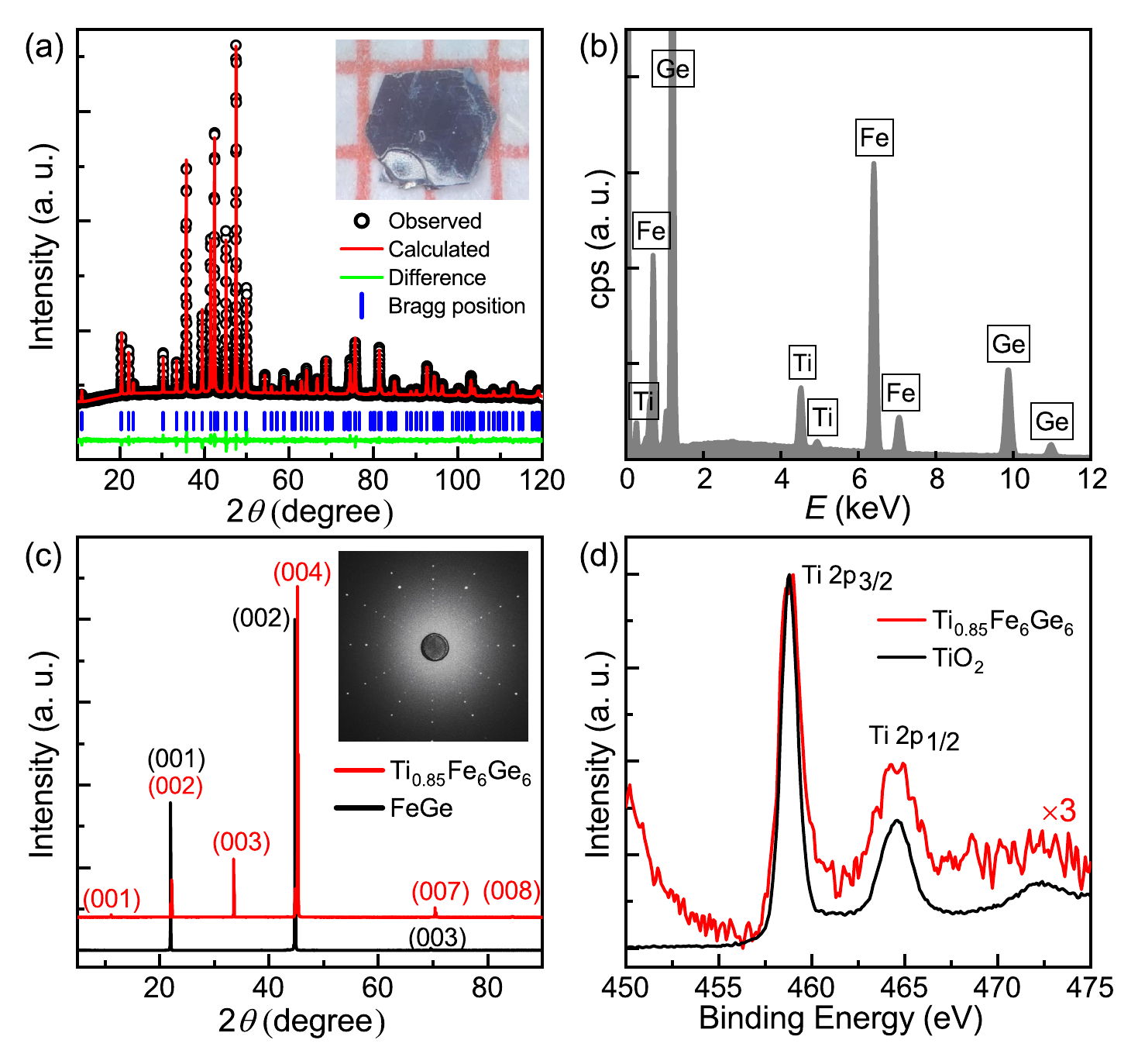}}
\caption{(a) Rietveld refinement fit to powder XRD pattern of \ce{Ti_{0.85}Fe6Ge6}. The inset shows the photograph of a typical \ce{Ti_{0.85}Fe6Ge6} single crystal on a millimeter grid. (b) EDX spectrum for a \ce{Ti_{0.85}Fe6Ge6} crystal. (c) X-ray diffraction data on the (0 0 $l$) surface of the \ce{Ti_{0.85}Fe6Ge6} and FeGe crystals. The inset displays the x-ray Laue diffraction pattern of a \ce{Ti_{0.85}Fe6Ge6} crystal. (d) XPS of Ti $2p$ spectra for a \ce{Ti_{0.85}Fe6Ge6} crystal and \ce{TiO2} powder.
}
\label{fig2}
\end{figure}

The powder XRD pattern [Fig.\@ \ref{fig2}(a)] and the EDX spectrum [Fig.\@ \ref{fig2}(b)] confirm phase purity of the \ce{Ti_{0.85}Fe6Ge6} crystals. The XRD pattern on the (0 0 $l$) lattice plane of a single crystal and the Laue diffraction pattern [Fig.\@ \ref{fig2}(c)] indicate the excellent crystalline nature of the as-grown single crystal with a preferred [001] orientation. From Fig.\@ \ref{fig2}(c),\@ a shift of the (0 0 $l$) reflections to slightly higher $2 \theta$ angles accompanied by the appearance of peaks with odd $l$ (in the unit cell doubled along $c$) can be observed, consistent with the structure modulation along the $c$ direction \cite{mazet_neutron_2000}. 

The choice of \ce{TiFe6Ge6} among the $AM_6X_6$ materials is mainly due to the requirement of the small ion radius at the $A$ site.\@ However, the ion radius sensitively depends on the valence, i.e., 0.67 {\AA} for \ce{Ti^3+} and 0.605 {\AA} for \ce{Ti^4+} in  VI-Coord. Shannon radius \cite{shannon_revised_1976}. XPS measurements were performed on a \ce{Ti_{0.85}Fe6Ge6} crystal and \ce{TiO2} powders to probe the valence of Ti in \ce{Ti_{0.8}Fe6Ge6}, where \ce{TiO2} serves as the reference. As shown in Fig. \ref{fig2}(d), the \ce{Ti_{0.85}Fe6Ge6} profile closely resembles that of \ce{TiO2}. Moreover, the binding energies for Ti $2p_{3/2}$  (458.8 eV) and Ti $2p_{1/2}$ (464.6 eV) peaks are close to the theoretical values of \ce{Ti^4+} \cite{sanjines_electronic_1994}.\@ These results confirm the \ce{Ti^4+} valence in \ce{Ti_{0.85}Fe6Ge6}, consistent with our initial motivation.

\subsection{Magnetic Properties}

\begin{figure}
\centerline{\includegraphics[width=7cm]{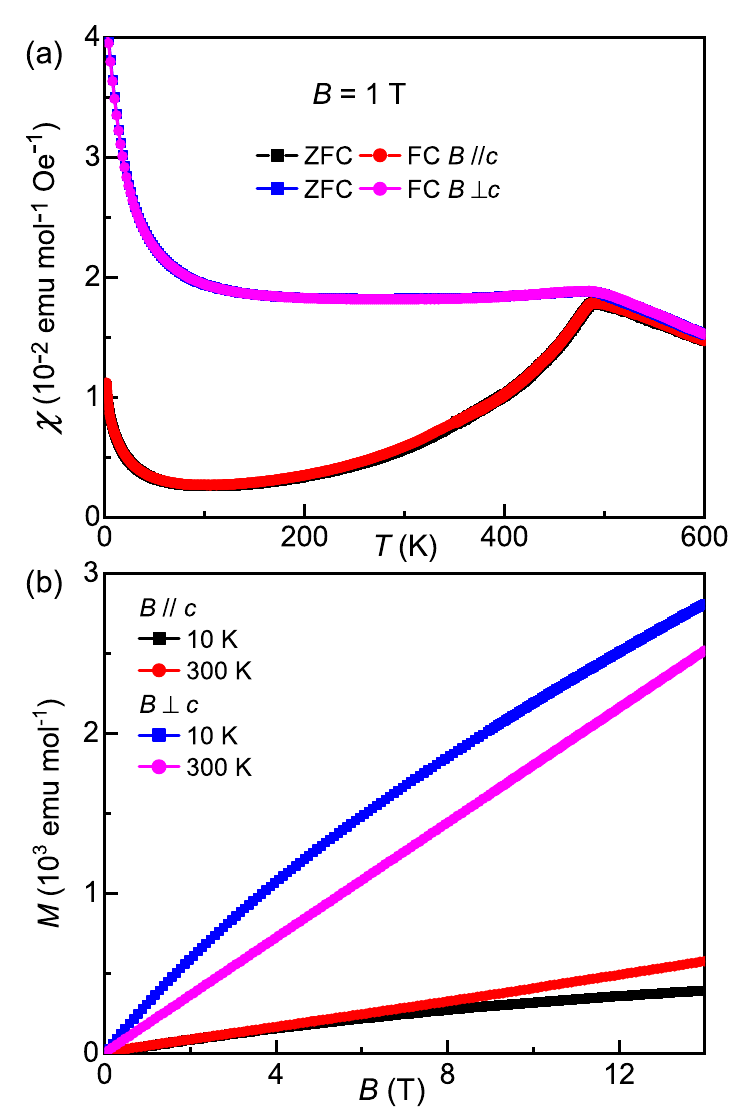}}
\caption{(a) The temperature dependence of zero-field-cooled (ZFC) and field-cooled (FC) magnetic susceptibilities measured with a 1 T magnetic field applied parallel and perpendicular to the $c$ axis of a \ce{Ti_{0.85}Fe6Ge6} crystal. (b) Field-dependent magnetization was collected at 2 and 300 K with $B \parallel c$ and $B \perp c$, respectively.  
 }
\label{fig3}
\end{figure}

Figure \ref{fig3}(a) shows the temperature dependence of the magnetic susceptibilities measured with $B \parallel c$ [$\chi_\parallel(T)$] and $B \perp c$ [$\chi_\perp(T)$] in a temperature range of 2$-$600 K. The general behavior of the $\chi(T)$ curves is consistent with that reported on \ce{TiFe6Ge6} polycrystals and similar to that of FeGe single crystal, indicative of a similar magnetic structure \cite{nishihara_magnetic_1999,mazet_macroscopic_2001,beckman_susceptibility_1972}. There is no obvious bifurcation between the susceptibilities measured in zero-field-cooled (ZFC) and field-cooled mode (FC). A magnetic transition can be identified at $T_\mathrm{N} =$ 488 K, which can be assigned to the antiferromagnetic ordering \cite{nishihara_magnetic_1999}. With the temperature cooling across $T_\mathrm{N}$, $\chi_\parallel(T)$ decreases quickly while the $\chi_\perp(T)$ curve remains flat until the Curie-Weiss-like upturn below 150 K. $\chi_\parallel(T) < \chi_\perp(T)$ indicates that the easy axis is along the $c$ direction. The strong Curie-Weiss tail has also been observed in polycrystals \cite{nishihara_magnetic_1999,mazet_macroscopic_2001}, which might come from clusters associated with defects or disorders. No transition can be identified below $T_\mathrm{N}$, implying the lack of any additional ordering (CDW or otherwise) at lower temperatures. According to our annealing experiments on FeGe \cite{wu_annealing-tunable_2024}, the CDW transition strongly depends on the disorder on the Ge1-site. Given the $\sim$15\% Ti vacancy, the randomly distributed Ti and Ge1 dimers might also prevent the formation of long-range CDW order. Therefore, the absence of a transition below $T_\mathrm{N}$ is not sufficient to rule out CDW order based on magnetization measurements.

As shown in Fig. \ref{fig3}(b), the isothermal magnetization $M(B)$ exhibits a Brillouin-type nonlinear paramagnetic behavior at 10 K and a linear behavior at 300 K, implying a dominant antiferromagnetic interaction with a significant paramagnetic contribution. The behavior of $\chi(T)$ and $M(B)$ curves are in good agreement with each other and consistent with the A-type easy axis antiferromagnetism revealed by neutron diffraction measurements, i.e., the magnetic moments of Fe are parallel to the $c$ axis, ferromagnetically arranged within the \ce{Fe3Ge} layer, and antiferromagnetically arranged between adjacent \ce{Fe3Ge} layer \cite{nishihara_magnetic_1999,mazet_neutron_2000}. 

The antiferromagnetic transition temperature $T_\mathrm{N} =$ 488 K for the \ce{Ti_{0.85}Fe6Ge6} crystal is remarkably higher than the 410 K in the FeGe single crystal and slightly lower than the $\sim$ 510 K in \ce{TiFe6Ge6} polycrystals \cite{nishihara_magnetic_1999,mazet_macroscopic_2001}. The discrepancy between $T_\mathrm{N}$ in single- and poly-crystalline samples can be attributed to the $\sim$15\% Ti deficiency in the single crystals (see Section A).\@ Compared to FeGe, the increase in $T_\mathrm{N}$ for \ce{Ti_{0.85}Fe6Ge6} can be ascribed to the reduction of interlayer distance caused by the Ti stuffing and the formation of dimers, where the interlayer coupling mainly determines $T_\mathrm{N}$ for an A-type AFM.   

\subsection{Transport Properties}

\begin{figure}[b]
\centerline{\includegraphics[width=9cm]{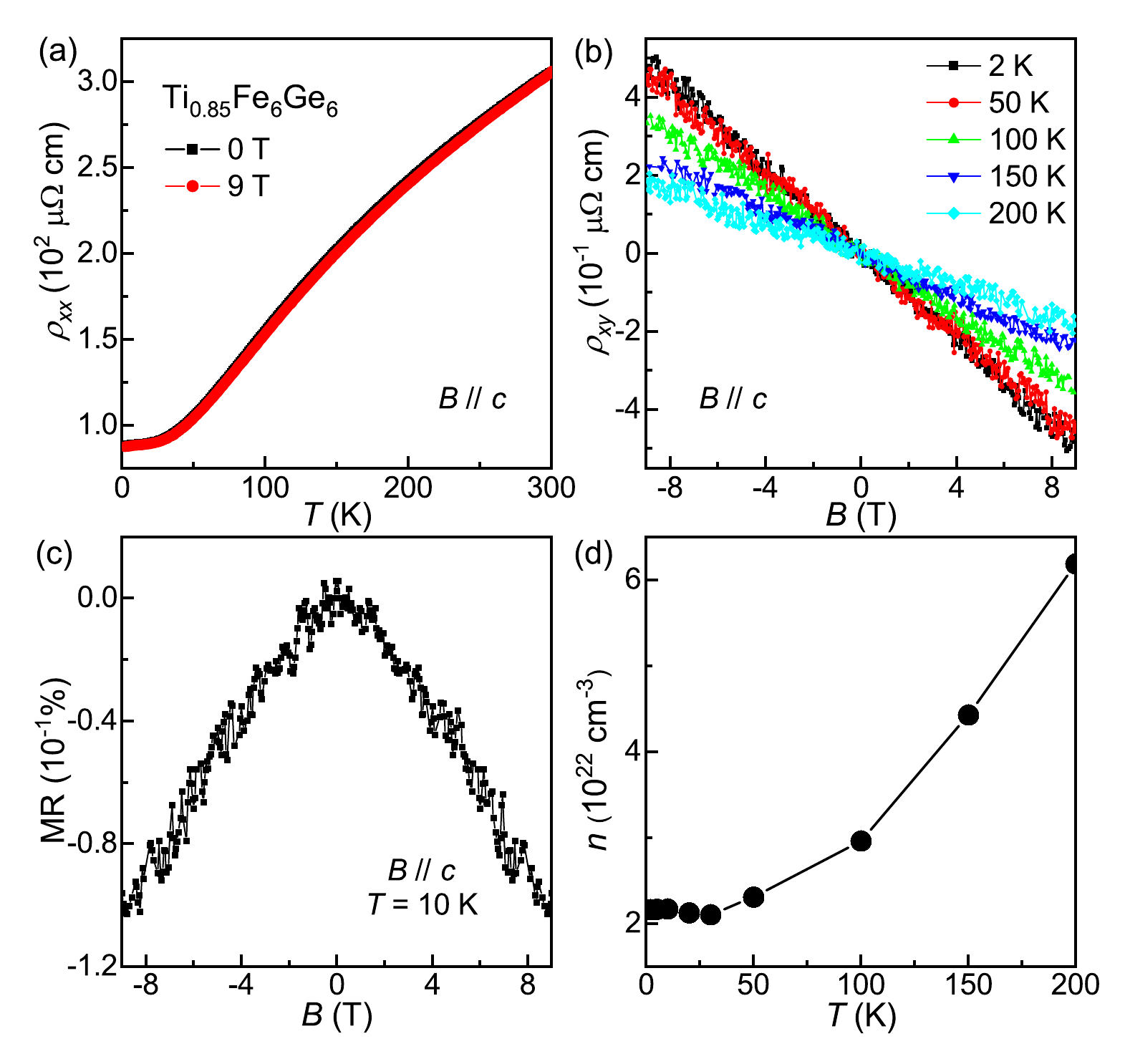}}
\caption{(a) Temperature dependence of in-plane resistivity measured with $B =$ 0 and 9 T ($B \parallel c$) for a \ce{Ti_{0.8}Fe6Ge6} crystal. (b) Magnetic field dependence of magnetoresistance measured at $T =$ 10 K with $B \parallel c$. (c) Magnetic field dependence of Hall resistivity $\rho_{xy}(B)$ at various temperatures with $B \parallel c$. (d) Carrier density of electrons obtained from fitting of $\rho_{xy}(B)$ curves in (c).}
\label{fig4}
\end{figure}

Temperature dependence of in-plane resistivity $\rho(T)$ measured with $B =$ 0 and 9 T is plotted in Fig.\@ \ref{fig4}(a).\@ $\rho$(0 T) and $\rho$(9 T) curves almost overlap each other in the whole temperature range and exhibit a metallic behavior with a sublinear temperature dependence, manifested as a resistivity curve with a slight convex profile in the temperature range of 50 $\sim$ 300 K.\@ The room-temperature resistivity is 305 \unit{\mu\Omega.cm} and the residual resistivity ratio $\rho$(300 K)/$\rho$(2 K) = 3.5.\@ The value of room-temperature resistivity is comparable to other $AM_6X_6$ compounds such as \ce{SmV6Sn6}, \ce{$R$Mn6Sn6} ($R$ = Gd--Tm, Lu), and \ce{$R$Cr6Ge6} ($R$ = Gd--Tm) \cite{huang_anisotropic_2023,ma_rare_2021,yang_crystal_2024} but about an order of magnitude larger than for \ce{$R$V6Sn6} ($R$ = Y, Gd--Tm) \cite{pokharel_electronic_2021,lee_anisotropic_2022}.\@ However, the sublinear resistivity behavior in \ce{Ti_{0.85}Fe6Ge6} differs from the linear resistivity observed in these $AM_6X_6$ compounds. The sublinear resistivity behavior was reported to be a universal behavior in vanadium kagome materials hosting charge density waves, such as \ce{ScV6Sn6} and \ce{$A$V3Sb5} ($A$ = K, Rb, and Cs) \cite{mozaffari_universal_2024,OrtizCsVSb}, which was interpreted in the context of interplay of dispersive bands and VHSs/flat bands \cite{peshcherenko_sublinear_2024,ye_hopping_2024}. 
Based on the electronic structure calculations presented in Section F, it can be observed that the electronic structure of \ce{TiFe6Ge6} differs from that of \ce{Ni3In} and \ce{ScV6Sn6} \cite{peshcherenko_sublinear_2024,ye_hopping_2024}, where most of the kagome-derived bands are gapped and located far away from the Fermi level ($E_\mathrm{F}$), suggesting a different origin.\@ The difference is further demonstrated by the varying magnetic field response of resistivity, i.e, weak for \ce{Ti_{0.85}Fe6Ge6} while relatively strong for \ce{Ni3In} and \ce{ScV6Sn6} \cite{ye_hopping_2024,mozaffari_universal_2024}.\@ Another explanation of the sublinear resistivity is the incoherence-coherence crossover behavior as observed in layered materials such as \ce{$A$Fe2As2} ($A$ = K, Rb, Cs),\@ \ce{Sr2RuO4},\@ and \ce{YbMnBi2} \cite{xiang_incoherencecoherence_2016,gutman_anomalous_2007,wang_magnetotransport_2016}.\@ Considering the gapping of bands around the $k_z =$ 0 plane in the Brillouin zone (see section F), the incoherence-coherence crossover is likely responsible for the sublinear resistivity in \ce{Ti_{0.85}Fe6Ge6}. 

A negligibly small MR in \ce{Ti_{0.85}Fe6Ge6} can be inferred from the nearly overlapped $\rho$(0 T) and $\rho$(9 T) curves [Fig. \ref{fig4}(a)].\@ The field dependence of the resistivity $\rho(B)$ was measured to further illustrate the MR behavior. The obtained MR, defined as $\mathrm{MR}=[\rho(B)-\rho(0)] / \rho(0) \times 100 \%$, is plotted in Fig. \ref{fig4}(b), showing a negative MR with a maximum value of 0.1\% at 10 K and 9 T, in good agreement with the $\rho(T)$ data. The negative MR usually originates from the spin alignment effect in magnetic materials. The negligibly small MR in \ce{Ti_{0.85}Fe6Ge6} can be ascribed to the existence of tiny alignable moments in the rigid antiferromagnetic matrix \cite{yamada_hiroshi_magnetoresistance_1973}.\@ These alignable moments may be linked to deficiencies, disorders, or impurities, as supported by the magnetization measurements (see Fig. \ref{fig3}).\@ 

As shown in Fig.\@ \ref{fig4}(c),\@ the Hall resistivity $\rho_{xy}(B)$ exhibits a linear behavior with a negative slope in the temperature range of 2 $\sim$ 200 K, suggesting the transport is dominated by electron-type carriers. The carrier density obtained from the single-band linear fit of $\rho_{xy}(B)$ curves is plotted in Fig.\@ \ref{fig4}(d),\@ which decreases from \SI{6.2e22}{cm^{-3}} to \SI{2.1e22}{cm^{-3}} with temperature decreasing from 200 K to 2 K.\@ Both the carrier-type and the carrier density at $T >$ 110 K for \ce{Ti_{0.85}Fe6Ge6} are comparable to those in FeGe \cite{teng_discovery_2022}. This result indicates that Ti doping does not significantly change the carrier density in FeGe, in strong contrast with the doping of 4 electrons expected by inserting a \ce{Ti^{4+}} ion. This result can be understood in terms of the bonding character of Ti, where the extra electrons introduced by Ti doping are localized around the \ce{Ti-{Ge1b}} covalent bonds. The detailed explanation based on the DOS and COHP analysis can be found in Section H.

\subsection{Specific Heat}

\begin{figure}
\centerline{\includegraphics[width=9cm]{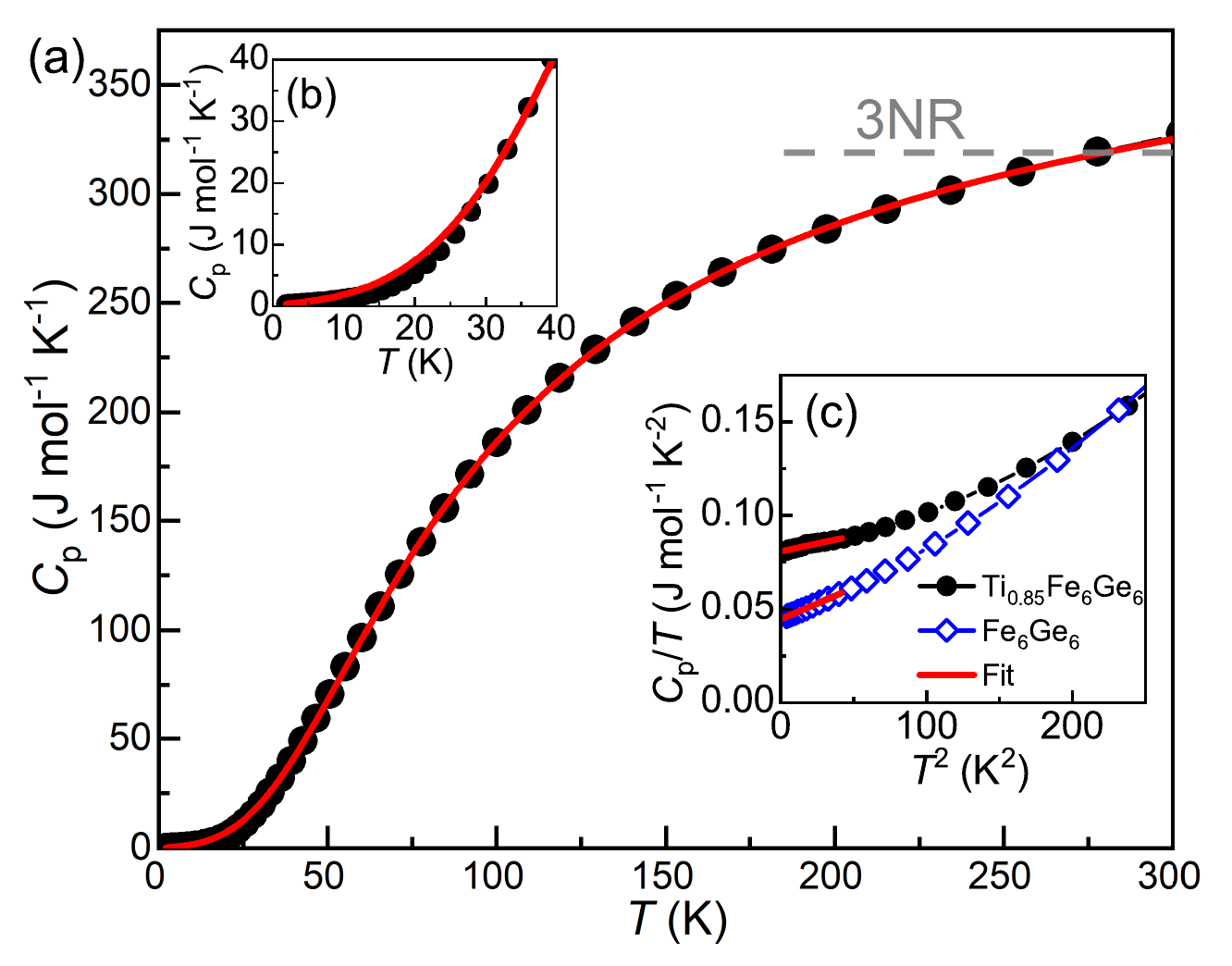}}
\caption{(a) Specific heat $C_\mathrm{p}(T)$ ($B =$ 0 T) as a function of temperature for a \ce{Ti_{0.85}Fe6Ge6} crystal. The red line is fit to the data using the Debye-Einstein model (see text). The horizontal dash line represents the Dulong-Petit limit. (b) The enlarged view of the low temperature part of (a). (c) Plots of $C_\mathrm{P}/T$ vs $T^2$ for \ce{Ti_{0.85}Fe6Ge6} and \ce{Fe6Ge6}. The red lines are the linear fit to the data in a $T^2$ range of 2 $\sim$ 40 \ce{K^2}.}
\label{fig5}
\end{figure}

Specific heat measurements were performed to probe the thermodynamic properties of \ce{Ti_{0.85}Fe6Ge6}, and the result is presented in Fig.\@ \ref{fig5}.\@ No phase transition can be observed in the temperature range between 2 and 300 K. The value of room-temperature specific heat is slightly larger than the Dulong-Petit limit of 3$NR$, where $N$ = 12.85 is the atomic number per chemical formula and $R$ = 8.314 \unit{J.mol^{-1}.K^{-1}} is the universal gas constant. This suggests that there is a minor magnetic contribution to $C_\mathrm{p}(T)$ data \cite{xia_doping-induced_2023}.\@ To quantitatively understand the electronic and phonon contributions to the total specific heat, the data were fitted by the Debye-Einstein model \cite{prakash_ferromagnetic_2016}:
\begin{equation}
\begin{aligned} C_\mathrm{el+ph}(T) = & \gamma T+\alpha9NR\left(\frac{T}{\theta_\mathrm{D}}\right)^{3} \int_{0}^{\theta_\mathrm{D} / T} \frac{x^{4} e^{x}}{\left(e^{x}-1\right)^{2}} d x \\ &+(1-\alpha)3NR\frac{\left(\theta_\mathrm{E} / T\right)^{2} e^{\theta_\mathrm{E} / T}}{\left(e^{\theta_\mathrm{E} / T}-1\right)^{2}} \end{aligned}
\end{equation}
where $\gamma$ is the Sommerfeld coefficient, and $\theta_\mathrm{D}$ and $\theta_\mathrm{E}$ are the Debye and Einstein temperatures, respectively. The coefficients $\alpha$ and $1 - \alpha$ denote the contribution of the Debye and Einstein terms to the phonon heat capacity, respectively. The best fit to the $C_\mathrm{p}(T)$ data gives rise to the following parameters: $\alpha =$ 0.87, $\gamma =$ \SI{0.11}{J.mol^{-1}.K^{-2}}, $\theta_\mathrm{D} =$ 326 K, and $\theta_\mathrm{E} =$ 624 K. The Debye temperature is comparable to 317 K in \ce{ScV6Sn6} \cite{arachchige_charge_2022}, but smaller than 427 K in FeGe \cite{meier_flat_2020}. Both the Debye and the Einstein temperatures are relatively high, which is indicative of strong bonding in \ce{Ti_{0.85}Fe6Ge6}.

It is noteworthy that $\gamma =$ \SI{0.11}{J.mol^{-1}.K^{-2}} deduced from the Debye-Einstein model for \ce{Ti_{0.85}Fe6Ge6} is much larger than \SI{7.6}{mJmol^{-1}K^{-2}} reported for FeGe \cite{meier_flat_2020}, which, according to $\gamma = \frac{\pi^2}{3}k_\mathrm{B}^2N(E_\mathrm{F})$, where $N(E_\mathrm{F})$ is the density of state at the Fermi level, contradicts the lower DOS($E_\mathrm{F}$) calculated in Fig. \ref{fig6}. From the enlarged view in Fig. \ref{fig5}(b), it can be seen that the low-temperature $C_\mathrm{P}(T)$ cannot be well described by the Debye-Einstein model, possibly overestimating the value of $\gamma$. To directly compare the $\gamma$ value between \ce{Ti_{0.85}Fe6Ge6} and FeGe, $C_\mathrm{P}/T$ vs $T^2$ curves are plotted in Fig. \ref{fig5}(c) for \ce{Ti_{0.85}Fe6Ge6} and \ce{Fe6Ge6}, where the chemical formula of FeGe is renormalized to \ce{Fe6Ge6} to have a comparable unit cell volume with \ce{Ti_{0.85}Fe6Ge6}. From Fig. \ref{fig5}(c) and Fig. S3 in Supplemental Material \cite{SM}, one can see that $C_\mathrm{P}/T$ vs $T^2$ curves are parallel to each other at high temperatures, while the bifurcation appears below 200 \unit{K^2}. Fitting the Debye model $C_\mathrm{p}/T = \gamma + {\beta}T^2$ to the low-temperature data gives rise to $\gamma =$ \SI{0.044}{J.mol^{-1}.K^{-2}} for \ce{Fe6Ge6}, and $\gamma =$ \SI{0.080}{J.mol^{-1}.K^{-2}} for \ce{Ti_{0.85}Fe6Ge6}. Although the $\gamma$ extracted by the low-temperature Debye model for \ce{Ti_{0.85}Fe6Ge6} is significantly smaller than that deduced by the Debye-Einstein model, it is still nearly two times larger than that in \ce{Fe6Ge6}. This result indicates that the contradiction between $\gamma$ and DOS is intrinsic and does not depend on the fitting. 

In fact, from the Hall measurements in Fig. \ref{fig4}(d), one can see that the high-temperature carrier density of \ce{Ti_{0.85}Fe6Ge6} is comparable to that of FeGe, indicative of the consistency between $\gamma$ and DOS at high temperatures. Therefore, the contradiction should stem from low-temperature issues, for example, the presence of CDW and spin-canting transitions in FeGe and the strong Curie-Weiss-like tail in \ce{Ti_{0.85}Fe6Ge6}. From Fig. \ref{fig5}(c), it can be seen that the bifurcation temperature for the $C_\mathrm{P}/T$ vs $T^2$ curves approximately coincides with the Curie-Weiss-like tail in magnetic susceptibility. This phenomenon suggests that the mechanism associated with the Curie-Weiss-like tail, i.e., magnetism, disorder, or impurity, should primarily account for the contradiction between $\gamma$ and DOS. Additionally, the formation of the CDW order in FeGe might also play an important role, which reduces the $\gamma$ value by opening a CDW gap around $E_\mathrm{F}$. This explanation is evidenced by the sudden decrease of $d\rho_{xy}/dB$ at $T_\mathrm{CDW}$ for FeGe \cite{teng_discovery_2022}, and consistent with the absence of CDW transition in \ce{Ti_{0.85}Fe6Ge6}.  

\subsection{Electronic Structure Calculation}

\begin{figure}
\centerline{\includegraphics[width=9cm]{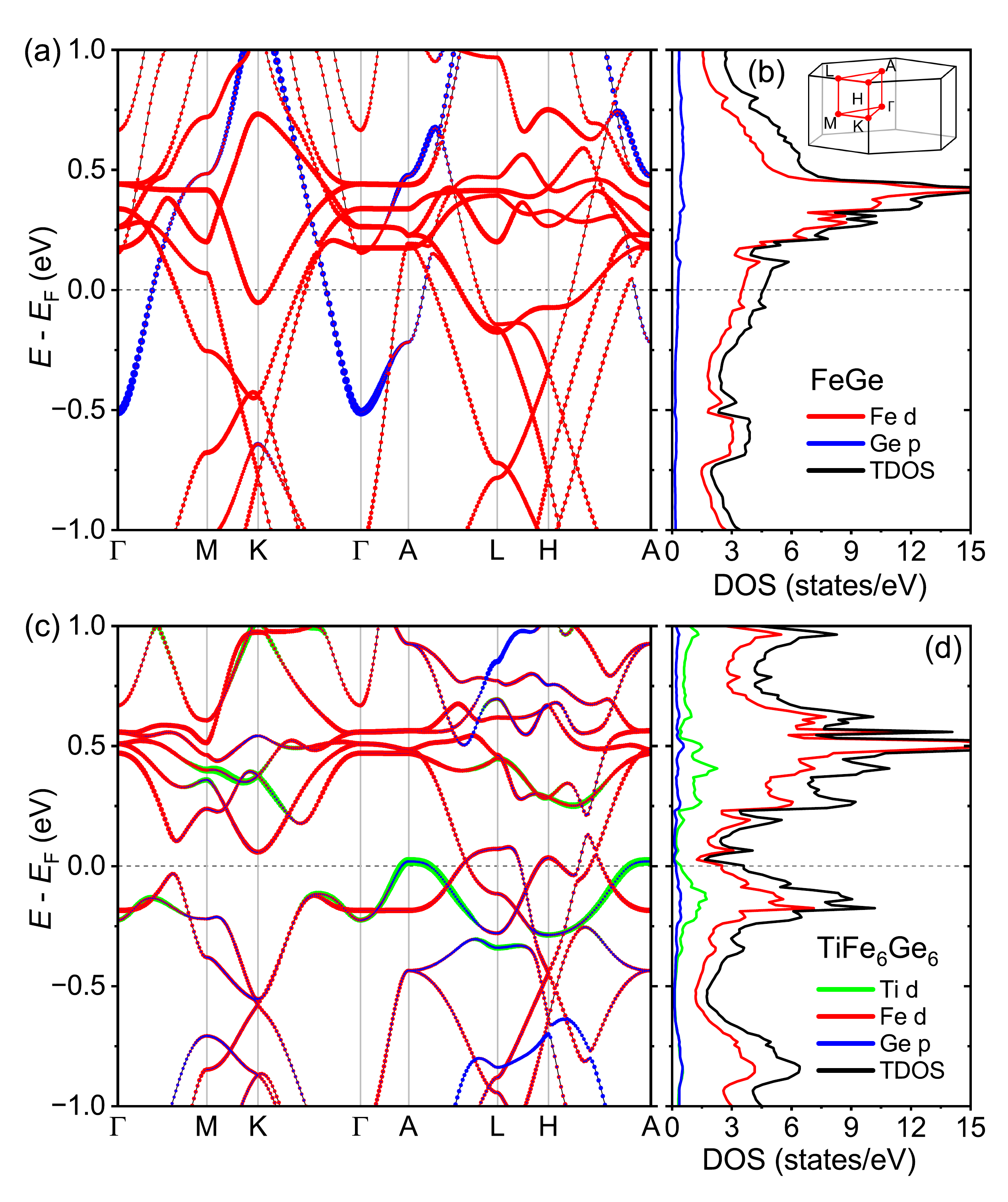}}
\caption{The orbital-projected band structure (a) and corresponding DOS (b) for FeGe. The inset of (b) shows the Brillouin zone and high-symmetry lines. The orbital-projected band structure (c) and corresponding DOS (d) for \ce{TiFe6Ge6}.}
\label{fig6}
\end{figure}

To understand the experimental results and the reason why a CDW is absent in \ce{Ti_{0.85}Fe6Ge6}, electronic structures of both FeGe and stoichiometric \ce{TiFe6Ge6} were calculated based on density functional theory, where the possible effect of Ti deficiency will be discussed in Section I.\@ Shown in Fig.\@ \ref{fig6}(a) is the orbital-projected band structure of FeGe,\@ where the bands around the Fermi level ($E_\mathrm{F}$) are dominated by the kagome lattice-derived Fe-$d$ orbitals and consistent with previous reports \cite{wu_electron_2023,shao_intertwining_2023,teng_magnetism_2023}.\@ Due to the A-type magnetic structure (ferromagnetic intralayer interaction and antiferromagnetic interlayer interaction), the bands of FeGe are split into spin-majority and spin-minority bands, resulting in multiple VHSs at the M/L point, the Dirac cones at the K/H points, and the flat bands in the range $E =$ 0.25 $\sim$ 0.5 eV. In the case of \ce{LiFe6Ge6}, Li-ion barely affects the band dispersion of FeGe except by providing one extra valence electron, and the band structure of \ce{LiFe6Ge6} can be obtained by an approximate rigid downward shift of FeGe bands by 0.12 eV \cite{wang_encoding_2024}. However, as shown in Fig.\@ \ref{fig6}(c),\@ the band dispersion of FeGe (and \ce{LiFe6Ge6}) is significantly modified in the case of Ti stuffing. The bands move upward/downward from $E_\mathrm{F}$, forming an indirect gap about 90 meV along the $\Gamma$--M--K--$\Gamma$ line ($k_z =$ 0 plane). For the A--L--H--A line ($k_z = \pi$ plane), although most bands are gapped, some bands are still crossing $E_\mathrm{F}$.\@ The primary feature for bands at the $k_z = \pi$ plane is the appearance of a band with a dominant Ti-$d$ orbital character around the $A$ point. The Ti stuffing-induced electronic structure change can be better recognized from the perspective of the density of states (DOSs). As shown in Figs.\@ \ref{fig6}(b) and \ref{fig6}(d),\@ it can be seen that, instead of the rigid shift expected by carrier doping, the states of \ce{TiFe6Ge6} move away from $E_\mathrm{F}$, forming a minimum at $E_\mathrm{F}$ and a strong peak at $E =$ -0.15 eV.

The band structure of \ce{TiFe6Ge6} is quite different from the sister compounds \ce{LiFe6Ge6} and \ce{LuFe6Ge6} \cite{wang_encoding_2024,sinha_twisting_2021},\@ and the established kagome CDW material \ce{ScV6Sn6} and \ce{LuNb6Sn6} \cite{cao_competing_2023,ortiz_stability_2025}.\@ There are two key differences that can be immediately concluded from the comparison of the band structure between these compounds: (1) Only carrier-doping induced rigid shifts of FeGe-derived bands were observed in the sister compounds \ce{LiFe6Ge6} and \ce{LuFe6Ge6}; (2) Compared with Ti, the orbital contribution from the filler atoms (Li, Sc, Lu) to the DOS around $E_\mathrm{F}$ was very weak \cite{wang_encoding_2024,cao_competing_2023,sinha_twisting_2021}. 

\begin{figure}
\centerline{\includegraphics[width=8.3cm]{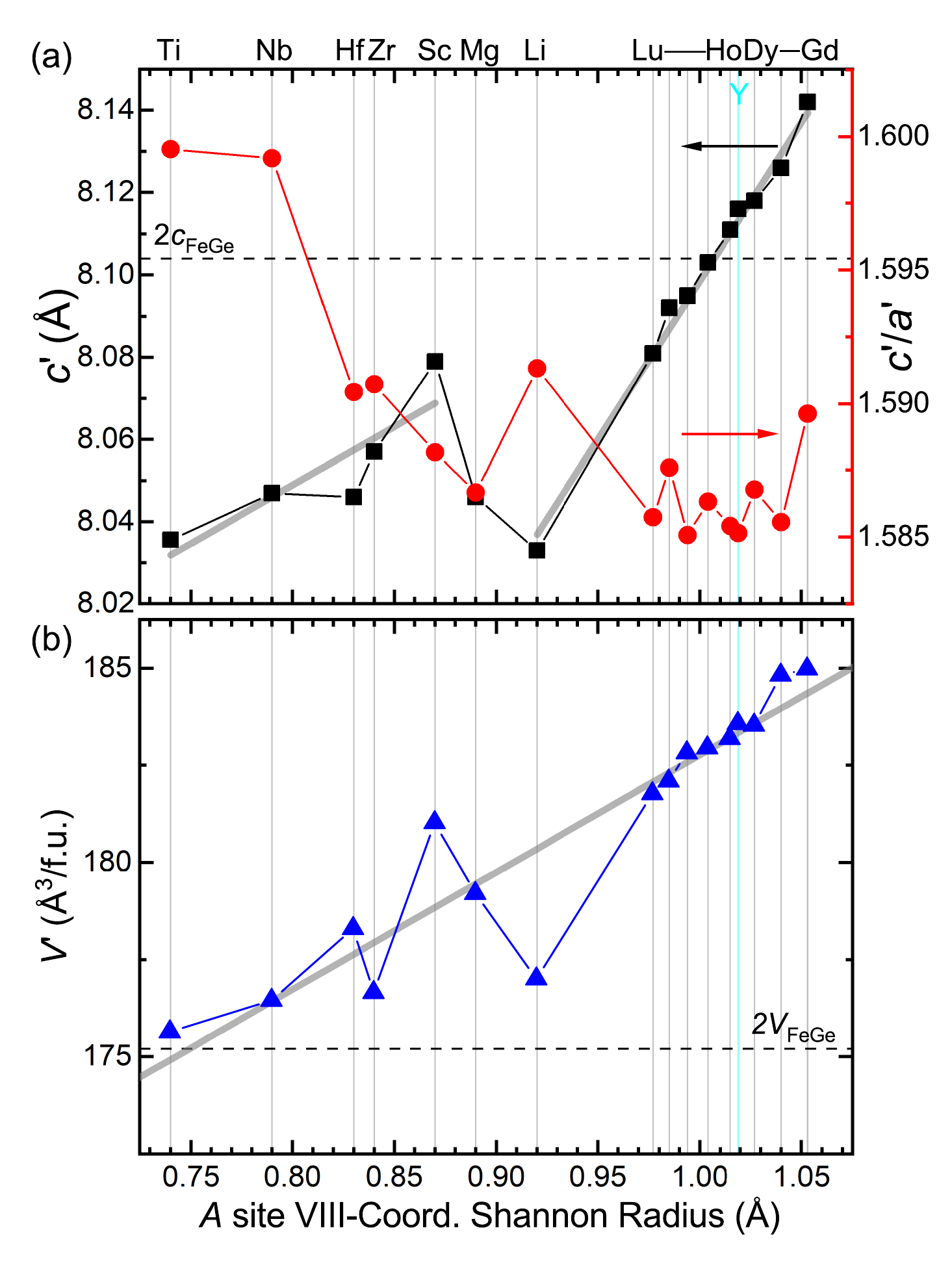}}
\caption{The evolution of the lattice parameters ($c^{\prime}$, $c^{\prime}/a^{\prime}$, and $V^{\prime}$) for all known \ce{$A$Fe6Ge6} compounds as a function of the ion radius of the filler atom $A$. Although \ce{$A$Fe6Ge6} compounds have a similar crystal structure, the unit cells are different due to the difference in detailed symmetry.  For the convenience of comparison, the lattice parameters are reduced to a \ce{HfFe6Ge6}-type unit cell, where $c^{\prime}$, $a^{\prime}$, and $V^{\prime}$ corresponding to 2$c_\mathrm{FeGe}$ (dash line in (a)), $a_\mathrm{FeGe}$, and 2$V_\mathrm{FeGe}$. Grey lines are guides to the eyes. The cyan line represents the Yttrium element which inserts between Ho and Dy, and breaks the sequence of Gd--Lu.
 }
\label{fig7}
\end{figure}

\begin{table*}
  \caption{Lattice parameters of known \ce{$R$Fe6Ge6} compounds. } 
  \centering
  \label{table3}  
  \setlength{\tabcolsep}{2.4mm}{
  \begin{tabular}{lllllllllllc}
    \hline\hline
     Compound & S.G. &$r_{\mathrm{ion}}$(\AA) & $a$ (\AA) & $b$ (\AA) & $c$ (\AA) &  $a^{\prime}$(\AA) & $c^{\prime}$(\AA) &$V^{\prime}$({\AA}$^3$) & $c^{\prime}/a^{\prime}$  &$T_\mathrm{N}$(K) &Ref. \\
    \hline
    FeGe & $P6/mmm$&---& 4.996 &---& 4.052 &  4.996 &8.104 &175.2 &1.622 &410& \cite{wu_annealing-tunable_2024} \\
    \ce{Ti_{0.85}Fe6Ge6} &  $P6/mmm$ &0.74 & 5.024 &---& 8.036 &5.024&8.036& 175.6 &1.600&488&  our  \\
     \ce{NbFe6Ge6} &  $P6/mmm$ &0.79 & 5.032 &---& 8.047 & 5.032&8.047&176.5&1.599&561&  \cite{mazet_neutron_2000}  \\
     \ce{HfFe6Ge6} &  $P6/mmm$ &0.83 & 5.059 &---& 8.046 & 5.059&8.046&178.3 &1.590& 453& \cite{mazet_neutron_2000}  \\
     \ce{ZrFe6Ge6} &  $P6/mmm$ &0.84 & 5.065 &---& 8.057 & 5.065&8.057&176.7 &1.591& 508& \cite{mazet_neutron_2000}  \\
     \ce{ScFe6Ge6} & $P6/mmm$&0.87 & 5.087 &---& 8.079 & 5.087&8.079&181.0 &1.588&497&   \cite{venturini_crystallographic_1992}   \\
     \ce{MgFe6Ge6} &  $P6/mmm$ &0.89 & 5.071 &---& 8.046 & 5.071&8.046&179.2 &1.587&501&  \cite{mazet_magnetic_2013}   \\
     \ce{LiFe6Ge6} &  $P6/mmm$ &0.92 & 8.744 &---& 8.033 & 5.048&8.033&177.0 &1.591& 575& \cite{welk_zur_1976}   \\
     \ce{LuFe6Ge6} & $P6/mmm$&0.977 & 5.096 &---& 8.081 & 5.096&8.081&181.8 &1.586& 471&  \cite{venturini_crystallographic_1992}   \\
     \ce{YbFe6Ge6} & $P6/mmm$& 0.985 & 5.097 &---& 8.092 & 5.097&8.092&182.1 &1.588& 475& \cite{venturini_crystallographic_1992}   \\
     \ce{TmFe6Ge6} &$Immm$& 0.994 & 8.095 &26.53& 5.107 & 5.107&8.095&182.8 &1.585& 465& \cite{venturini_crystallographic_1992}   \\
     \ce{ErFe6Ge6} & $Immm$&1.004 & 8.103 &26.52& 5.108 & 5.108&8.103&183.0 &1.586& 470& \cite{venturini_crystallographic_1992}   \\
     \ce{HoFe6Ge6} & $Cmcm$&1.015 & 8.111 &17.66& 5.116 &5.116&8.111& 183.2 &1.585& 473& \cite{venturini_crystallographic_1992}   \\
     \ce{YFe6Ge6} &$Cmcm$& 1.019 & 8.116 &17.67& 5.12 & 5.12&8.116&183.6 &1.585& 477& \cite{venturini_crystallographic_1992}   \\
     \ce{DyFe6Ge6} & $Cmcm$&1.027 & 8.118 &17.68& 5.116 & 5.116&8.118&183.5 &1.587& 475& \cite{venturini_crystallographic_1992}   \\
     \ce{TbFe6Ge6} & $Cmcm$&1.04 & 8.126 &17.76& 5.125 & 5.125&8.126& 184.8 &1.586&480&  \cite{venturini_crystallographic_1992}  \\
     \ce{GdFe6Ge6} &  $P6/mmm$&1.053 & 5.122 &---& 4.071 & 5.122&8.142&185.0 &1.590 & 482&\cite{cadogan_155gd_2007}   \\
    \hline\hline
  \end{tabular}}\\
$a^{\prime}$, $c^{\prime}$, $V^{\prime}$ are reduced lattice parameters and volumes, corresponding to a \ce{HfFe6Ge6}-type unit cell.  
\end{table*}

\subsection{Structural Trends of \texorpdfstring{\ce{$A$Fe6Ge6}}{}}

To further explore the doping mechanism of the filler atoms, we construct phase diagrams of lattice parameters ($a^{\prime}$, $c^{\prime}$, and $V^{\prime}$) vs $A$ site ion radius [$r(A)$] for known \ce{$A$Fe6Ge6} compounds, where the $a^{\prime}$, $c^{\prime}$, and $V^{\prime}$ are the reduced lattice parameters corresponding to a \ce{HfFe6Ge6}-type unit cell. The results are plotted in Fig.\@ \ref{fig7} and the corresponding data are listed in Table \ref{table3}. For $c^\prime$ vs $r(A)$ data shown in Fig.\@ \ref{fig7}(a),\@ it is evident that the elements on the left part of the phase diagram [$r < r$(Li), containing Ti and other transition metals such as Nb, Hf, Zr, and Sc] and elements on the right of the phase diagram [$r \geq r$(Li), containing Li and rare earth such as Y, and Gd-Lu] follow different trends, which can be described by linear behaviors with different slopes. As shown by the right $y$-axis of Fig.\@ \ref{fig7}(a),\@ with $r(A)$ increasing from $r$(Ti) to $r$(Mg), $c^{\prime}/a^{\prime}$ quickly decreases, and becomes flat when $r \geq r$(Mg). Note that Li and Gd deviate from this trend. Interestingly, although both $c^{\prime}$ and $c^{\prime}/a^{\prime}$ show a strong $r(A)$ dependence, the $V^{\prime}$ vs $r(A)$ points generally fall on a straight line [Fig.\@ \ref{fig7}(b)].
These results indicate that the volume of \ce{$A$Fe6Ge6} is primarily determined by the ionic radius of $A$. In contrast, the interlayer distance of the kagome planes and the $c^{\prime}/a^{\prime}$ ratio are more significantly influenced by chemical factors other than the ionic radius. It is noteworthy that on the right part of Fig.\@ \ref{fig7} are elements with small electronegativity and no $d$ electrons, while on the left part of Fig. \ref{fig7} are elements with relatively large electronegativity and $d$ electrons. Since electronegativity and $d$ electrons can significantly influence bonding in materials, the differences in electronic structure between \ce{LiFe6Ge6} and \ce{TiFe6Ge6} may be explained in terms of chemical bonding. Indeed, the covalent interaction between $R$ $nd-$Fe $3d$ electrons was reported to be responsible for the increase in Fe moment and Fe hyperfine fields in \ce{$R$Fe6Ge6} series \cite{mazet_covalent_2003}.

\subsection{DOS and COHP analysis}
\begin{figure*}
\centerline{\includegraphics[width=18cm]{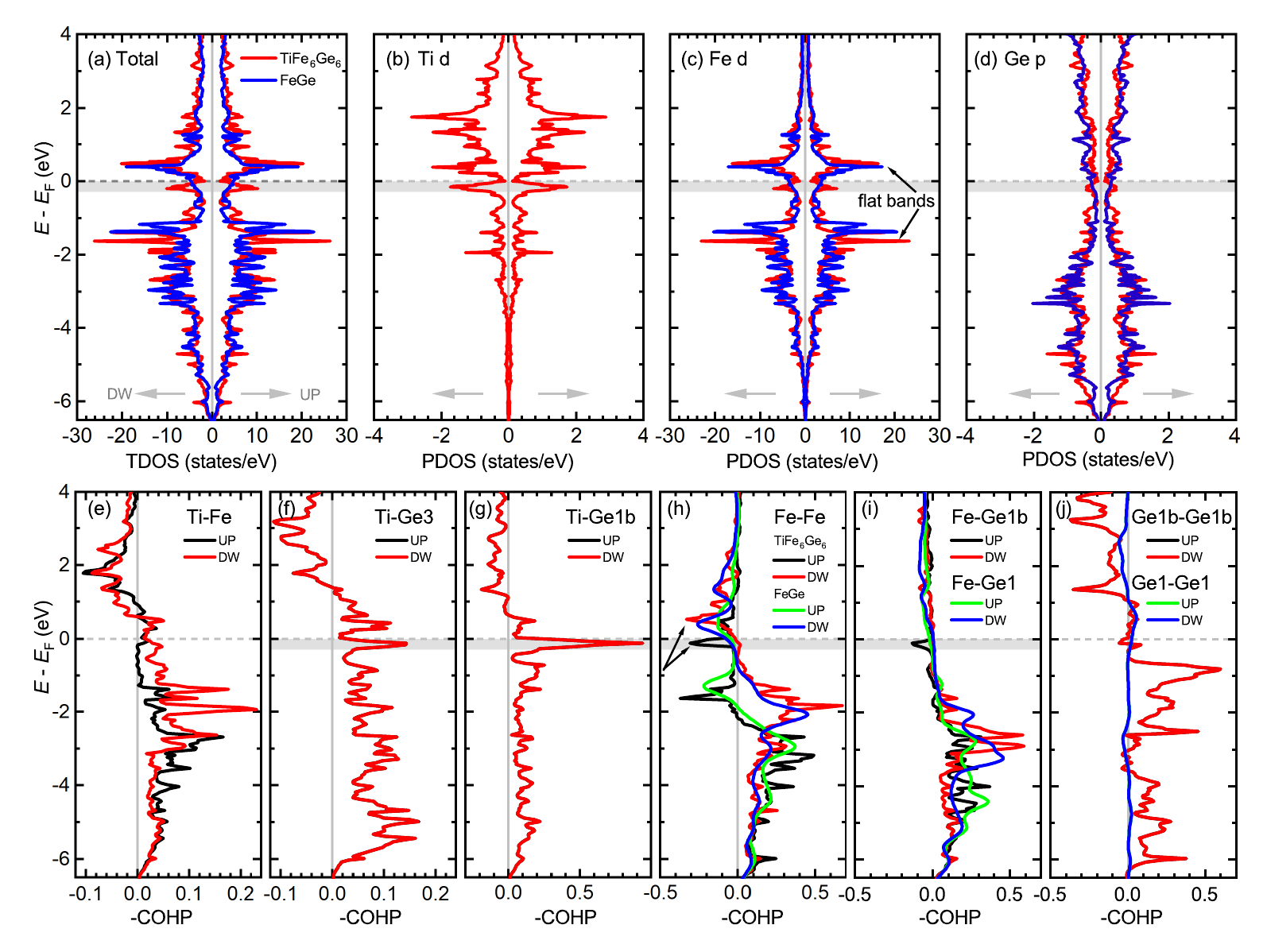}}
\caption{ (a) Total density of states (TDOS) for FeGe and \ce{TiFe6Ge6}. Partial density of states (PDOS) for (b) Ti $d$, (c) Fe $d$, (d) Ge $p$ orbitals. Panels (a-d) use the same legend which is shown in (a). In panels (a-d), red lines and blue lines represent the data for FeGe and \ce{TiFe6Ge6}; $\rightarrow$ and $\leftarrow$ denote the spin-up (UP) and spin-down (DW) states, respectively. Crystal orbital Hamilton population curves for bonding of (a) \ce{Ti-Fe}, (b) \ce{Ti-{Ge3}}, (c) \ce{Ti-{Ge1b}}, (d) \ce{Fe-Fe}, (e) \ce{Fe-{Ge1b}} for \ce{TiFe6Ge6} and \ce{Fe-{Ge1}} for FeGe, (f) \ce{{Ge1b}-{Ge1b}} for \ce{TiFe6Ge6} and \ce{{Ge1}-{Ge1}} for FeGe. The crystallographic atomic labels can be found in Fig. \ref{fig1}.  Panels (e-j) use the same legend as shown in (h), where black and red lines represent the data for \ce{TiFe6Ge6}, green and blue lines represent the data for FeGe; note that no spin splitting is observed in panels (f), (g), and (j). The shaded areas mark the peak at $E =$ -0.15 eV.
 }
\label{fig8}
\end{figure*}

Figure \ref{fig8}(a-d) shows the total and partial density of states (DOSs) on a larger energy scale for \ce{TiFe6Ge6} (red lines) and FeGe (blue lines), respectively. One can see that the DOS of the spin-up and spin-down electrons are equal for both FeGe and \ce{TiFe6Ge6}, which is the consequence of the preservation of the $\mathcal{P}\mathcal{T}$ symmetry (i.e., the combination of space inversion $\mathcal{P}$ and time-reversal $\mathcal{T}$) in the A-type antiferromagnetic phase. The Ti doping-induced change can be better visualized by taking the sharp peaks as the anchor points, as marked by the arrows in Fig.\@ \ref{fig8}(c).\@ These sharp peaks correspond to the kagome lattice-derived flat bands (the upper peak around 0.45 eV for the spin minority sublattice, and the lower peak around -1.5 eV for the spin majority sublattice) \cite{teng_magnetism_2023}. Compared with FeGe, the upper peaks in \ce{TiFe6Ge6} shift upward by 0.1 eV while the lower peak shifts downward by 0.25 eV, i.e., the spin splitting between spin minority and spin majority flat bands is increased from 1.8 eV in FeGe to 2.1 eV in \ce{TiFe6Ge6}. This result is consistent with the increase in $T_\mathrm{N}$ from 410 K in FeGe to 488 K in \ce{Ti_{0.85}Fe6Ge6} and the absence of a notable carrier-doping-induced rigid shift. Accompanied by the decrease of the TDOS($E_\mathrm{F}$) from 4.6 states/eV in FeGe to 3.6 states/eV in \ce{TiFe6Ge6}, a sharp peak appears at $E =$ -0.15 eV. Interestingly, the area of TDOS enclosed by these sharp peaks is integrated to be 3.7 electrons per chemical formula, close to 4 electrons expected to be doped by \ce{Ti^{4+}}.\@ This result indicates that, in contrast to the delocalized electrons observed in \ce{LiFe6Ge6} and \ce{LuFe6Ge6} \cite{wang_encoding_2024,sinha_twisting_2021},\@ the excess electrons contributed by Ti atoms in \ce{TiFe6Ge6} are localized around $E = -0.15$ eV---and thus spatially confined around the Ti sites---resembling the behavior observed in LuGe \cite{freccero_excess_2021}.\@ This localization provides a convincing explanation for the absence of a rigid band shift in \ce{TiFe6Ge6}.\@ As shown in Figs.\@ \ref{fig8}(b-d),\@ the PDOS of Ti, Fe, and Ge exhibit a similar peak at $E =$ -0.15 eV, indicating that it might arise from the chemical bonds involving the Ti state.

The bonding in \ce{TiFe6Ge6} and FeGe is further analyzed by the crystal orbital Hamilton population (COHP) method. The values for the integrated crystal orbital Hamilton population (ICOHP) for possible bonds in \ce{TiFe6Ge6} and FeGe are presented in Tables S1 and S2 in the Supplemental Material \cite{SM},\@ and the main results are plotted in Figs.\@ \ref{fig8}(e-j).\@ From Figs.\@ \ref{fig8}(e-j),\@ it can be seen that the Fermi level is located in the bonding regions for \ce{Ti-{Ge3}}, and \ce{Ti-{Ge1b}}, antibonding regions for \ce{Fe-Fe}, \ce{Fe-{Ge1b}}/\ce{Fe-{Ge1}}. Spin splitting can be observed for bonds associated with Fe [see Figs. \ref{fig8}(e), \ref{fig8}(h), \ref{fig8}(i)].\@ Sharp peaks near -0.15 eV,\@ which correspond to the same energy observed in the DOS,\@ are also clearly visible in the COHP curves for the \ce{Ti-{Ge3}},\@ \ce{Ti-{Ge1b}},\@ \ce{Fe-Fe},\@ and \ce{Fe-{Ge1b}}/\ce{Fe-{Ge1}} bonds.\@ These features are indicated by the shaded region below $E_\mathrm{F}$. The peak height for the \ce{Ti-{Ge1b}} bond is 0.94, much higher than that of \ce{Ti-{Ge3}} (0.14), \ce{Fe-Fe} (-0.31), and \ce{Fe-{Ge1b}} (-0.13), indicative of the strong covalent bonding between Ti and Ge1b.\@ Therefore, the excess electrons introduced by \ce{Ti^4+} are mainly localized in \ce{Ti-{Ge1b}} bonds, rather than spreading out across the entire crystal, providing a natural explanation for the absence of a carrier-doping-induced rigid shift in the electronic structure.

The interest in the CDW in FeGe is mainly due to its correlation with magnetism. The Ti doping-induced change in magnetism can be understood in terms of the chemical bonds associated with Fe. As shown in Fig.\@ \ref{fig8}(h),\@ the broad antibonding peaks for \ce{Fe-Fe} bonds in FeGe are replaced by narrow peaks in \ce{TiFe6Ge6}, which exhibit enhanced spin splitting in energy, as indicated by the arrows in Fig.\@ \ref{fig8}(h).\@ In the itinerant magnetism scenario, the magnetic ordering with ferromagnetic nearest-neighbor interaction correlates with the presence of strongly antibonding states at the Fermi level, where the system is electronically stabilized by spin splitting instead of a structural transition  \cite{landrum_orbital_2000,landrum_ferromagnetism_1999}.\@ Notably, the Fermi level crosses the antibonding peak of the COHP curve of \ce{TiFe6Ge6}, resulting in a remarkable increase in antibonding interaction between \ce{Fe-Fe} within the kagome layer. The enhancement in antibonding interaction strength can also be inferred from the spin splitting in ICOHP, where the values for the spin up/down ICOHP data are increased from 1.11/1.29 eV per bond in FeGe to 1.07/1.33 eV per bond in \ce{TiFe6Ge6} [Table S1 and S2 in Supplemental Material \cite{SM}].\@ This result suggests that spin polarization should be significantly enhanced in \ce{TiFe6Ge6},\@ which is corroborated by the increase of the calculated magnetic moment from 1.50 in FeGe to 1.97 in \ce{TiFe6Ge6} and manifested as the depletion of the density of states around $E_\mathrm{F}$ in the electronic structure. Thus, the absence of a CDW in \ce{TiFe6Ge6} can be readily understood within the framework of the magnetic energy-saving model: the magnetic energy in \ce{TiFe6Ge6} is already lower than that of the CDW phase in FeGe, leaving no additional magnetic energy available to compensate for the structural distortion required by the CDW transition.

\subsection{Discussion}

It is now particularly intriguing to scrutinize the rattling chain model by incorporating the chemical bonding characteristics we have identified above: In the CDW phase of $AM_6X_6$ compounds, the primary feature of the structural distortion is the ordering of $X$–$X$ dimers \cite{ortiz_stability_2025}. The $X-X$ dimer and adjacent $A$ atoms can be regarded as the $A-X-X-A$ chains located in the channels in the hexagonal voids in the kagome and honeycomb layers \cite{meier_tiny_2023}.\@ The precondition for the occurrence of a CDW is the small size of $A$ atoms, which could provide extra room for the atoms in the chains to rattle.  In the case of \ce{TiFe6Ge6},\@ from the COHP curves shown in Fig.\@ \ref{fig8}(j),\@ the clear signature for the formation of bonds between \ce{{Ge1b}-{Ge1b}} atoms in \ce{TiFe6Ge6} can be observed, in contrast with no bonding between \ce{{Ge1}-{Ge1}} atoms in FeGe.  The bond distance for \ce{{Ge1b}-{Ge1b}} in \ce{TiFe6Ge6} is 2.619 {\AA}, which is shorter than 2.669 {\AA} for the Ge-Ge dimer in the CDW phase of FeGe. This result suggests that \ce{TiFe6Ge6} represents the fully (100 \%) dimerized phase of FeGe, whereas the CDW phase of FeGe can be regarded as a partially (1/4) dimerized state, thereby preventing a phase transition driven solely by dimer formation.

Our analysis of \ce{TiFe6Ge6} above demonstrates the existence of bonds between \ce{Ti-{Ge1b}}, \ce{{Ge1b}-{Ge1b}}, and \ce{Ti-{Ge3}} atoms, suggesting that the \ce{Ti-{Ge1b}-{Ge1b}-Ti} structure forms a stiff chain that cannot ''rattle" due to the strong bonding interactions within the chain. Additionally, the \ce{Ti-{Ge3}} bonds can serve as the anchor that secure the \ce{Ti-{Ge1b}-{Ge1b}-Ti} chain to the \ce{Fe6Ge6} host. This result underscores that, beyond ionic radii, the bonding characteristics of the filler atoms must also be taken into account when exploring CDW phenomena in $AM_6X_6$ compounds.

Since the CDW transition in FeGe is highly sensitive to annealing-induced disorder, and significant Ti deficiencies have been identified in \ce{Ti_{0.85}Fe6Ge6}, it is natural to question whether a CDW transition could occur in stoichiometric \ce{TiFe6Ge6}. In the following, we demonstrate that the absence of CDW order in \ce{TiFe6Ge6} is intrinsic, as supported by existing CDW models, and is not attributable to Ti deficiency: (1) From the perspective of the rattling chain model, the CDW transition is suppressed by the formation of 85 \% stiff \ce{Ti-{Ge1b}-{Ge1b}-Ti} chains. The stoichiometric compound \ce{TiFe6Ge6}, with 100 \% stiff chains, therefore represents a more stable phase. In contrast, in the prototypical rattling-chain material \ce{ScV6Sn6}, the CDW is rapidly suppressed when Sc is replaced by larger atoms such as Y or Lu \cite{meier_tiny_2023}, whereas substitution with smaller atoms favors the CDW. By analogy, treating Ti deficiency as vacancy ($\square$) doping, i.e., \ce{Ti_{1-x}$\square$_xFe6Ge6}, suggests that Ti deficiency promotes the CDW transition. Indeed, in the extreme Ti-deficient limit, the CDW reemerges in \ce{Ti_{1-x}$\square$_xFe6Ge6}. (2) Within the framework of the magnetic saving model, the CDW transition in FeGe is governed by the competition between magnetic energy gain and the structural energy penalty linked to the dimerization of one quarter of the Ge1 atoms. In \ce{Ti_{0.85}Fe6Ge6}, where 85 \% of Ge1 atoms form dimers, the system resides in a regime where the magnetic energy gain outweighs the structural cost, resulting in stability. The ideal stoichiometric phase, \ce{TiFe6Ge6}, which exhibits a higher $T_\mathrm{N}$, should therefore be even more stable than \ce{Ti_{0.85}Fe6Ge6}.

In our attempts to grow stoichiometric \ce{TiFe6Ge6} single crystals, we obtained nearly fully Ti-occupied \ce{Ti_{0.95}Fe6Ge6}. However, compared to \ce{Ti_{0.85}Fe6Ge6}, the crystals exhibit a needle-like rather than plate-like morphology, with significantly reduced yield and crystal size. Across the available \ce{Ti_xFe6Ge6} compositions ($0.6 < x < 0.95$), crystals with $x \sim 0.8$ are energetically favored during CVT growth, typically yielding larger crystals with higher growth efficiency. This observation is consistent with the magnetic energy saving model, which predicts multiple energetic minima arising from the competition between magnetic energy gain and structural distortion \cite{wang_enhanced_2023,wang_encoding_2024}. In this framework, \ce{Ti_{0.85}Fe6Ge6}, with $\sim$80 \% Ge1 dimers, may represent another local energetic minimum, complementing the 25 \% Ge1-dimer state associated with the CDW phase in FeGe, thereby facilitating the preferential growth of Ti-deficient crystals.

\section{Conclusion}
In summary, we have successfully grown \ce{Ti_{0.85}Fe6Ge6} single crystals using the chemical transport method. The magnetization is consistent with A-type antiferromagnetism with an ordering temperature of 488 K. A metallic resistivity with a negligibly small magnetoresistance and a signature of incoherence-coherence crossover is observed, which is dominated by the hole-type carrier. No signature of a CDW can be identified in structure, magnetization, and transport measurements. DFT calculation reveals an electronic structure very different from the sister compounds, i.e., the DOS is pushed away from the Fermi level, and a gap opens around the Fermi level, accompanied by the appearance of a band with Ti orbital character. The absence of the CDW and the distinctive electronic structure can be well understood in terms of chemical bonds within \ce{Ti_{0.85}Fe6Ge6}.\@ Our results highlight the pivotal role of chemical bonds in the emergence of CDWs in kagome materials, offering crucial supplementary insights to the established rattling chain model.

\begin{acknowledgements}
We thank Yajun Yan for her helpful discussions. The work at Chongqing University was supported by the Natural Science Foundation of China (No. 12474142) and the New Chongqing Youth Innovative Talent Project (No. 2024NSCQ-qncxX0474). We would like to thank Guiwen Wang, Yan Liu, and Xiangnan Gong at the Analytical and Testing Center of Chongqing University for their assistance with measurements as well as Siegmar Roth and Andre Beck at the Institute for Quantum Materials and Technologies, Karlsruhe Institute of Technology for technical support of x-ray diffraction.
\end{acknowledgements}


\begin{thebibliography}{69}%
\makeatletter
\providecommand \@ifxundefined [1]{%
 \@ifx{#1\undefined}
}%
\providecommand \@ifnum [1]{%
 \ifnum #1\expandafter \@firstoftwo
 \else \expandafter \@secondoftwo
 \fi
}%
\providecommand \@ifx [1]{%
 \ifx #1\expandafter \@firstoftwo
 \else \expandafter \@secondoftwo
 \fi
}%
\providecommand \natexlab [1]{#1}%
\providecommand \enquote  [1]{``#1''}%
\providecommand \bibnamefont  [1]{#1}%
\providecommand \bibfnamefont [1]{#1}%
\providecommand \citenamefont [1]{#1}%
\providecommand \href@noop [0]{\@secondoftwo}%
\providecommand \href [0]{\begingroup \@sanitize@url \@href}%
\providecommand \@href[1]{\@@startlink{#1}\@@href}%
\providecommand \@@href[1]{\endgroup#1\@@endlink}%
\providecommand \@sanitize@url [0]{\catcode `\\12\catcode `\$12\catcode
  `\&12\catcode `\#12\catcode `\^12\catcode `\_12\catcode `\%12\relax}%
\providecommand \@@startlink[1]{}%
\providecommand \@@endlink[0]{}%
\providecommand \url  [0]{\begingroup\@sanitize@url \@url }%
\providecommand \@url [1]{\endgroup\@href {#1}{\urlprefix }}%
\providecommand \urlprefix  [0]{URL }%
\providecommand \Eprint [0]{\href }%
\providecommand \doibase [0]{https://doi.org/}%
\providecommand \selectlanguage [0]{\@gobble}%
\providecommand \bibinfo  [0]{\@secondoftwo}%
\providecommand \bibfield  [0]{\@secondoftwo}%
\providecommand \translation [1]{[#1]}%
\providecommand \BibitemOpen [0]{}%
\providecommand \bibitemStop [0]{}%
\providecommand \bibitemNoStop [0]{.\EOS\space}%
\providecommand \EOS [0]{\spacefactor3000\relax}%
\providecommand \BibitemShut  [1]{\csname bibitem#1\endcsname}%
\let\auto@bib@innerbib\@empty
\bibitem [{\citenamefont {Yin}\ \emph {et~al.}(2022)\citenamefont {Yin},
  \citenamefont {Lian},\ and\ \citenamefont {Hasan}}]{yin_topological_2022}%
  \BibitemOpen
  \bibfield  {author} {\bibinfo {author} {\bibfnamefont {J.-X.}\ \bibnamefont
  {Yin}}, \bibinfo {author} {\bibfnamefont {B.}~\bibnamefont {Lian}},\ and\
  \bibinfo {author} {\bibfnamefont {M.~Z.}\ \bibnamefont {Hasan}},\ }\bibfield
  {title} {\bibinfo {title} {Topological kagome magnets and superconductors},\
  }\href {https://doi.org/10.1038/s41586-022-05516-0} {\bibfield  {journal}
  {\bibinfo  {journal} {Nature}\ }\textbf {\bibinfo {volume} {612}},\ \bibinfo
  {pages} {647} (\bibinfo {year} {2022})}\BibitemShut {NoStop}%
\bibitem [{\citenamefont {Wang}\ \emph {et~al.}(2023)\citenamefont {Wang},
  \citenamefont {Wu}, \citenamefont {McCandless}, \citenamefont {Chan},\ and\
  \citenamefont {Ali}}]{wang_quantum_2023}%
  \BibitemOpen
  \bibfield  {author} {\bibinfo {author} {\bibfnamefont {Y.}~\bibnamefont
  {Wang}}, \bibinfo {author} {\bibfnamefont {H.}~\bibnamefont {Wu}}, \bibinfo
  {author} {\bibfnamefont {G.~T.}\ \bibnamefont {McCandless}}, \bibinfo
  {author} {\bibfnamefont {J.~Y.}\ \bibnamefont {Chan}},\ and\ \bibinfo
  {author} {\bibfnamefont {M.~N.}\ \bibnamefont {Ali}},\ }\bibfield  {title}
  {\bibinfo {title} {Quantum states and intertwining phases in kagome
  materials},\ }\href {https://doi.org/10.1038/s42254-023-00635-7} {\bibfield
  {journal} {\bibinfo  {journal} {Nature Reviews Physics}\ }\textbf {\bibinfo
  {volume} {5}},\ \bibinfo {pages} {635} (\bibinfo {year} {2023})}\BibitemShut
  {NoStop}%
\bibitem [{\citenamefont {Wilson}\ and\ \citenamefont
  {Ortiz}(2024)}]{wilson_av3sb5_2024}%
  \BibitemOpen
  \bibfield  {author} {\bibinfo {author} {\bibfnamefont {S.~D.}\ \bibnamefont
  {Wilson}}\ and\ \bibinfo {author} {\bibfnamefont {B.~R.}\ \bibnamefont
  {Ortiz}},\ }\bibfield  {title} {\bibinfo {title} {{AV$_3$Sb$_5$} kagome
  superconductors},\ }\href {https://doi.org/10.1038/s41578-024-00677-y}
  {\bibfield  {journal} {\bibinfo  {journal} {Nature Reviews Materials}\
  }\textbf {\bibinfo {volume} {9}},\ \bibinfo {pages} {420} (\bibinfo {year}
  {2024})}\BibitemShut {NoStop}%
\bibitem [{\citenamefont {Wang}\ \emph {et~al.}(2024)\citenamefont {Wang},
  \citenamefont {Lei}, \citenamefont {Qi},\ and\ \citenamefont
  {Felser}}]{wang_topological_2024}%
  \BibitemOpen
  \bibfield  {author} {\bibinfo {author} {\bibfnamefont {Q.}~\bibnamefont
  {Wang}}, \bibinfo {author} {\bibfnamefont {H.}~\bibnamefont {Lei}}, \bibinfo
  {author} {\bibfnamefont {Y.}~\bibnamefont {Qi}},\ and\ \bibinfo {author}
  {\bibfnamefont {C.}~\bibnamefont {Felser}},\ }\bibfield  {title} {\bibinfo
  {title} {Topological {Quantum} {Materials} with {Kagome} {Lattice}},\ }\href
  {https://doi.org/10.1021/accountsmr.3c00291} {\bibfield  {journal} {\bibinfo
  {journal} {Accounts of Materials Research}\ }\textbf {\bibinfo {volume}
  {5}},\ \bibinfo {pages} {786} (\bibinfo {year} {2024})}\BibitemShut {NoStop}%
\bibitem [{\citenamefont {Xu}\ \emph {et~al.}(2023)\citenamefont {Xu},
  \citenamefont {Yin}, \citenamefont {Qu},\ and\ \citenamefont
  {Jia}}]{xu_quantum_2023}%
  \BibitemOpen
  \bibfield  {author} {\bibinfo {author} {\bibfnamefont {X.}~\bibnamefont
  {Xu}}, \bibinfo {author} {\bibfnamefont {J.-X.}\ \bibnamefont {Yin}},
  \bibinfo {author} {\bibfnamefont {Z.}~\bibnamefont {Qu}},\ and\ \bibinfo
  {author} {\bibfnamefont {S.}~\bibnamefont {Jia}},\ }\bibfield  {title}
  {\bibinfo {title} {Quantum interactions in topological {R166} kagome
  magnet},\ }\href {https://doi.org/10.1088/1361-6633/acfd3d} {\bibfield
  {journal} {\bibinfo  {journal} {Reports on Progress in Physics}\ }\textbf
  {\bibinfo {volume} {86}},\ \bibinfo {pages} {114502} (\bibinfo {year}
  {2023})}\BibitemShut {NoStop}%
\bibitem [{\citenamefont {Jiang}\ \emph {et~al.}(2023)\citenamefont {Jiang},
  \citenamefont {Wu}, \citenamefont {Yin}, \citenamefont {Wang}, \citenamefont
  {Hasan}, \citenamefont {Wilson}, \citenamefont {Chen},\ and\ \citenamefont
  {Hu}}]{jiang_kagome_2023}%
  \BibitemOpen
  \bibfield  {author} {\bibinfo {author} {\bibfnamefont {K.}~\bibnamefont
  {Jiang}}, \bibinfo {author} {\bibfnamefont {T.}~\bibnamefont {Wu}}, \bibinfo
  {author} {\bibfnamefont {J.-X.}\ \bibnamefont {Yin}}, \bibinfo {author}
  {\bibfnamefont {Z.}~\bibnamefont {Wang}}, \bibinfo {author} {\bibfnamefont
  {M.~Z.}\ \bibnamefont {Hasan}}, \bibinfo {author} {\bibfnamefont {S.~D.}\
  \bibnamefont {Wilson}}, \bibinfo {author} {\bibfnamefont {X.}~\bibnamefont
  {Chen}},\ and\ \bibinfo {author} {\bibfnamefont {J.}~\bibnamefont {Hu}},\
  }\bibfield  {title} {\bibinfo {title} {Kagome superconductors
  {$A$V$_3$Sb$_5$} ({$A$} = {K}, {Rb}, {Cs})},\ }\href
  {https://doi.org/10.1093/nsr/nwac199} {\bibfield  {journal} {\bibinfo
  {journal} {National Science Review}\ }\textbf {\bibinfo {volume} {10}},\
  \bibinfo {pages} {nwac199} (\bibinfo {year} {2023})}\BibitemShut {NoStop}%
\bibitem [{\citenamefont {Ortiz}\ \emph {et~al.}(2020)\citenamefont {Ortiz},
  \citenamefont {Teicher}, \citenamefont {Hu}, \citenamefont {Zuo},
  \citenamefont {Sarte}, \citenamefont {Schueller}, \citenamefont {Abeykoon},
  \citenamefont {Krogstad}, \citenamefont {Rosenkranz}, \citenamefont {Osborn},
  \citenamefont {Seshadri}, \citenamefont {Balents}, \citenamefont {He},\ and\
  \citenamefont {Wilson}}]{OrtizCsVSb}%
  \BibitemOpen
  \bibfield  {author} {\bibinfo {author} {\bibfnamefont {B.~R.}\ \bibnamefont
  {Ortiz}}, \bibinfo {author} {\bibfnamefont {S.~M.~L.}\ \bibnamefont
  {Teicher}}, \bibinfo {author} {\bibfnamefont {Y.}~\bibnamefont {Hu}},
  \bibinfo {author} {\bibfnamefont {J.~L.}\ \bibnamefont {Zuo}}, \bibinfo
  {author} {\bibfnamefont {P.~M.}\ \bibnamefont {Sarte}}, \bibinfo {author}
  {\bibfnamefont {E.~C.}\ \bibnamefont {Schueller}}, \bibinfo {author}
  {\bibfnamefont {A.~M.~M.}\ \bibnamefont {Abeykoon}}, \bibinfo {author}
  {\bibfnamefont {M.~J.}\ \bibnamefont {Krogstad}}, \bibinfo {author}
  {\bibfnamefont {S.}~\bibnamefont {Rosenkranz}}, \bibinfo {author}
  {\bibfnamefont {R.}~\bibnamefont {Osborn}}, \bibinfo {author} {\bibfnamefont
  {R.}~\bibnamefont {Seshadri}}, \bibinfo {author} {\bibfnamefont
  {L.}~\bibnamefont {Balents}}, \bibinfo {author} {\bibfnamefont
  {J.}~\bibnamefont {He}},\ and\ \bibinfo {author} {\bibfnamefont {S.~D.}\
  \bibnamefont {Wilson}},\ }\bibfield  {title} {\bibinfo {title} {\ce{CsV3Sb5}:
  A ${\mathbb{z}}_{2}$ topological kagome metal with a superconducting ground
  state},\ }\href {https://doi.org/10.1103/PhysRevLett.125.247002} {\bibfield
  {journal} {\bibinfo  {journal} {Phys. Rev. Lett.}\ }\textbf {\bibinfo
  {volume} {125}},\ \bibinfo {pages} {247002} (\bibinfo {year}
  {2020})}\BibitemShut {NoStop}%
\bibitem [{\citenamefont {Arachchige}\ \emph {et~al.}(2022)\citenamefont
  {Arachchige}, \citenamefont {Meier}, \citenamefont {Marshall}, \citenamefont
  {Matsuoka}, \citenamefont {Xue}, \citenamefont {McGuire}, \citenamefont
  {Hermann}, \citenamefont {Cao},\ and\ \citenamefont
  {Mandrus}}]{arachchige_charge_2022}%
  \BibitemOpen
  \bibfield  {author} {\bibinfo {author} {\bibfnamefont {H.~W.~S.}\
  \bibnamefont {Arachchige}}, \bibinfo {author} {\bibfnamefont {W.~R.}\
  \bibnamefont {Meier}}, \bibinfo {author} {\bibfnamefont {M.}~\bibnamefont
  {Marshall}}, \bibinfo {author} {\bibfnamefont {T.}~\bibnamefont {Matsuoka}},
  \bibinfo {author} {\bibfnamefont {R.}~\bibnamefont {Xue}}, \bibinfo {author}
  {\bibfnamefont {M.~A.}\ \bibnamefont {McGuire}}, \bibinfo {author}
  {\bibfnamefont {R.~P.}\ \bibnamefont {Hermann}}, \bibinfo {author}
  {\bibfnamefont {H.}~\bibnamefont {Cao}},\ and\ \bibinfo {author}
  {\bibfnamefont {D.}~\bibnamefont {Mandrus}},\ }\bibfield  {title} {\bibinfo
  {title} {Charge density wave in kagome lattice intermetallic
  {ScV$_6$Sn$_6$}},\ }\href {https://doi.org/10.1103/PhysRevLett.129.216402}
  {\bibfield  {journal} {\bibinfo  {journal} {Phys. Rev. Lett.}\ }\textbf
  {\bibinfo {volume} {129}},\ \bibinfo {pages} {216402} (\bibinfo {year}
  {2022})}\BibitemShut {NoStop}%
\bibitem [{\citenamefont {Teng}\ \emph {et~al.}(2022)\citenamefont {Teng},
  \citenamefont {Chen}, \citenamefont {Ye}, \citenamefont {Rosenberg},
  \citenamefont {Liu}, \citenamefont {Yin}, \citenamefont {Jiang},
  \citenamefont {Oh}, \citenamefont {Hasan}, \citenamefont {Neubauer},
  \citenamefont {Gao}, \citenamefont {Xie}, \citenamefont {Hashimoto},
  \citenamefont {Lu}, \citenamefont {Jozwiak}, \citenamefont {Bostwick},
  \citenamefont {Rotenberg}, \citenamefont {Birgeneau}, \citenamefont {Chu},
  \citenamefont {Yi},\ and\ \citenamefont {Dai}}]{teng_discovery_2022}%
  \BibitemOpen
  \bibfield  {author} {\bibinfo {author} {\bibfnamefont {X.}~\bibnamefont
  {Teng}}, \bibinfo {author} {\bibfnamefont {L.}~\bibnamefont {Chen}}, \bibinfo
  {author} {\bibfnamefont {F.}~\bibnamefont {Ye}}, \bibinfo {author}
  {\bibfnamefont {E.}~\bibnamefont {Rosenberg}}, \bibinfo {author}
  {\bibfnamefont {Z.}~\bibnamefont {Liu}}, \bibinfo {author} {\bibfnamefont
  {J.-X.}\ \bibnamefont {Yin}}, \bibinfo {author} {\bibfnamefont {Y.-X.}\
  \bibnamefont {Jiang}}, \bibinfo {author} {\bibfnamefont {J.~S.}\ \bibnamefont
  {Oh}}, \bibinfo {author} {\bibfnamefont {M.~Z.}\ \bibnamefont {Hasan}},
  \bibinfo {author} {\bibfnamefont {K.~J.}\ \bibnamefont {Neubauer}}, \bibinfo
  {author} {\bibfnamefont {B.}~\bibnamefont {Gao}}, \bibinfo {author}
  {\bibfnamefont {Y.}~\bibnamefont {Xie}}, \bibinfo {author} {\bibfnamefont
  {M.}~\bibnamefont {Hashimoto}}, \bibinfo {author} {\bibfnamefont
  {D.}~\bibnamefont {Lu}}, \bibinfo {author} {\bibfnamefont {C.}~\bibnamefont
  {Jozwiak}}, \bibinfo {author} {\bibfnamefont {A.}~\bibnamefont {Bostwick}},
  \bibinfo {author} {\bibfnamefont {E.}~\bibnamefont {Rotenberg}}, \bibinfo
  {author} {\bibfnamefont {R.~J.}\ \bibnamefont {Birgeneau}}, \bibinfo {author}
  {\bibfnamefont {J.-H.}\ \bibnamefont {Chu}}, \bibinfo {author} {\bibfnamefont
  {M.}~\bibnamefont {Yi}},\ and\ \bibinfo {author} {\bibfnamefont
  {P.}~\bibnamefont {Dai}},\ }\bibfield  {title} {\bibinfo {title} {Discovery
  of charge density wave in a kagome lattice antiferromagnet},\ }\href
  {https://doi.org/10.1038/s41586-022-05034-z} {\bibfield  {journal} {\bibinfo
  {journal} {Nature}\ }\textbf {\bibinfo {volume} {609}},\ \bibinfo {pages}
  {490} (\bibinfo {year} {2022})}\BibitemShut {NoStop}%
\bibitem [{\citenamefont {Lee}\ \emph {et~al.}(2024)\citenamefont {Lee},
  \citenamefont {Won}, \citenamefont {Kim}, \citenamefont {Yoo}, \citenamefont
  {Park}, \citenamefont {Denlinger}, \citenamefont {Jozwiak}, \citenamefont
  {Bostwick}, \citenamefont {Rotenberg}, \citenamefont {Comin}, \citenamefont
  {Kang},\ and\ \citenamefont {Park}}]{lee_nature_2024}%
  \BibitemOpen
  \bibfield  {author} {\bibinfo {author} {\bibfnamefont {S.}~\bibnamefont
  {Lee}}, \bibinfo {author} {\bibfnamefont {C.}~\bibnamefont {Won}}, \bibinfo
  {author} {\bibfnamefont {J.}~\bibnamefont {Kim}}, \bibinfo {author}
  {\bibfnamefont {J.}~\bibnamefont {Yoo}}, \bibinfo {author} {\bibfnamefont
  {S.}~\bibnamefont {Park}}, \bibinfo {author} {\bibfnamefont {J.}~\bibnamefont
  {Denlinger}}, \bibinfo {author} {\bibfnamefont {C.}~\bibnamefont {Jozwiak}},
  \bibinfo {author} {\bibfnamefont {A.}~\bibnamefont {Bostwick}}, \bibinfo
  {author} {\bibfnamefont {E.}~\bibnamefont {Rotenberg}}, \bibinfo {author}
  {\bibfnamefont {R.}~\bibnamefont {Comin}}, \bibinfo {author} {\bibfnamefont
  {M.}~\bibnamefont {Kang}},\ and\ \bibinfo {author} {\bibfnamefont {J.-H.}\
  \bibnamefont {Park}},\ }\bibfield  {title} {\bibinfo {title} {Nature of
  charge density wave in kagome metal {ScV$_6$Sn$_6$}},\ }\href
  {https://doi.org/10.1038/s41535-024-00620-y} {\bibfield  {journal} {\bibinfo
  {journal} {npj Quantum Materials}\ }\textbf {\bibinfo {volume} {9}},\
  \bibinfo {pages} {15} (\bibinfo {year} {2024})}\BibitemShut {NoStop}%
\bibitem [{\citenamefont {Tan}\ and\ \citenamefont
  {Yan}(2023)}]{tan_abundant_2023}%
  \BibitemOpen
  \bibfield  {author} {\bibinfo {author} {\bibfnamefont {H.}~\bibnamefont
  {Tan}}\ and\ \bibinfo {author} {\bibfnamefont {B.}~\bibnamefont {Yan}},\
  }\bibfield  {title} {\bibinfo {title} {Abundant {Lattice} {Instability} in
  {Kagome} {Metal} {ScV$_6$Sn$_6$}},\ }\href
  {https://doi.org/10.1103/PhysRevLett.130.266402} {\bibfield  {journal}
  {\bibinfo  {journal} {Physical Review Letters}\ }\textbf {\bibinfo {volume}
  {130}},\ \bibinfo {pages} {266402} (\bibinfo {year} {2023})}\BibitemShut
  {NoStop}%
\bibitem [{\citenamefont {Cao}\ \emph {et~al.}(2023)\citenamefont {Cao},
  \citenamefont {Xu}, \citenamefont {Fukui}, \citenamefont {Manjo},
  \citenamefont {Dong}, \citenamefont {Shi}, \citenamefont {Liu}, \citenamefont
  {Cao},\ and\ \citenamefont {Song}}]{cao_competing_2023}%
  \BibitemOpen
  \bibfield  {author} {\bibinfo {author} {\bibfnamefont {S.}~\bibnamefont
  {Cao}}, \bibinfo {author} {\bibfnamefont {C.}~\bibnamefont {Xu}}, \bibinfo
  {author} {\bibfnamefont {H.}~\bibnamefont {Fukui}}, \bibinfo {author}
  {\bibfnamefont {T.}~\bibnamefont {Manjo}}, \bibinfo {author} {\bibfnamefont
  {Y.}~\bibnamefont {Dong}}, \bibinfo {author} {\bibfnamefont {M.}~\bibnamefont
  {Shi}}, \bibinfo {author} {\bibfnamefont {Y.}~\bibnamefont {Liu}}, \bibinfo
  {author} {\bibfnamefont {C.}~\bibnamefont {Cao}},\ and\ \bibinfo {author}
  {\bibfnamefont {Y.}~\bibnamefont {Song}},\ }\bibfield  {title} {\bibinfo
  {title} {Competing charge-density wave instabilities in the kagome metal
  {ScV$_6$Sn$_6$}},\ }\href {https://doi.org/10.1038/s41467-023-43454-1}
  {\bibfield  {journal} {\bibinfo  {journal} {Nature Communications}\ }\textbf
  {\bibinfo {volume} {14}},\ \bibinfo {pages} {7671} (\bibinfo {year}
  {2023})}\BibitemShut {NoStop}%
\bibitem [{\citenamefont {Teng}\ \emph {et~al.}(2023)\citenamefont {Teng},
  \citenamefont {Oh}, \citenamefont {Tan}, \citenamefont {Chen}, \citenamefont
  {Huang}, \citenamefont {Gao}, \citenamefont {Yin}, \citenamefont {Chu},
  \citenamefont {Hashimoto}, \citenamefont {Lu}, \citenamefont {Jozwiak},
  \citenamefont {Bostwick}, \citenamefont {Rotenberg}, \citenamefont
  {Granroth}, \citenamefont {Yan}, \citenamefont {Birgeneau}, \citenamefont
  {Dai},\ and\ \citenamefont {Yi}}]{teng_magnetism_2023}%
  \BibitemOpen
  \bibfield  {author} {\bibinfo {author} {\bibfnamefont {X.}~\bibnamefont
  {Teng}}, \bibinfo {author} {\bibfnamefont {J.~S.}\ \bibnamefont {Oh}},
  \bibinfo {author} {\bibfnamefont {H.}~\bibnamefont {Tan}}, \bibinfo {author}
  {\bibfnamefont {L.}~\bibnamefont {Chen}}, \bibinfo {author} {\bibfnamefont
  {J.}~\bibnamefont {Huang}}, \bibinfo {author} {\bibfnamefont
  {B.}~\bibnamefont {Gao}}, \bibinfo {author} {\bibfnamefont {J.-X.}\
  \bibnamefont {Yin}}, \bibinfo {author} {\bibfnamefont {J.-H.}\ \bibnamefont
  {Chu}}, \bibinfo {author} {\bibfnamefont {M.}~\bibnamefont {Hashimoto}},
  \bibinfo {author} {\bibfnamefont {D.}~\bibnamefont {Lu}}, \bibinfo {author}
  {\bibfnamefont {C.}~\bibnamefont {Jozwiak}}, \bibinfo {author} {\bibfnamefont
  {A.}~\bibnamefont {Bostwick}}, \bibinfo {author} {\bibfnamefont
  {E.}~\bibnamefont {Rotenberg}}, \bibinfo {author} {\bibfnamefont {G.~E.}\
  \bibnamefont {Granroth}}, \bibinfo {author} {\bibfnamefont {B.}~\bibnamefont
  {Yan}}, \bibinfo {author} {\bibfnamefont {R.~J.}\ \bibnamefont {Birgeneau}},
  \bibinfo {author} {\bibfnamefont {P.}~\bibnamefont {Dai}},\ and\ \bibinfo
  {author} {\bibfnamefont {M.}~\bibnamefont {Yi}},\ }\bibfield  {title}
  {\bibinfo {title} {Magnetism and charge density wave order in kagome
  {FeGe}},\ }\href {https://doi.org/10.1038/s41567-023-01985-w} {\bibfield
  {journal} {\bibinfo  {journal} {Nature Physics}\ }\textbf {\bibinfo {volume}
  {19}},\ \bibinfo {pages} {814} (\bibinfo {year} {2023})}\BibitemShut
  {NoStop}%
\bibitem [{\citenamefont {Shao}\ \emph {et~al.}(2023)\citenamefont {Shao},
  \citenamefont {Yin}, \citenamefont {Belopolski}, \citenamefont {You},
  \citenamefont {Hou}, \citenamefont {Chen}, \citenamefont {Jiang},
  \citenamefont {Hossain}, \citenamefont {Yahyavi}, \citenamefont {Hsu},
  \citenamefont {Feng}, \citenamefont {Bansil}, \citenamefont {Hasan},\ and\
  \citenamefont {Chang}}]{shao_intertwining_2023}%
  \BibitemOpen
  \bibfield  {author} {\bibinfo {author} {\bibfnamefont {S.}~\bibnamefont
  {Shao}}, \bibinfo {author} {\bibfnamefont {J.-X.}\ \bibnamefont {Yin}},
  \bibinfo {author} {\bibfnamefont {I.}~\bibnamefont {Belopolski}}, \bibinfo
  {author} {\bibfnamefont {J.-Y.}\ \bibnamefont {You}}, \bibinfo {author}
  {\bibfnamefont {T.}~\bibnamefont {Hou}}, \bibinfo {author} {\bibfnamefont
  {H.}~\bibnamefont {Chen}}, \bibinfo {author} {\bibfnamefont {Y.}~\bibnamefont
  {Jiang}}, \bibinfo {author} {\bibfnamefont {M.~S.}\ \bibnamefont {Hossain}},
  \bibinfo {author} {\bibfnamefont {M.}~\bibnamefont {Yahyavi}}, \bibinfo
  {author} {\bibfnamefont {C.-H.}\ \bibnamefont {Hsu}}, \bibinfo {author}
  {\bibfnamefont {Y.~P.}\ \bibnamefont {Feng}}, \bibinfo {author}
  {\bibfnamefont {A.}~\bibnamefont {Bansil}}, \bibinfo {author} {\bibfnamefont
  {M.~Z.}\ \bibnamefont {Hasan}},\ and\ \bibinfo {author} {\bibfnamefont
  {G.}~\bibnamefont {Chang}},\ }\bibfield  {title} {\bibinfo {title}
  {Intertwining of {Magnetism} and {Charge} {Ordering} in {Kagome} {FeGe}},\
  }\href {https://doi.org/10.1021/acsnano.3c00229} {\bibfield  {journal}
  {\bibinfo  {journal} {ACS Nano}\ }\textbf {\bibinfo {volume} {17}},\ \bibinfo
  {pages} {10164} (\bibinfo {year} {2023})}\BibitemShut {NoStop}%
\bibitem [{\citenamefont {Wu}\ \emph {et~al.}(2023)\citenamefont {Wu},
  \citenamefont {Hu}, \citenamefont {Fan}, \citenamefont {Wang},\ and\
  \citenamefont {Wan}}]{wu_electron_2023}%
  \BibitemOpen
  \bibfield  {author} {\bibinfo {author} {\bibfnamefont {L.}~\bibnamefont
  {Wu}}, \bibinfo {author} {\bibfnamefont {Y.}~\bibnamefont {Hu}}, \bibinfo
  {author} {\bibfnamefont {D.}~\bibnamefont {Fan}}, \bibinfo {author}
  {\bibfnamefont {D.}~\bibnamefont {Wang}},\ and\ \bibinfo {author}
  {\bibfnamefont {X.}~\bibnamefont {Wan}},\ }\bibfield  {title} {\bibinfo
  {title} {Electron-{Correlation}-{Induced} {Charge} {Density} {Wave} in
  {FeGe}},\ }\href {https://doi.org/10.1088/0256-307X/40/11/117103} {\bibfield
  {journal} {\bibinfo  {journal} {Chinese Physics Letters}\ }\textbf {\bibinfo
  {volume} {40}},\ \bibinfo {pages} {117103} (\bibinfo {year}
  {2023})}\BibitemShut {NoStop}%
\bibitem [{\citenamefont {Ma}\ \emph {et~al.}(2024)\citenamefont {Ma},
  \citenamefont {Yin}, \citenamefont {Zahid~Hasan},\ and\ \citenamefont
  {Liu}}]{ma_theory_2024}%
  \BibitemOpen
  \bibfield  {author} {\bibinfo {author} {\bibfnamefont {H.-Y.}\ \bibnamefont
  {Ma}}, \bibinfo {author} {\bibfnamefont {J.-X.}\ \bibnamefont {Yin}},
  \bibinfo {author} {\bibfnamefont {M.}~\bibnamefont {Zahid~Hasan}},\ and\
  \bibinfo {author} {\bibfnamefont {J.}~\bibnamefont {Liu}},\ }\bibfield
  {title} {\bibinfo {title} {Theory for {Charge} {Density} {Wave} and
  {Orbital}-{Flux} {State} in {Antiferromagnetic} {Kagome} {Metal} {FeGe}},\
  }\href {https://doi.org/10.1088/0256-307X/41/4/047103} {\bibfield  {journal}
  {\bibinfo  {journal} {Chinese Physics Letters}\ }\textbf {\bibinfo {volume}
  {41}},\ \bibinfo {pages} {047103} (\bibinfo {year} {2024})}\BibitemShut
  {NoStop}%
\bibitem [{\citenamefont {Wang}(2023)}]{wang_enhanced_2023}%
  \BibitemOpen
  \bibfield  {author} {\bibinfo {author} {\bibfnamefont {Y.}~\bibnamefont
  {Wang}},\ }\bibfield  {title} {\bibinfo {title} {Enhanced spin-polarization
  via partial {Ge}-dimerization as the driving force of the charge density wave
  in {FeGe}},\ }\href {https://doi.org/10.1103/PhysRevMaterials.7.104006}
  {\bibfield  {journal} {\bibinfo  {journal} {Phys. Rev. Mater.}\ }\textbf
  {\bibinfo {volume} {7}},\ \bibinfo {pages} {104006} (\bibinfo {year}
  {2023})}\BibitemShut {NoStop}%
\bibitem [{\citenamefont {Zhang}\ \emph {et~al.}(2024)\citenamefont {Zhang},
  \citenamefont {Ji}, \citenamefont {Xu},\ and\ \citenamefont
  {Xiang}}]{zhang_electronic_2024}%
  \BibitemOpen
  \bibfield  {author} {\bibinfo {author} {\bibfnamefont {B.}~\bibnamefont
  {Zhang}}, \bibinfo {author} {\bibfnamefont {J.}~\bibnamefont {Ji}}, \bibinfo
  {author} {\bibfnamefont {C.}~\bibnamefont {Xu}},\ and\ \bibinfo {author}
  {\bibfnamefont {H.}~\bibnamefont {Xiang}},\ }\bibfield  {title} {\bibinfo
  {title} {Electronic and magnetic origins of unconventional charge density
  wave in kagome {FeGe}},\ }\href {https://doi.org/10.1103/PhysRevB.110.125139}
  {\bibfield  {journal} {\bibinfo  {journal} {Physical Review B}\ }\textbf
  {\bibinfo {volume} {110}},\ \bibinfo {pages} {125139} (\bibinfo {year}
  {2024})}\BibitemShut {NoStop}%
\bibitem [{\citenamefont {Wu}\ \emph {et~al.}(2024)\citenamefont {Wu},
  \citenamefont {Mi}, \citenamefont {Zhang}, \citenamefont {Wang},
  \citenamefont {Maraytta}, \citenamefont {Zhou}, \citenamefont {He},
  \citenamefont {Merz}, \citenamefont {Chai},\ and\ \citenamefont
  {Wang}}]{wu_annealing-tunable_2024}%
  \BibitemOpen
  \bibfield  {author} {\bibinfo {author} {\bibfnamefont {X.}~\bibnamefont
  {Wu}}, \bibinfo {author} {\bibfnamefont {X.}~\bibnamefont {Mi}}, \bibinfo
  {author} {\bibfnamefont {L.}~\bibnamefont {Zhang}}, \bibinfo {author}
  {\bibfnamefont {C.-W.}\ \bibnamefont {Wang}}, \bibinfo {author}
  {\bibfnamefont {N.}~\bibnamefont {Maraytta}}, \bibinfo {author}
  {\bibfnamefont {X.}~\bibnamefont {Zhou}}, \bibinfo {author} {\bibfnamefont
  {M.}~\bibnamefont {He}}, \bibinfo {author} {\bibfnamefont {M.}~\bibnamefont
  {Merz}}, \bibinfo {author} {\bibfnamefont {Y.}~\bibnamefont {Chai}},\ and\
  \bibinfo {author} {\bibfnamefont {A.}~\bibnamefont {Wang}},\ }\bibfield
  {title} {\bibinfo {title} {Annealing-{Tunable} {Charge} {Density} {Wave} in
  the {Magnetic} {Kagome} {Material} {FeGe}},\ }\href
  {https://doi.org/10.1103/PhysRevLett.132.256501} {\bibfield  {journal}
  {\bibinfo  {journal} {Physical Review Letters}\ }\textbf {\bibinfo {volume}
  {132}},\ \bibinfo {pages} {256501} (\bibinfo {year} {2024})}\BibitemShut
  {NoStop}%
\bibitem [{\citenamefont {Miao}\ \emph {et~al.}(2023)\citenamefont {Miao},
  \citenamefont {Zhang}, \citenamefont {Li}, \citenamefont {Fabbris},
  \citenamefont {Said}, \citenamefont {Tartaglia}, \citenamefont {Yilmaz},
  \citenamefont {Vescovo}, \citenamefont {Yin}, \citenamefont {Murakami},
  \citenamefont {Feng}, \citenamefont {Jiang}, \citenamefont {Wu},
  \citenamefont {Wang}, \citenamefont {Okamoto}, \citenamefont {Wang},\ and\
  \citenamefont {Lee}}]{miao_signature_2023}%
  \BibitemOpen
  \bibfield  {author} {\bibinfo {author} {\bibfnamefont {H.}~\bibnamefont
  {Miao}}, \bibinfo {author} {\bibfnamefont {T.~T.}\ \bibnamefont {Zhang}},
  \bibinfo {author} {\bibfnamefont {H.~X.}\ \bibnamefont {Li}}, \bibinfo
  {author} {\bibfnamefont {G.}~\bibnamefont {Fabbris}}, \bibinfo {author}
  {\bibfnamefont {A.~H.}\ \bibnamefont {Said}}, \bibinfo {author}
  {\bibfnamefont {R.}~\bibnamefont {Tartaglia}}, \bibinfo {author}
  {\bibfnamefont {T.}~\bibnamefont {Yilmaz}}, \bibinfo {author} {\bibfnamefont
  {E.}~\bibnamefont {Vescovo}}, \bibinfo {author} {\bibfnamefont {J.-X.}\
  \bibnamefont {Yin}}, \bibinfo {author} {\bibfnamefont {S.}~\bibnamefont
  {Murakami}}, \bibinfo {author} {\bibfnamefont {X.~L.}\ \bibnamefont {Feng}},
  \bibinfo {author} {\bibfnamefont {K.}~\bibnamefont {Jiang}}, \bibinfo
  {author} {\bibfnamefont {X.~L.}\ \bibnamefont {Wu}}, \bibinfo {author}
  {\bibfnamefont {A.~F.}\ \bibnamefont {Wang}}, \bibinfo {author}
  {\bibfnamefont {S.}~\bibnamefont {Okamoto}}, \bibinfo {author} {\bibfnamefont
  {Y.~L.}\ \bibnamefont {Wang}},\ and\ \bibinfo {author} {\bibfnamefont
  {H.~N.}\ \bibnamefont {Lee}},\ }\bibfield  {title} {\bibinfo {title}
  {Signature of spin-phonon coupling driven charge density wave in a kagome
  magnet},\ }\href {https://doi.org/10.1038/s41467-023-41957-5} {\bibfield
  {journal} {\bibinfo  {journal} {Nature Communications}\ }\textbf {\bibinfo
  {volume} {14}},\ \bibinfo {pages} {6183} (\bibinfo {year}
  {2023})}\BibitemShut {NoStop}%
\bibitem [{\citenamefont {Chen}\ \emph {et~al.}(2024)\citenamefont {Chen},
  \citenamefont {Wu}, \citenamefont {Zhou}, \citenamefont {Zhang},
  \citenamefont {Yin}, \citenamefont {Li}, \citenamefont {Li}, \citenamefont
  {Gong}, \citenamefont {He}, \citenamefont {Chai}, \citenamefont {Zhou},
  \citenamefont {Wang}, \citenamefont {Wang}, \citenamefont {Yan},\ and\
  \citenamefont {Feng}}]{chen_discovery_2024}%
  \BibitemOpen
  \bibfield  {author} {\bibinfo {author} {\bibfnamefont {Z.}~\bibnamefont
  {Chen}}, \bibinfo {author} {\bibfnamefont {X.}~\bibnamefont {Wu}}, \bibinfo
  {author} {\bibfnamefont {S.}~\bibnamefont {Zhou}}, \bibinfo {author}
  {\bibfnamefont {J.}~\bibnamefont {Zhang}}, \bibinfo {author} {\bibfnamefont
  {R.}~\bibnamefont {Yin}}, \bibinfo {author} {\bibfnamefont {Y.}~\bibnamefont
  {Li}}, \bibinfo {author} {\bibfnamefont {M.}~\bibnamefont {Li}}, \bibinfo
  {author} {\bibfnamefont {J.}~\bibnamefont {Gong}}, \bibinfo {author}
  {\bibfnamefont {M.}~\bibnamefont {He}}, \bibinfo {author} {\bibfnamefont
  {Y.}~\bibnamefont {Chai}}, \bibinfo {author} {\bibfnamefont {X.}~\bibnamefont
  {Zhou}}, \bibinfo {author} {\bibfnamefont {Y.}~\bibnamefont {Wang}}, \bibinfo
  {author} {\bibfnamefont {A.}~\bibnamefont {Wang}}, \bibinfo {author}
  {\bibfnamefont {Y.-J.}\ \bibnamefont {Yan}},\ and\ \bibinfo {author}
  {\bibfnamefont {D.-L.}\ \bibnamefont {Feng}},\ }\bibfield  {title} {\bibinfo
  {title} {{Discovery} of a long-ranged charge order with
  {1/4}{Ge1}-dimerization in an antiferromagnetic {Kagome} metal},\ }\href
  {https://doi.org/10.1038/s41467-024-50661-x} {\bibfield  {journal} {\bibinfo
  {journal} {Nature Communications}\ }\textbf {\bibinfo {volume} {15}},\
  \bibinfo {pages} {6262} (\bibinfo {year} {2024})},\ \bibinfo {note}
  {publisher: Nature Publishing Group}\BibitemShut {NoStop}%
\bibitem [{\citenamefont {Pokharel}\ \emph {et~al.}(2023)\citenamefont
  {Pokharel}, \citenamefont {Ortiz}, \citenamefont {Kautzsch}, \citenamefont
  {Gomez~Alvarado}, \citenamefont {Mallayya}, \citenamefont {Wu}, \citenamefont
  {Kim}, \citenamefont {Ruff}, \citenamefont {Sarker},\ and\ \citenamefont
  {Wilson}}]{pokharel_frustrated_2023}%
  \BibitemOpen
  \bibfield  {author} {\bibinfo {author} {\bibfnamefont {G.}~\bibnamefont
  {Pokharel}}, \bibinfo {author} {\bibfnamefont {B.~R.}\ \bibnamefont {Ortiz}},
  \bibinfo {author} {\bibfnamefont {L.}~\bibnamefont {Kautzsch}}, \bibinfo
  {author} {\bibfnamefont {S.~J.}\ \bibnamefont {Gomez~Alvarado}}, \bibinfo
  {author} {\bibfnamefont {K.}~\bibnamefont {Mallayya}}, \bibinfo {author}
  {\bibfnamefont {G.}~\bibnamefont {Wu}}, \bibinfo {author} {\bibfnamefont
  {E.-A.}\ \bibnamefont {Kim}}, \bibinfo {author} {\bibfnamefont {J.~P.~C.}\
  \bibnamefont {Ruff}}, \bibinfo {author} {\bibfnamefont {S.}~\bibnamefont
  {Sarker}},\ and\ \bibinfo {author} {\bibfnamefont {S.~D.}\ \bibnamefont
  {Wilson}},\ }\bibfield  {title} {\bibinfo {title} {Frustrated charge order
  and cooperative distortions in {ScV$_6$Sn$_6$}},\ }\href
  {https://doi.org/10.1103/PhysRevMaterials.7.104201} {\bibfield  {journal}
  {\bibinfo  {journal} {Physical Review Materials}\ }\textbf {\bibinfo {volume}
  {7}},\ \bibinfo {pages} {104201} (\bibinfo {year} {2023})}\BibitemShut
  {NoStop}%
\bibitem [{\citenamefont {Meier}\ \emph {et~al.}(2023)\citenamefont {Meier},
  \citenamefont {Madhogaria}, \citenamefont {Mozaffari}, \citenamefont
  {Marshall}, \citenamefont {Graf}, \citenamefont {McGuire}, \citenamefont
  {Arachchige}, \citenamefont {Allen}, \citenamefont {Driver}, \citenamefont
  {Cao},\ and\ \citenamefont {Mandrus}}]{meier_tiny_2023}%
  \BibitemOpen
  \bibfield  {author} {\bibinfo {author} {\bibfnamefont {W.~R.}\ \bibnamefont
  {Meier}}, \bibinfo {author} {\bibfnamefont {R.~P.}\ \bibnamefont
  {Madhogaria}}, \bibinfo {author} {\bibfnamefont {S.}~\bibnamefont
  {Mozaffari}}, \bibinfo {author} {\bibfnamefont {M.}~\bibnamefont {Marshall}},
  \bibinfo {author} {\bibfnamefont {D.~E.}\ \bibnamefont {Graf}}, \bibinfo
  {author} {\bibfnamefont {M.~A.}\ \bibnamefont {McGuire}}, \bibinfo {author}
  {\bibfnamefont {H.~W.~S.}\ \bibnamefont {Arachchige}}, \bibinfo {author}
  {\bibfnamefont {C.~L.}\ \bibnamefont {Allen}}, \bibinfo {author}
  {\bibfnamefont {J.}~\bibnamefont {Driver}}, \bibinfo {author} {\bibfnamefont
  {H.}~\bibnamefont {Cao}},\ and\ \bibinfo {author} {\bibfnamefont
  {D.}~\bibnamefont {Mandrus}},\ }\bibfield  {title} {\bibinfo {title} {Tiny
  {Sc} {Allows} the {Chains} to {Rattle}: {Impact} of {Lu} and {Y} {Doping} on
  the {Charge}-{Density} {Wave} in {ScV}$_{\textrm{6}}${Sn}$_{\textrm{6}}$},\
  }\href {https://doi.org/10.1021/jacs.3c06394} {\bibfield  {journal} {\bibinfo
   {journal} {Journal of the American Chemical Society}\ }\textbf {\bibinfo
  {volume} {145}},\ \bibinfo {pages} {20943} (\bibinfo {year}
  {2023})}\BibitemShut {NoStop}%
\bibitem [{\citenamefont {Ortiz}\ \emph {et~al.}(2025)\citenamefont {Ortiz},
  \citenamefont {Meier}, \citenamefont {Pokharel}, \citenamefont {Chamorro},
  \citenamefont {Yang}, \citenamefont {Mozaffari}, \citenamefont {Thaler},
  \citenamefont {Gomez~Alvarado}, \citenamefont {Zhang}, \citenamefont
  {Parker}, \citenamefont {Samolyuk}, \citenamefont {Paddison}, \citenamefont
  {Yan}, \citenamefont {Ye}, \citenamefont {Sarker}, \citenamefont {Wilson},
  \citenamefont {Miao}, \citenamefont {Mandrus},\ and\ \citenamefont
  {McGuire}}]{ortiz_stability_2025}%
  \BibitemOpen
  \bibfield  {author} {\bibinfo {author} {\bibfnamefont {B.~R.}\ \bibnamefont
  {Ortiz}}, \bibinfo {author} {\bibfnamefont {W.~R.}\ \bibnamefont {Meier}},
  \bibinfo {author} {\bibfnamefont {G.}~\bibnamefont {Pokharel}}, \bibinfo
  {author} {\bibfnamefont {J.}~\bibnamefont {Chamorro}}, \bibinfo {author}
  {\bibfnamefont {F.}~\bibnamefont {Yang}}, \bibinfo {author} {\bibfnamefont
  {S.}~\bibnamefont {Mozaffari}}, \bibinfo {author} {\bibfnamefont
  {A.}~\bibnamefont {Thaler}}, \bibinfo {author} {\bibfnamefont {S.~J.}\
  \bibnamefont {Gomez~Alvarado}}, \bibinfo {author} {\bibfnamefont
  {H.}~\bibnamefont {Zhang}}, \bibinfo {author} {\bibfnamefont {D.~S.}\
  \bibnamefont {Parker}}, \bibinfo {author} {\bibfnamefont {G.~D.}\
  \bibnamefont {Samolyuk}}, \bibinfo {author} {\bibfnamefont {J.~A.~M.}\
  \bibnamefont {Paddison}}, \bibinfo {author} {\bibfnamefont {J.}~\bibnamefont
  {Yan}}, \bibinfo {author} {\bibfnamefont {F.}~\bibnamefont {Ye}}, \bibinfo
  {author} {\bibfnamefont {S.}~\bibnamefont {Sarker}}, \bibinfo {author}
  {\bibfnamefont {S.~D.}\ \bibnamefont {Wilson}}, \bibinfo {author}
  {\bibfnamefont {H.}~\bibnamefont {Miao}}, \bibinfo {author} {\bibfnamefont
  {D.}~\bibnamefont {Mandrus}},\ and\ \bibinfo {author} {\bibfnamefont {M.~A.}\
  \bibnamefont {McGuire}},\ }\bibfield  {title} {\bibinfo {title} {Stability
  {Frontiers} in the {$AM_6X_6$} {Kagome} {Metals}: {The} \ce{$Ln$Nb6Sn6}
  (\textit{{Ln}}: {Ce}–{Lu},{Y}) {Family} and {Density}-{Wave} {Transition}
  in \ce{LuNb6Sn6}},\ }\href {https://doi.org/10.1021/jacs.4c16347} {\bibfield
  {journal} {\bibinfo  {journal} {Journal of the American Chemical Society}\
  }\textbf {\bibinfo {volume} {147}},\ \bibinfo {pages} {5279} (\bibinfo {year}
  {2025})}\BibitemShut {NoStop}%
\bibitem [{\citenamefont {Nishihara}\ \emph {et~al.}(1999)\citenamefont
  {Nishihara}, \citenamefont {Akimitsu}, \citenamefont {Hori}, \citenamefont
  {Niida}, \citenamefont {Ohoyama}, \citenamefont {Ohashi}, \citenamefont
  {Yamaguchi},\ and\ \citenamefont {Nakagawa}}]{nishihara_magnetic_1999}%
  \BibitemOpen
  \bibfield  {author} {\bibinfo {author} {\bibfnamefont {R.}~\bibnamefont
  {Nishihara}}, \bibinfo {author} {\bibfnamefont {M.}~\bibnamefont {Akimitsu}},
  \bibinfo {author} {\bibfnamefont {T.}~\bibnamefont {Hori}}, \bibinfo {author}
  {\bibfnamefont {H.}~\bibnamefont {Niida}}, \bibinfo {author} {\bibfnamefont
  {K.}~\bibnamefont {Ohoyama}}, \bibinfo {author} {\bibfnamefont
  {M.}~\bibnamefont {Ohashi}}, \bibinfo {author} {\bibfnamefont
  {Y.}~\bibnamefont {Yamaguchi}},\ and\ \bibinfo {author} {\bibfnamefont
  {Y.}~\bibnamefont {Nakagawa}},\ }\bibfield  {title} {\bibinfo {title}
  {Magnetic properties of hp13 type {TiFe$_6$Ge$_6$} alloy},\ }\href
  {https://doi.org/10.1016/S0304-8853(98)00908-1} {\bibfield  {journal}
  {\bibinfo  {journal} {Journal of Magnetism and Magnetic Materials}\ }\textbf
  {\bibinfo {volume} {196-197}},\ \bibinfo {pages} {665} (\bibinfo {year}
  {1999})}\BibitemShut {NoStop}%
\bibitem [{\citenamefont {Mazet}\ \emph {et~al.}(2000)\citenamefont {Mazet},
  \citenamefont {Isnard},\ and\ \citenamefont {Malaman}}]{mazet_neutron_2000}%
  \BibitemOpen
  \bibfield  {author} {\bibinfo {author} {\bibfnamefont {T.}~\bibnamefont
  {Mazet}}, \bibinfo {author} {\bibfnamefont {O.}~\bibnamefont {Isnard}},\ and\
  \bibinfo {author} {\bibfnamefont {B.}~\bibnamefont {Malaman}},\ }\bibfield
  {title} {\bibinfo {title} {Neutron diffraction and \ce{^57Fe} {Mössbauer}
  study of the \ce{HfFe6Ge6}-type \ce{$R$Fe6Ge6} compounds ({$R$}={Sc}, {Ti},
  {Zr}, {Hf}, {Nb})},\ }\href {https://doi.org/10.1016/S0038-1098(00)00003-X}
  {\bibfield  {journal} {\bibinfo  {journal} {Solid State Communications}\
  }\textbf {\bibinfo {volume} {114}},\ \bibinfo {pages} {91} (\bibinfo {year}
  {2000})}\BibitemShut {NoStop}%
\bibitem [{Cry()}]{CrysAlis}%
  \BibitemOpen
  \href@noop {} {}\bibinfo {howpublished} {Rigaku Oxford Diffraction Ltd,
  CrysAlisPro software system, version 1.171.44, Rigaku Corporation, Wroclaw,
  Poland, Rigaku Oxford Diffraction, Yarnton, Oxfordshire, E 2015
  CrysAlisPro}\BibitemShut {NoStop}%
\bibitem [{\citenamefont {Petříček}\ \emph {et~al.}(2014)\citenamefont
  {Petříček}, \citenamefont {Dušek},\ and\ \citenamefont
  {Palatinus}}]{Vaclav_229_2014}%
  \BibitemOpen
  \bibfield  {author} {\bibinfo {author} {\bibfnamefont {V.}~\bibnamefont
  {Petříček}}, \bibinfo {author} {\bibfnamefont {M.}~\bibnamefont
  {Dušek}},\ and\ \bibinfo {author} {\bibfnamefont {L.}~\bibnamefont
  {Palatinus}},\ }\bibfield  {title} {\bibinfo {title} {Crystallographic
  {Computing} {System} {JANA2006}: {General} features},\ }\href
  {https://doi.org/doi:10.1515/zkri-2014-1737} {\bibfield  {journal} {\bibinfo
  {journal} {Zeitschrift für Kristallographie - Crystalline Materials}\
  }\textbf {\bibinfo {volume} {229}},\ \bibinfo {pages} {345} (\bibinfo {year}
  {2014})}\BibitemShut {NoStop}%
\bibitem [{\citenamefont {Kresse}\ and\ \citenamefont
  {Furthm{\"u}ller}(1996)}]{vasp1996}%
  \BibitemOpen
  \bibfield  {author} {\bibinfo {author} {\bibfnamefont {G.}~\bibnamefont
  {Kresse}}\ and\ \bibinfo {author} {\bibfnamefont {J.}~\bibnamefont
  {Furthm{\"u}ller}},\ }\bibfield  {title} {\bibinfo {title} {Efficient
  iterative schemes for{\emph{ab initio}} total-energy calculations using a
  plane-wave basis set},\ }\href {https://doi.org/10.1103/PhysRevB.54.11169}
  {\bibfield  {journal} {\bibinfo  {journal} {Physical Review B}\ }\textbf
  {\bibinfo {volume} {54}},\ \bibinfo {pages} {11169} (\bibinfo {year}
  {1996})}\BibitemShut {NoStop}%
\bibitem [{\citenamefont {Perdew}\ \emph {et~al.}(1996)\citenamefont {Perdew},
  \citenamefont {Burke},\ and\ \citenamefont {Ernzerhof}}]{pbe1996}%
  \BibitemOpen
  \bibfield  {author} {\bibinfo {author} {\bibfnamefont {J.~P.}\ \bibnamefont
  {Perdew}}, \bibinfo {author} {\bibfnamefont {K.}~\bibnamefont {Burke}},\ and\
  \bibinfo {author} {\bibfnamefont {M.}~\bibnamefont {Ernzerhof}},\ }\bibfield
  {title} {\bibinfo {title} {Generalized {{Gradient Approximation Made
  Simple}}},\ }\href {https://doi.org/10.1103/PhysRevLett.77.3865} {\bibfield
  {journal} {\bibinfo  {journal} {Physical Review Letters}\ }\textbf {\bibinfo
  {volume} {77}},\ \bibinfo {pages} {3865} (\bibinfo {year}
  {1996})}\BibitemShut {NoStop}%
\bibitem [{\citenamefont {Wang}\ \emph {et~al.}(2021)\citenamefont {Wang},
  \citenamefont {Xu}, \citenamefont {Liu}, \citenamefont {Tang},\ and\
  \citenamefont {Geng}}]{VASPKIT2021}%
  \BibitemOpen
  \bibfield  {author} {\bibinfo {author} {\bibfnamefont {V.}~\bibnamefont
  {Wang}}, \bibinfo {author} {\bibfnamefont {N.}~\bibnamefont {Xu}}, \bibinfo
  {author} {\bibfnamefont {J.-C.}\ \bibnamefont {Liu}}, \bibinfo {author}
  {\bibfnamefont {G.}~\bibnamefont {Tang}},\ and\ \bibinfo {author}
  {\bibfnamefont {W.-T.}\ \bibnamefont {Geng}},\ }\bibfield  {title} {\bibinfo
  {title} {{{VASPKIT}}: {{A}} user-friendly interface facilitating
  high-throughput computing and analysis using {{VASP}} code},\ }\href
  {https://doi.org/10.1016/j.cpc.2021.108033} {\bibfield  {journal} {\bibinfo
  {journal} {Computer Physics Communications}\ }\textbf {\bibinfo {volume}
  {267}},\ \bibinfo {pages} {108033} (\bibinfo {year} {2021})}\BibitemShut
  {NoStop}%
\bibitem [{\citenamefont {Dronskowski}\ and\ \citenamefont
  {Bloechl}(1993)}]{cohp1993}%
  \BibitemOpen
  \bibfield  {author} {\bibinfo {author} {\bibfnamefont {R.}~\bibnamefont
  {Dronskowski}}\ and\ \bibinfo {author} {\bibfnamefont {P.~E.}\ \bibnamefont
  {Bloechl}},\ }\bibfield  {title} {\bibinfo {title} {Crystal orbital
  {{Hamilton}} populations ({{COHP}}): Energy-resolved visualization of
  chemical bonding in solids based on density-functional calculations},\ }\href
  {https://doi.org/10.1021/j100135a014} {\bibfield  {journal} {\bibinfo
  {journal} {The Journal of Physical Chemistry}\ }\textbf {\bibinfo {volume}
  {97}},\ \bibinfo {pages} {8617} (\bibinfo {year} {1993})}\BibitemShut
  {NoStop}%
\bibitem [{\citenamefont {Deringer}\ \emph {et~al.}(2011)\citenamefont
  {Deringer}, \citenamefont {Tchougr{\'e}eff},\ and\ \citenamefont
  {Dronskowski}}]{cohp2011}%
  \BibitemOpen
  \bibfield  {author} {\bibinfo {author} {\bibfnamefont {V.~L.}\ \bibnamefont
  {Deringer}}, \bibinfo {author} {\bibfnamefont {A.~L.}\ \bibnamefont
  {Tchougr{\'e}eff}},\ and\ \bibinfo {author} {\bibfnamefont {R.}~\bibnamefont
  {Dronskowski}},\ }\bibfield  {title} {\bibinfo {title} {Crystal {{Orbital
  Hamilton Population}} ({{COHP}}) {{Analysis As Projected}} from {{Plane-Wave
  Basis Sets}}},\ }\href {https://doi.org/10.1021/jp202489s} {\bibfield
  {journal} {\bibinfo  {journal} {The Journal of Physical Chemistry A}\
  }\textbf {\bibinfo {volume} {115}},\ \bibinfo {pages} {5461} (\bibinfo {year}
  {2011})}\BibitemShut {NoStop}%
\bibitem [{\citenamefont {Maintz}\ \emph {et~al.}(2013)\citenamefont {Maintz},
  \citenamefont {Deringer}, \citenamefont {Tchougr{\'e}eff},\ and\
  \citenamefont {Dronskowski}}]{cohp2013}%
  \BibitemOpen
  \bibfield  {author} {\bibinfo {author} {\bibfnamefont {S.}~\bibnamefont
  {Maintz}}, \bibinfo {author} {\bibfnamefont {V.~L.}\ \bibnamefont
  {Deringer}}, \bibinfo {author} {\bibfnamefont {A.~L.}\ \bibnamefont
  {Tchougr{\'e}eff}},\ and\ \bibinfo {author} {\bibfnamefont {R.}~\bibnamefont
  {Dronskowski}},\ }\bibfield  {title} {\bibinfo {title} {Analytic projection
  from plane-wave and {{PAW}} wavefunctions and application to chemical-bonding
  analysis in solids},\ }\href {https://doi.org/10.1002/jcc.23424} {\bibfield
  {journal} {\bibinfo  {journal} {Journal of Computational Chemistry}\ }\textbf
  {\bibinfo {volume} {34}},\ \bibinfo {pages} {2557} (\bibinfo {year}
  {2013})}\BibitemShut {NoStop}%
\bibitem [{\citenamefont {Maintz}\ \emph
  {et~al.}(2016{\natexlab{a}})\citenamefont {Maintz}, \citenamefont {Deringer},
  \citenamefont {Tchougr{\'e}eff},\ and\ \citenamefont
  {Dronskowski}}]{cohp2016_1}%
  \BibitemOpen
  \bibfield  {author} {\bibinfo {author} {\bibfnamefont {S.}~\bibnamefont
  {Maintz}}, \bibinfo {author} {\bibfnamefont {V.~L.}\ \bibnamefont
  {Deringer}}, \bibinfo {author} {\bibfnamefont {A.~L.}\ \bibnamefont
  {Tchougr{\'e}eff}},\ and\ \bibinfo {author} {\bibfnamefont {R.}~\bibnamefont
  {Dronskowski}},\ }\bibfield  {title} {\bibinfo {title} {{{LOBSTER}}: {{A}}
  tool to extract chemical bonding from plane-wave based {{DFT}}},\ }\href
  {https://doi.org/10.1002/jcc.24300} {\bibfield  {journal} {\bibinfo
  {journal} {Journal of Computational Chemistry}\ }\textbf {\bibinfo {volume}
  {37}},\ \bibinfo {pages} {1030} (\bibinfo {year}
  {2016}{\natexlab{a}})}\BibitemShut {NoStop}%
\bibitem [{\citenamefont {Maintz}\ \emph
  {et~al.}(2016{\natexlab{b}})\citenamefont {Maintz}, \citenamefont {Esser},\
  and\ \citenamefont {Dronskowski}}]{cohp2016_2}%
  \BibitemOpen
  \bibfield  {author} {\bibinfo {author} {\bibfnamefont {S.}~\bibnamefont
  {Maintz}}, \bibinfo {author} {\bibfnamefont {M.}~\bibnamefont {Esser}},\ and\
  \bibinfo {author} {\bibfnamefont {R.}~\bibnamefont {Dronskowski}},\
  }\bibfield  {title} {\bibinfo {title} {Efficient {{Rotation}} of {{Local
  Basis Functions Using Real Spherical Harmonics}}},\ }\href
  {https://doi.org/10.5506/APhysPolB.47.1165} {\bibfield  {journal} {\bibinfo
  {journal} {Acta Physica Polonica B}\ }\textbf {\bibinfo {volume} {47}},\
  \bibinfo {pages} {1165} (\bibinfo {year} {2016}{\natexlab{b}})}\BibitemShut
  {NoStop}%
\bibitem [{\citenamefont {Nelson}\ \emph {et~al.}(2020)\citenamefont {Nelson},
  \citenamefont {Ertural}, \citenamefont {George}, \citenamefont {Deringer},
  \citenamefont {Hautier},\ and\ \citenamefont {Dronskowski}}]{cohp2020}%
  \BibitemOpen
  \bibfield  {author} {\bibinfo {author} {\bibfnamefont {R.}~\bibnamefont
  {Nelson}}, \bibinfo {author} {\bibfnamefont {C.}~\bibnamefont {Ertural}},
  \bibinfo {author} {\bibfnamefont {J.}~\bibnamefont {George}}, \bibinfo
  {author} {\bibfnamefont {V.~L.}\ \bibnamefont {Deringer}}, \bibinfo {author}
  {\bibfnamefont {G.}~\bibnamefont {Hautier}},\ and\ \bibinfo {author}
  {\bibfnamefont {R.}~\bibnamefont {Dronskowski}},\ }\bibfield  {title}
  {\bibinfo {title} {{{LOBSTER}}: {{Local}} orbital projections, atomic
  charges, and chemical-bonding analysis from projector-augmented-wave-based
  density-functional theory},\ }\href {https://doi.org/10.1002/jcc.26353}
  {\bibfield  {journal} {\bibinfo  {journal} {Journal of Computational
  Chemistry}\ }\textbf {\bibinfo {volume} {41}},\ \bibinfo {pages} {1931}
  (\bibinfo {year} {2020})}\BibitemShut {NoStop}%
\bibitem [{\citenamefont {Olenich}\ \emph {et~al.}(1981)\citenamefont
  {Olenich}, \citenamefont {Akselrud},\ and\ \citenamefont
  {R.}}]{etde_5753394}%
  \BibitemOpen
  \bibfield  {author} {\bibinfo {author} {\bibfnamefont {R.~R.}\ \bibnamefont
  {Olenich}}, \bibinfo {author} {\bibfnamefont {L.~G.}\ \bibnamefont
  {Akselrud}},\ and\ \bibinfo {author} {\bibfnamefont {Y.~Y.}\ \bibnamefont
  {R.}},\ }\bibfield  {title} {\bibinfo {title} {Crystal structure of ternary
  germanides \ce{$R$Fe6Ge6} ({$R$} = {Sc, Ti, Zr, Hf, Nd}) and \ce{$R$Co6Ge6}
  ({$R$} = {Ti, Zr, Hf})},\ }\href@noop {} {\bibfield  {journal} {\bibinfo
  {journal} {Dopov. Akad. Nauk Ukr. RSR, Ser. A}\ }\textbf {\bibinfo {volume}
  {2}},\ \bibinfo {pages} {84} (\bibinfo {year} {1981})}\BibitemShut {NoStop}%
\bibitem [{SM(2024)}]{SM}%
  \BibitemOpen
  \href@noop {} {}\bibinfo {howpublished} {See Supporting Information at [URL
  will be inserted by publisher] for the EDX mapping, structural trend,
  specific heat for \ce{Ti_{0.85}Fe6Ge6}, and tables of ICOHP values for
  possible bonds in \ce{TiFe6Ge6} and FeGe} (\bibinfo {year}
  {2024})\BibitemShut {NoStop}%
\bibitem [{\citenamefont {Meier}\ \emph {et~al.}(2020)\citenamefont {Meier},
  \citenamefont {Du}, \citenamefont {Okamoto}, \citenamefont {Mohanta},
  \citenamefont {May}, \citenamefont {McGuire}, \citenamefont {Bridges},
  \citenamefont {Samolyuk},\ and\ \citenamefont {Sales}}]{meier_flat_2020}%
  \BibitemOpen
  \bibfield  {author} {\bibinfo {author} {\bibfnamefont {W.~R.}\ \bibnamefont
  {Meier}}, \bibinfo {author} {\bibfnamefont {M.-H.}\ \bibnamefont {Du}},
  \bibinfo {author} {\bibfnamefont {S.}~\bibnamefont {Okamoto}}, \bibinfo
  {author} {\bibfnamefont {N.}~\bibnamefont {Mohanta}}, \bibinfo {author}
  {\bibfnamefont {A.~F.}\ \bibnamefont {May}}, \bibinfo {author} {\bibfnamefont
  {M.~A.}\ \bibnamefont {McGuire}}, \bibinfo {author} {\bibfnamefont {C.~A.}\
  \bibnamefont {Bridges}}, \bibinfo {author} {\bibfnamefont {G.~D.}\
  \bibnamefont {Samolyuk}},\ and\ \bibinfo {author} {\bibfnamefont {B.~C.}\
  \bibnamefont {Sales}},\ }\bibfield  {title} {\bibinfo {title} {Flat bands in
  the {CoSn}-type compounds},\ }\href
  {https://doi.org/10.1103/PhysRevB.102.075148} {\bibfield  {journal} {\bibinfo
   {journal} {Physical Review B}\ }\textbf {\bibinfo {volume} {102}},\ \bibinfo
  {pages} {075148} (\bibinfo {year} {2020})}\BibitemShut {NoStop}%
\bibitem [{\citenamefont {Venturini}(2006)}]{venturini_filling_2006}%
  \BibitemOpen
  \bibfield  {author} {\bibinfo {author} {\bibfnamefont {G.}~\bibnamefont
  {Venturini}},\ }\bibfield  {title} {\bibinfo {title} {Filling the {CoSn}
  host-cell: the \ce{HfFe6Ge6}-type and the related structures},\ }\href
  {https://doi.org/doi:10.1524/zkri.2006.221.5-7.511} {\bibfield  {journal}
  {\bibinfo  {journal} {Zeitschrift für Kristallographie - Crystalline
  Materials}\ }\textbf {\bibinfo {volume} {221}},\ \bibinfo {pages} {511}
  (\bibinfo {year} {2006})}\BibitemShut {NoStop}%
\bibitem [{\citenamefont {Shannon}(1976)}]{shannon_revised_1976}%
  \BibitemOpen
  \bibfield  {author} {\bibinfo {author} {\bibfnamefont {R.~D.}\ \bibnamefont
  {Shannon}},\ }\bibfield  {title} {\bibinfo {title} {Revised effective ionic
  radii and systematic studies of interatomic distances in halides and
  chalcogenides},\ }\href {https://doi.org/10.1107/S0567739476001551}
  {\bibfield  {journal} {\bibinfo  {journal} {Acta Crystallographica Section
  A}\ }\textbf {\bibinfo {volume} {32}},\ \bibinfo {pages} {751} (\bibinfo
  {year} {1976})}\BibitemShut {NoStop}%
\bibitem [{\citenamefont {Sanjinés}\ \emph {et~al.}(1994)\citenamefont
  {Sanjinés}, \citenamefont {Tang}, \citenamefont {Berger}, \citenamefont
  {Gozzo}, \citenamefont {Margaritondo},\ and\ \citenamefont
  {Lévy}}]{sanjines_electronic_1994}%
  \BibitemOpen
  \bibfield  {author} {\bibinfo {author} {\bibfnamefont {R.}~\bibnamefont
  {Sanjinés}}, \bibinfo {author} {\bibfnamefont {H.}~\bibnamefont {Tang}},
  \bibinfo {author} {\bibfnamefont {H.}~\bibnamefont {Berger}}, \bibinfo
  {author} {\bibfnamefont {F.}~\bibnamefont {Gozzo}}, \bibinfo {author}
  {\bibfnamefont {G.}~\bibnamefont {Margaritondo}},\ and\ \bibinfo {author}
  {\bibfnamefont {F.}~\bibnamefont {Lévy}},\ }\bibfield  {title} {\bibinfo
  {title} {Electronic structure of anatase \ce{TiO2} oxide},\ }\href
  {https://doi.org/10.1063/1.356190} {\bibfield  {journal} {\bibinfo  {journal}
  {Journal of Applied Physics}\ }\textbf {\bibinfo {volume} {75}},\ \bibinfo
  {pages} {2945} (\bibinfo {year} {1994})}\BibitemShut {NoStop}%
\bibitem [{\citenamefont {Mazet}\ and\ \citenamefont
  {Malaman}(2001)}]{mazet_macroscopic_2001}%
  \BibitemOpen
  \bibfield  {author} {\bibinfo {author} {\bibfnamefont {T.}~\bibnamefont
  {Mazet}}\ and\ \bibinfo {author} {\bibfnamefont {B.}~\bibnamefont
  {Malaman}},\ }\bibfield  {title} {\bibinfo {title} {Macroscopic magnetic
  properties of the \ce{HfFe6Ge6}-type \ce{$R$Fe6X6} ({X} = {Ge} or {Sn})
  compounds involving a non-magnetic {$R$} metal},\ }\href
  {https://doi.org/10.1016/S0925-8388(01)01378-0} {\bibfield  {journal}
  {\bibinfo  {journal} {Journal of Alloys and Compounds}\ }\textbf {\bibinfo
  {volume} {325}},\ \bibinfo {pages} {67} (\bibinfo {year} {2001})}\BibitemShut
  {NoStop}%
\bibitem [{\citenamefont {Beckman}\ \emph {et~al.}(1972)\citenamefont
  {Beckman}, \citenamefont {Carrander}, \citenamefont {Lundgren},\ and\
  \citenamefont {Richardson}}]{beckman_susceptibility_1972}%
  \BibitemOpen
  \bibfield  {author} {\bibinfo {author} {\bibfnamefont {O.}~\bibnamefont
  {Beckman}}, \bibinfo {author} {\bibfnamefont {K.}~\bibnamefont {Carrander}},
  \bibinfo {author} {\bibfnamefont {L.}~\bibnamefont {Lundgren}},\ and\
  \bibinfo {author} {\bibfnamefont {M.}~\bibnamefont {Richardson}},\ }\bibfield
   {title} {\bibinfo {title} {Susceptibility {Measurements} and {Magnetic}
  {Ordering} of {Hexagonal} {FeGe}},\ }\href
  {https://doi.org/10.1088/0031-8949/6/2-3/009} {\bibfield  {journal} {\bibinfo
   {journal} {Physica Scripta}\ }\textbf {\bibinfo {volume} {6}},\ \bibinfo
  {pages} {151} (\bibinfo {year} {1972})}\BibitemShut {NoStop}%
\bibitem [{\citenamefont {Huang}\ \emph {et~al.}(2023)\citenamefont {Huang},
  \citenamefont {Cui}, \citenamefont {Huang}, \citenamefont {Huo},
  \citenamefont {Liu}, \citenamefont {Li}, \citenamefont {Liang}, \citenamefont
  {Chen}, \citenamefont {Sun}, \citenamefont {Shen}, \citenamefont {Zhang},\
  and\ \citenamefont {Wang}}]{huang_anisotropic_2023}%
  \BibitemOpen
  \bibfield  {author} {\bibinfo {author} {\bibfnamefont {X.}~\bibnamefont
  {Huang}}, \bibinfo {author} {\bibfnamefont {Z.}~\bibnamefont {Cui}}, \bibinfo
  {author} {\bibfnamefont {C.}~\bibnamefont {Huang}}, \bibinfo {author}
  {\bibfnamefont {M.}~\bibnamefont {Huo}}, \bibinfo {author} {\bibfnamefont
  {H.}~\bibnamefont {Liu}}, \bibinfo {author} {\bibfnamefont {J.}~\bibnamefont
  {Li}}, \bibinfo {author} {\bibfnamefont {F.}~\bibnamefont {Liang}}, \bibinfo
  {author} {\bibfnamefont {L.}~\bibnamefont {Chen}}, \bibinfo {author}
  {\bibfnamefont {H.}~\bibnamefont {Sun}}, \bibinfo {author} {\bibfnamefont
  {B.}~\bibnamefont {Shen}}, \bibinfo {author} {\bibfnamefont {Y.}~\bibnamefont
  {Zhang}},\ and\ \bibinfo {author} {\bibfnamefont {M.}~\bibnamefont {Wang}},\
  }\bibfield  {title} {\bibinfo {title} {Anisotropic magnetism and electronic
  properties of the kagome metal {SmV}$_6${Sn}$_6$},\ }\href
  {https://doi.org/10.1103/PhysRevMaterials.7.054403} {\bibfield  {journal}
  {\bibinfo  {journal} {Physical Review Materials}\ }\textbf {\bibinfo {volume}
  {7}},\ \bibinfo {pages} {054403} (\bibinfo {year} {2023})}\BibitemShut
  {NoStop}%
\bibitem [{\citenamefont {Ma}\ \emph {et~al.}(2021)\citenamefont {Ma},
  \citenamefont {Xu}, \citenamefont {Yin}, \citenamefont {Yang}, \citenamefont
  {Zhou}, \citenamefont {Cheng}, \citenamefont {Huang}, \citenamefont {Qu},
  \citenamefont {Wang}, \citenamefont {Hasan},\ and\ \citenamefont
  {Jia}}]{ma_rare_2021}%
  \BibitemOpen
  \bibfield  {author} {\bibinfo {author} {\bibfnamefont {W.}~\bibnamefont
  {Ma}}, \bibinfo {author} {\bibfnamefont {X.}~\bibnamefont {Xu}}, \bibinfo
  {author} {\bibfnamefont {J.-X.}\ \bibnamefont {Yin}}, \bibinfo {author}
  {\bibfnamefont {H.}~\bibnamefont {Yang}}, \bibinfo {author} {\bibfnamefont
  {H.}~\bibnamefont {Zhou}}, \bibinfo {author} {\bibfnamefont {Z.-J.}\
  \bibnamefont {Cheng}}, \bibinfo {author} {\bibfnamefont {Y.}~\bibnamefont
  {Huang}}, \bibinfo {author} {\bibfnamefont {Z.}~\bibnamefont {Qu}}, \bibinfo
  {author} {\bibfnamefont {F.}~\bibnamefont {Wang}}, \bibinfo {author}
  {\bibfnamefont {M.~Z.}\ \bibnamefont {Hasan}},\ and\ \bibinfo {author}
  {\bibfnamefont {S.}~\bibnamefont {Jia}},\ }\bibfield  {title} {\bibinfo
  {title} {Rare {Earth} {Engineering} in {$R$}{Mn}$_6${Sn}$_6$ ({$R$} =
  {Gd}-{Tm}, {Lu}) {Topological} {Kagome} {Magnets}},\ }\href
  {https://doi.org/10.1103/PhysRevLett.126.246602} {\bibfield  {journal}
  {\bibinfo  {journal} {Physical Review Letters}\ }\textbf {\bibinfo {volume}
  {126}},\ \bibinfo {pages} {246602} (\bibinfo {year} {2021})}\BibitemShut
  {NoStop}%
\bibitem [{\citenamefont {Yang}\ \emph {et~al.}(2024)\citenamefont {Yang},
  \citenamefont {Zeng}, \citenamefont {He}, \citenamefont {Xu}, \citenamefont
  {Du},\ and\ \citenamefont {Qu}}]{yang_crystal_2024}%
  \BibitemOpen
  \bibfield  {author} {\bibinfo {author} {\bibfnamefont {X.}~\bibnamefont
  {Yang}}, \bibinfo {author} {\bibfnamefont {Q.}~\bibnamefont {Zeng}}, \bibinfo
  {author} {\bibfnamefont {M.}~\bibnamefont {He}}, \bibinfo {author}
  {\bibfnamefont {X.}~\bibnamefont {Xu}}, \bibinfo {author} {\bibfnamefont
  {H.}~\bibnamefont {Du}},\ and\ \bibinfo {author} {\bibfnamefont
  {Z.}~\bibnamefont {Qu}},\ }\bibfield  {title} {\bibinfo {title} {Crystal
  growth, magnetic and electrical transport properties of the kagome magnet
  ${R}${Cr}$_6${Ge}$_6$ (${R}$ = {Gd}–{Tm})},\ }\href
  {https://doi.org/10.1088/1674-1056/ad3dcf} {\bibfield  {journal} {\bibinfo
  {journal} {Chinese Physics B}\ }\textbf {\bibinfo {volume} {33}},\ \bibinfo
  {pages} {077501} (\bibinfo {year} {2024})}\BibitemShut {NoStop}%
\bibitem [{\citenamefont {Pokharel}\ \emph {et~al.}(2021)\citenamefont
  {Pokharel}, \citenamefont {Teicher}, \citenamefont {Ortiz}, \citenamefont
  {Sarte}, \citenamefont {Wu}, \citenamefont {Peng}, \citenamefont {He},
  \citenamefont {Seshadri},\ and\ \citenamefont
  {Wilson}}]{pokharel_electronic_2021}%
  \BibitemOpen
  \bibfield  {author} {\bibinfo {author} {\bibfnamefont {G.}~\bibnamefont
  {Pokharel}}, \bibinfo {author} {\bibfnamefont {S.~M.~L.}\ \bibnamefont
  {Teicher}}, \bibinfo {author} {\bibfnamefont {B.~R.}\ \bibnamefont {Ortiz}},
  \bibinfo {author} {\bibfnamefont {P.~M.}\ \bibnamefont {Sarte}}, \bibinfo
  {author} {\bibfnamefont {G.}~\bibnamefont {Wu}}, \bibinfo {author}
  {\bibfnamefont {S.}~\bibnamefont {Peng}}, \bibinfo {author} {\bibfnamefont
  {J.}~\bibnamefont {He}}, \bibinfo {author} {\bibfnamefont {R.}~\bibnamefont
  {Seshadri}},\ and\ \bibinfo {author} {\bibfnamefont {S.~D.}\ \bibnamefont
  {Wilson}},\ }\bibfield  {title} {\bibinfo {title} {Electronic properties of
  the topological kagome metals \ce{YV6Sn6} and \ce{GdV6Sn6}},\ }\href
  {https://doi.org/10.1103/PhysRevB.104.235139} {\bibfield  {journal} {\bibinfo
   {journal} {Physical Review B}\ }\textbf {\bibinfo {volume} {104}},\ \bibinfo
  {pages} {235139} (\bibinfo {year} {2021})}\BibitemShut {NoStop}%
\bibitem [{\citenamefont {Lee}\ and\ \citenamefont
  {Mun}(2022)}]{lee_anisotropic_2022}%
  \BibitemOpen
  \bibfield  {author} {\bibinfo {author} {\bibfnamefont {J.}~\bibnamefont
  {Lee}}\ and\ \bibinfo {author} {\bibfnamefont {E.}~\bibnamefont {Mun}},\
  }\bibfield  {title} {\bibinfo {title} {Anisotropic magnetic property of
  single crystals \ce{$R$V6Sn6} ({$R$} = {Y} , {Gd}-{Tm} , {Lu})},\ }\href
  {https://doi.org/10.1103/PhysRevMaterials.6.083401} {\bibfield  {journal}
  {\bibinfo  {journal} {Physical Review Materials}\ }\textbf {\bibinfo {volume}
  {6}},\ \bibinfo {pages} {083401} (\bibinfo {year} {2022})}\BibitemShut
  {NoStop}%
\bibitem [{\citenamefont {Mozaffari}\ \emph {et~al.}(2024)\citenamefont
  {Mozaffari}, \citenamefont {Meier}, \citenamefont {Madhogaria}, \citenamefont
  {Peshcherenko}, \citenamefont {Kang}, \citenamefont {Villanova},
  \citenamefont {Arachchige}, \citenamefont {Zheng}, \citenamefont {Zhu},
  \citenamefont {Chen}, \citenamefont {Jenkins}, \citenamefont {Zhang},
  \citenamefont {Chan}, \citenamefont {Li}, \citenamefont {Yoon}, \citenamefont
  {Zhang},\ and\ \citenamefont {Mandrus}}]{mozaffari_universal_2024}%
  \BibitemOpen
  \bibfield  {author} {\bibinfo {author} {\bibfnamefont {S.}~\bibnamefont
  {Mozaffari}}, \bibinfo {author} {\bibfnamefont {W.~R.}\ \bibnamefont
  {Meier}}, \bibinfo {author} {\bibfnamefont {R.~P.}\ \bibnamefont
  {Madhogaria}}, \bibinfo {author} {\bibfnamefont {N.}~\bibnamefont
  {Peshcherenko}}, \bibinfo {author} {\bibfnamefont {S.-H.}\ \bibnamefont
  {Kang}}, \bibinfo {author} {\bibfnamefont {J.~W.}\ \bibnamefont {Villanova}},
  \bibinfo {author} {\bibfnamefont {H.~W.~S.}\ \bibnamefont {Arachchige}},
  \bibinfo {author} {\bibfnamefont {G.}~\bibnamefont {Zheng}}, \bibinfo
  {author} {\bibfnamefont {Y.}~\bibnamefont {Zhu}}, \bibinfo {author}
  {\bibfnamefont {K.-W.}\ \bibnamefont {Chen}}, \bibinfo {author}
  {\bibfnamefont {K.}~\bibnamefont {Jenkins}}, \bibinfo {author} {\bibfnamefont
  {D.}~\bibnamefont {Zhang}}, \bibinfo {author} {\bibfnamefont
  {A.}~\bibnamefont {Chan}}, \bibinfo {author} {\bibfnamefont {L.}~\bibnamefont
  {Li}}, \bibinfo {author} {\bibfnamefont {M.}~\bibnamefont {Yoon}}, \bibinfo
  {author} {\bibfnamefont {Y.}~\bibnamefont {Zhang}},\ and\ \bibinfo {author}
  {\bibfnamefont {D.~G.}\ \bibnamefont {Mandrus}},\ }\bibfield  {title}
  {\bibinfo {title} {Universal sublinear resistivity in vanadium kagome
  materials hosting charge density waves},\ }\href
  {https://doi.org/10.1103/PhysRevB.110.035135} {\bibfield  {journal} {\bibinfo
   {journal} {Physical Review B}\ }\textbf {\bibinfo {volume} {110}},\ \bibinfo
  {pages} {035135} (\bibinfo {year} {2024})}\BibitemShut {NoStop}%
\bibitem [{\citenamefont {Peshcherenko}\ \emph {et~al.}(2024)\citenamefont
  {Peshcherenko}, \citenamefont {Mao}, \citenamefont {Felser},\ and\
  \citenamefont {Zhang}}]{peshcherenko_sublinear_2024}%
  \BibitemOpen
  \bibfield  {author} {\bibinfo {author} {\bibfnamefont {N.}~\bibnamefont
  {Peshcherenko}}, \bibinfo {author} {\bibfnamefont {N.}~\bibnamefont {Mao}},
  \bibinfo {author} {\bibfnamefont {C.}~\bibnamefont {Felser}},\ and\ \bibinfo
  {author} {\bibfnamefont {Y.}~\bibnamefont {Zhang}},\ }\href
  {https://doi.org/10.48550/arXiv.2404.11612} {\bibinfo {title} {Sublinear
  transport in {Kagome} metals: {Interplay} of {Dirac} cones and {Van} {Hove}
  singularities}} (\bibinfo {year} {2024}),\ \bibinfo {note} {arXiv:2404.11612
  [cond-mat]}\BibitemShut {NoStop}%
\bibitem [{\citenamefont {Ye}\ \emph {et~al.}(2024)\citenamefont {Ye},
  \citenamefont {Fang}, \citenamefont {Kang}, \citenamefont {Kaufmann},
  \citenamefont {Lee}, \citenamefont {John}, \citenamefont {Neves},
  \citenamefont {Zhao}, \citenamefont {Denlinger}, \citenamefont {Jozwiak},
  \citenamefont {Bostwick}, \citenamefont {Rotenberg}, \citenamefont {Kaxiras},
  \citenamefont {Bell}, \citenamefont {Janson}, \citenamefont {Comin},\ and\
  \citenamefont {Checkelsky}}]{ye_hopping_2024}%
  \BibitemOpen
  \bibfield  {author} {\bibinfo {author} {\bibfnamefont {L.}~\bibnamefont
  {Ye}}, \bibinfo {author} {\bibfnamefont {S.}~\bibnamefont {Fang}}, \bibinfo
  {author} {\bibfnamefont {M.}~\bibnamefont {Kang}}, \bibinfo {author}
  {\bibfnamefont {J.}~\bibnamefont {Kaufmann}}, \bibinfo {author}
  {\bibfnamefont {Y.}~\bibnamefont {Lee}}, \bibinfo {author} {\bibfnamefont
  {C.}~\bibnamefont {John}}, \bibinfo {author} {\bibfnamefont {P.~M.}\
  \bibnamefont {Neves}}, \bibinfo {author} {\bibfnamefont {S.~Y.~F.}\
  \bibnamefont {Zhao}}, \bibinfo {author} {\bibfnamefont {J.}~\bibnamefont
  {Denlinger}}, \bibinfo {author} {\bibfnamefont {C.}~\bibnamefont {Jozwiak}},
  \bibinfo {author} {\bibfnamefont {A.}~\bibnamefont {Bostwick}}, \bibinfo
  {author} {\bibfnamefont {E.}~\bibnamefont {Rotenberg}}, \bibinfo {author}
  {\bibfnamefont {E.}~\bibnamefont {Kaxiras}}, \bibinfo {author} {\bibfnamefont
  {D.~C.}\ \bibnamefont {Bell}}, \bibinfo {author} {\bibfnamefont
  {O.}~\bibnamefont {Janson}}, \bibinfo {author} {\bibfnamefont
  {R.}~\bibnamefont {Comin}},\ and\ \bibinfo {author} {\bibfnamefont {J.~G.}\
  \bibnamefont {Checkelsky}},\ }\bibfield  {title} {\bibinfo {title} {Hopping
  frustration-induced flat band and strange metallicity in a kagome metal},\
  }\href {https://doi.org/10.1038/s41567-023-02360-5} {\bibfield  {journal}
  {\bibinfo  {journal} {Nature Physics}\ }\textbf {\bibinfo {volume} {20}},\
  \bibinfo {pages} {610} (\bibinfo {year} {2024})}\BibitemShut {NoStop}%
\bibitem [{\citenamefont {Xiang}\ \emph {et~al.}(2016)\citenamefont {Xiang},
  \citenamefont {Wang}, \citenamefont {Wang}, \citenamefont {Zhao},
  \citenamefont {Sun}, \citenamefont {Luo}, \citenamefont {Wu},\ and\
  \citenamefont {Chen}}]{xiang_incoherencecoherence_2016}%
  \BibitemOpen
  \bibfield  {author} {\bibinfo {author} {\bibfnamefont {Z.~J.}\ \bibnamefont
  {Xiang}}, \bibinfo {author} {\bibfnamefont {N.~Z.}\ \bibnamefont {Wang}},
  \bibinfo {author} {\bibfnamefont {A.~F.}\ \bibnamefont {Wang}}, \bibinfo
  {author} {\bibfnamefont {D.}~\bibnamefont {Zhao}}, \bibinfo {author}
  {\bibfnamefont {Z.~L.}\ \bibnamefont {Sun}}, \bibinfo {author} {\bibfnamefont
  {X.~G.}\ \bibnamefont {Luo}}, \bibinfo {author} {\bibfnamefont
  {T.}~\bibnamefont {Wu}},\ and\ \bibinfo {author} {\bibfnamefont {X.~H.}\
  \bibnamefont {Chen}},\ }\bibfield  {title} {\bibinfo {title}
  {Incoherence–coherence crossover and low-temperature {Fermi}-liquid-like
  behavior in \textit{{A}}{Fe}$_{\textrm{2}}${As}$_{\textrm{2}}$ (\textit{{A}}
  = {K}, {Rb}, {Cs}): evidence from electrical transport properties},\ }\href
  {https://doi.org/10.1088/0953-8984/28/42/425702} {\bibfield  {journal}
  {\bibinfo  {journal} {Journal of Physics: Condensed Matter}\ }\textbf
  {\bibinfo {volume} {28}},\ \bibinfo {pages} {425702} (\bibinfo {year}
  {2016})}\BibitemShut {NoStop}%
\bibitem [{\citenamefont {Gutman}\ and\ \citenamefont
  {Maslov}(2007)}]{gutman_anomalous_2007}%
  \BibitemOpen
  \bibfield  {author} {\bibinfo {author} {\bibfnamefont {D.~B.}\ \bibnamefont
  {Gutman}}\ and\ \bibinfo {author} {\bibfnamefont {D.~L.}\ \bibnamefont
  {Maslov}},\ }\bibfield  {title} {\bibinfo {title} {Anomalous $c$-{Axis}
  {Transport} in {Layered} {Metals}},\ }\href
  {https://doi.org/10.1103/PhysRevLett.99.196602} {\bibfield  {journal}
  {\bibinfo  {journal} {Physical Review Letters}\ }\textbf {\bibinfo {volume}
  {99}},\ \bibinfo {pages} {196602} (\bibinfo {year} {2007})}\BibitemShut
  {NoStop}%
\bibitem [{\citenamefont {Wang}\ \emph {et~al.}(2016)\citenamefont {Wang},
  \citenamefont {Zaliznyak}, \citenamefont {Ren}, \citenamefont {Wu},
  \citenamefont {Graf}, \citenamefont {Garlea}, \citenamefont {Warren},
  \citenamefont {Bozin}, \citenamefont {Zhu},\ and\ \citenamefont
  {Petrovic}}]{wang_magnetotransport_2016}%
  \BibitemOpen
  \bibfield  {author} {\bibinfo {author} {\bibfnamefont {A.}~\bibnamefont
  {Wang}}, \bibinfo {author} {\bibfnamefont {I.}~\bibnamefont {Zaliznyak}},
  \bibinfo {author} {\bibfnamefont {W.}~\bibnamefont {Ren}}, \bibinfo {author}
  {\bibfnamefont {L.}~\bibnamefont {Wu}}, \bibinfo {author} {\bibfnamefont
  {D.}~\bibnamefont {Graf}}, \bibinfo {author} {\bibfnamefont {V.~O.}\
  \bibnamefont {Garlea}}, \bibinfo {author} {\bibfnamefont {J.~B.}\
  \bibnamefont {Warren}}, \bibinfo {author} {\bibfnamefont {E.}~\bibnamefont
  {Bozin}}, \bibinfo {author} {\bibfnamefont {Y.}~\bibnamefont {Zhu}},\ and\
  \bibinfo {author} {\bibfnamefont {C.}~\bibnamefont {Petrovic}},\ }\bibfield
  {title} {\bibinfo {title} {Magnetotransport study of {Dirac} fermions in
  {YbMnBi$_2$} antiferromagnet},\ }\href
  {https://doi.org/10.1103/PhysRevB.94.165161} {\bibfield  {journal} {\bibinfo
  {journal} {Physical Review B}\ }\textbf {\bibinfo {volume} {94}},\ \bibinfo
  {pages} {165161} (\bibinfo {year} {2016})}\BibitemShut {NoStop}%
\bibitem [{\citenamefont {Yamada}\ and\ \citenamefont
  {Takada}(1973)}]{yamada_hiroshi_magnetoresistance_1973}%
  \BibitemOpen
  \bibfield  {author} {\bibinfo {author} {\bibfnamefont {H.}~\bibnamefont
  {Yamada}}\ and\ \bibinfo {author} {\bibfnamefont {S.}~\bibnamefont
  {Takada}},\ }\bibfield  {title} {\bibinfo {title} {{Magnetoresistance} of
  {Antiferromagnetic} {Metals} {Due} to s-d {Interaction}},\ }\href
  {https://doi.org/doi.org/10.1143/JPSJ.34.51} {\bibfield  {journal} {\bibinfo
  {journal} {Journal of the Physical Society of Japan}\ }\textbf {\bibinfo
  {volume} {34}},\ \bibinfo {pages} {51} (\bibinfo {year} {1973})}\BibitemShut
  {NoStop}%
\bibitem [{\citenamefont {Xia}\ \emph {et~al.}(2023)\citenamefont {Xia},
  \citenamefont {Wang}, \citenamefont {Liu}, \citenamefont {Zhang},
  \citenamefont {Wu}, \citenamefont {Zhang}, \citenamefont {Yang},
  \citenamefont {Yang}, \citenamefont {He}, \citenamefont {Chai}, \citenamefont
  {Zhou},\ and\ \citenamefont {Wang}}]{xia_doping-induced_2023}%
  \BibitemOpen
  \bibfield  {author} {\bibinfo {author} {\bibfnamefont {Y.}~\bibnamefont
  {Xia}}, \bibinfo {author} {\bibfnamefont {L.}~\bibnamefont {Wang}}, \bibinfo
  {author} {\bibfnamefont {Y.}~\bibnamefont {Liu}}, \bibinfo {author}
  {\bibfnamefont {L.}~\bibnamefont {Zhang}}, \bibinfo {author} {\bibfnamefont
  {X.}~\bibnamefont {Wu}}, \bibinfo {author} {\bibfnamefont {L.}~\bibnamefont
  {Zhang}}, \bibinfo {author} {\bibfnamefont {T.}~\bibnamefont {Yang}},
  \bibinfo {author} {\bibfnamefont {K.}~\bibnamefont {Yang}}, \bibinfo {author}
  {\bibfnamefont {M.}~\bibnamefont {He}}, \bibinfo {author} {\bibfnamefont
  {Y.}~\bibnamefont {Chai}}, \bibinfo {author} {\bibfnamefont {X.}~\bibnamefont
  {Zhou}},\ and\ \bibinfo {author} {\bibfnamefont {A.}~\bibnamefont {Wang}},\
  }\bibfield  {title} {\bibinfo {title} {Doping-induced spin reorientation and
  magnetic phase diagram of \ce{EuMn_{1-x}Zn_xSb2} ( 0 $\leq x \leq$ 1 )},\
  }\href {https://doi.org/10.1103/PhysRevB.107.184415} {\bibfield  {journal}
  {\bibinfo  {journal} {Physical Review B}\ }\textbf {\bibinfo {volume}
  {107}},\ \bibinfo {pages} {184415} (\bibinfo {year} {2023})}\BibitemShut
  {NoStop}%
\bibitem [{\citenamefont {Prakash}\ \emph {et~al.}(2016)\citenamefont
  {Prakash}, \citenamefont {Thamizhavel},\ and\ \citenamefont
  {Ramakrishnan}}]{prakash_ferromagnetic_2016}%
  \BibitemOpen
  \bibfield  {author} {\bibinfo {author} {\bibfnamefont {O.}~\bibnamefont
  {Prakash}}, \bibinfo {author} {\bibfnamefont {A.}~\bibnamefont
  {Thamizhavel}},\ and\ \bibinfo {author} {\bibfnamefont {S.}~\bibnamefont
  {Ramakrishnan}},\ }\bibfield  {title} {\bibinfo {title} {Ferromagnetic
  ordering of minority {Ce$^{3+}$} spins in a quasi-skutterudite
  {Ce$_3$Os$_4$Ge$_{13}$} single crystal},\ }\href
  {https://doi.org/10.1103/PhysRevB.93.064427} {\bibfield  {journal} {\bibinfo
  {journal} {Physical Review B}\ }\textbf {\bibinfo {volume} {93}},\ \bibinfo
  {pages} {064427} (\bibinfo {year} {2016})}\BibitemShut {NoStop}%
\bibitem [{\citenamefont {Wang}(2024)}]{wang_encoding_2024}%
  \BibitemOpen
  \bibfield  {author} {\bibinfo {author} {\bibfnamefont {Y.}~\bibnamefont
  {Wang}},\ }\bibfield  {title} {\bibinfo {title} {Encoding innumerable charge
  density waves of {FeGe} into polymorphs of {LiFe$_6$Ge$_6$}},\ }\href
  {https://doi.org/10.1007/s11433-024-2423-2} {\bibfield  {journal} {\bibinfo
  {journal} {Science China Physics, Mechanics \& Astronomy}\ }\textbf {\bibinfo
  {volume} {67}},\ \bibinfo {pages} {297011} (\bibinfo {year}
  {2024})}\BibitemShut {NoStop}%
\bibitem [{\citenamefont {Sinha}\ \emph {et~al.}(2021)\citenamefont {Sinha},
  \citenamefont {Vivanco}, \citenamefont {Wan}, \citenamefont {Siegler},
  \citenamefont {Stewart}, \citenamefont {Pogue}, \citenamefont {Pressley},
  \citenamefont {Berry}, \citenamefont {Wang}, \citenamefont {Johnson},
  \citenamefont {Chen}, \citenamefont {Tran}, \citenamefont {Phelan},\ and\
  \citenamefont {McQueen}}]{sinha_twisting_2021}%
  \BibitemOpen
  \bibfield  {author} {\bibinfo {author} {\bibfnamefont {M.}~\bibnamefont
  {Sinha}}, \bibinfo {author} {\bibfnamefont {H.~K.}\ \bibnamefont {Vivanco}},
  \bibinfo {author} {\bibfnamefont {C.}~\bibnamefont {Wan}}, \bibinfo {author}
  {\bibfnamefont {M.~A.}\ \bibnamefont {Siegler}}, \bibinfo {author}
  {\bibfnamefont {V.~J.}\ \bibnamefont {Stewart}}, \bibinfo {author}
  {\bibfnamefont {E.~A.}\ \bibnamefont {Pogue}}, \bibinfo {author}
  {\bibfnamefont {L.~A.}\ \bibnamefont {Pressley}}, \bibinfo {author}
  {\bibfnamefont {T.}~\bibnamefont {Berry}}, \bibinfo {author} {\bibfnamefont
  {Z.}~\bibnamefont {Wang}}, \bibinfo {author} {\bibfnamefont {I.}~\bibnamefont
  {Johnson}}, \bibinfo {author} {\bibfnamefont {M.}~\bibnamefont {Chen}},
  \bibinfo {author} {\bibfnamefont {T.~T.}\ \bibnamefont {Tran}}, \bibinfo
  {author} {\bibfnamefont {W.~A.}\ \bibnamefont {Phelan}},\ and\ \bibinfo
  {author} {\bibfnamefont {T.~M.}\ \bibnamefont {McQueen}},\ }\bibfield
  {title} {\bibinfo {title} {Twisting of {2D} {Kagomé} {Sheets} in {Layered}
  {Intermetallics}},\ }\href {https://doi.org/10.1021/acscentsci.1c00599}
  {\bibfield  {journal} {\bibinfo  {journal} {ACS Central Science}\ }\textbf
  {\bibinfo {volume} {7}},\ \bibinfo {pages} {1381} (\bibinfo {year}
  {2021})}\BibitemShut {NoStop}%
\bibitem [{\citenamefont {Venturini}\ \emph {et~al.}(1992)\citenamefont
  {Venturini}, \citenamefont {Welter},\ and\ \citenamefont
  {Malaman}}]{venturini_crystallographic_1992}%
  \BibitemOpen
  \bibfield  {author} {\bibinfo {author} {\bibfnamefont {G.}~\bibnamefont
  {Venturini}}, \bibinfo {author} {\bibfnamefont {R.}~\bibnamefont {Welter}},\
  and\ \bibinfo {author} {\bibfnamefont {B.}~\bibnamefont {Malaman}},\
  }\bibfield  {title} {\bibinfo {title} {Crystallographic data and magnetic
  properties of \ce{$RT$6Ge6} compounds ({$R$} = {Sc}, {Y}, {Nd}, {Sm},
  {Gd}-{Lu}; {$T$} = {Mn}, {Fe})},\ }\href
  {https://doi.org/10.1016/0925-8388(92)90558-Q} {\bibfield  {journal}
  {\bibinfo  {journal} {Journal of Alloys and Compounds}\ }\textbf {\bibinfo
  {volume} {185}},\ \bibinfo {pages} {99} (\bibinfo {year} {1992})}\BibitemShut
  {NoStop}%
\bibitem [{\citenamefont {Mazet}\ \emph {et~al.}(2013)\citenamefont {Mazet},
  \citenamefont {Ban}, \citenamefont {Sibille}, \citenamefont {Capelli},\ and\
  \citenamefont {Malaman}}]{mazet_magnetic_2013}%
  \BibitemOpen
  \bibfield  {author} {\bibinfo {author} {\bibfnamefont {T.}~\bibnamefont
  {Mazet}}, \bibinfo {author} {\bibfnamefont {V.}~\bibnamefont {Ban}}, \bibinfo
  {author} {\bibfnamefont {R.}~\bibnamefont {Sibille}}, \bibinfo {author}
  {\bibfnamefont {S.}~\bibnamefont {Capelli}},\ and\ \bibinfo {author}
  {\bibfnamefont {B.}~\bibnamefont {Malaman}},\ }\bibfield  {title} {\bibinfo
  {title} {Magnetic properties of \ce{MgFe6Ge6}},\ }\href
  {https://doi.org/10.1016/j.ssc.2013.01.027} {\bibfield  {journal} {\bibinfo
  {journal} {Solid State Communications}\ }\textbf {\bibinfo {volume} {159}},\
  \bibinfo {pages} {79} (\bibinfo {year} {2013})}\BibitemShut {NoStop}%
\bibitem [{\citenamefont {Welk}\ and\ \citenamefont
  {Schuster}(1976)}]{welk_zur_1976}%
  \BibitemOpen
  \bibfield  {author} {\bibinfo {author} {\bibfnamefont {E.}~\bibnamefont
  {Welk}}\ and\ \bibinfo {author} {\bibfnamefont {H.}~\bibnamefont
  {Schuster}},\ }\bibfield  {title} {\bibinfo {title} {Zur {Kenntnis} der
  {Phase} \ce{LiFe6Ge6}},\ }\href {https://doi.org/10.1002/zaac.19764240302}
  {\bibfield  {journal} {\bibinfo  {journal} {Zeitschrift für anorganische und
  allgemeine Chemie}\ }\textbf {\bibinfo {volume} {424}},\ \bibinfo {pages}
  {193} (\bibinfo {year} {1976})}\BibitemShut {NoStop}%
\bibitem [{\citenamefont {Cadogan}\ \emph {et~al.}(2007)\citenamefont
  {Cadogan}, \citenamefont {Ryan},\ and\ \citenamefont
  {Cashion}}]{cadogan_155gd_2007}%
  \BibitemOpen
  \bibfield  {author} {\bibinfo {author} {\bibfnamefont {J.~M.}\ \bibnamefont
  {Cadogan}}, \bibinfo {author} {\bibfnamefont {D.~H.}\ \bibnamefont {Ryan}},\
  and\ \bibinfo {author} {\bibfnamefont {J.~D.}\ \bibnamefont {Cashion}},\
  }\bibfield  {title} {\bibinfo {title} {\ce{^{155}Gd} {Mössbauer} study of
  \ce{GdFe6Ge6}},\ }\href {https://doi.org/10.1088/0953-8984/19/21/216204}
  {\bibfield  {journal} {\bibinfo  {journal} {Journal of Physics: Condensed
  Matter}\ }\textbf {\bibinfo {volume} {19}},\ \bibinfo {pages} {216204}
  (\bibinfo {year} {2007})}\BibitemShut {NoStop}%
\bibitem [{\citenamefont {Mazet}\ \emph {et~al.}(2003)\citenamefont {Mazet},
  \citenamefont {Tobola},\ and\ \citenamefont {Malaman}}]{mazet_covalent_2003}%
  \BibitemOpen
  \bibfield  {author} {\bibinfo {author} {\bibfnamefont {T.}~\bibnamefont
  {Mazet}}, \bibinfo {author} {\bibfnamefont {J.}~\bibnamefont {Tobola}},\ and\
  \bibinfo {author} {\bibfnamefont {B.}~\bibnamefont {Malaman}},\ }\bibfield
  {title} {\bibinfo {title} {Covalent magnetism in the \ce{$R$Fe6Ge6} series},\
  }\href {https://doi.org/10.1140/epjb/e2003-00155-x} {\bibfield  {journal}
  {\bibinfo  {journal} {The European Physical Journal B - Condensed Matter}\
  }\textbf {\bibinfo {volume} {33}},\ \bibinfo {pages} {183} (\bibinfo {year}
  {2003})}\BibitemShut {NoStop}%
\bibitem [{\citenamefont {Freccero}\ \emph {et~al.}(2021)\citenamefont
  {Freccero}, \citenamefont {Hübner}, \citenamefont {Prots}, \citenamefont
  {Schnelle}, \citenamefont {Schmidt}, \citenamefont {Wagner}, \citenamefont
  {Schwarz},\ and\ \citenamefont {Grin}}]{freccero_excess_2021}%
  \BibitemOpen
  \bibfield  {author} {\bibinfo {author} {\bibfnamefont {R.}~\bibnamefont
  {Freccero}}, \bibinfo {author} {\bibfnamefont {J.-M.}\ \bibnamefont
  {Hübner}}, \bibinfo {author} {\bibfnamefont {Y.}~\bibnamefont {Prots}},
  \bibinfo {author} {\bibfnamefont {W.}~\bibnamefont {Schnelle}}, \bibinfo
  {author} {\bibfnamefont {M.}~\bibnamefont {Schmidt}}, \bibinfo {author}
  {\bibfnamefont {F.~R.}\ \bibnamefont {Wagner}}, \bibinfo {author}
  {\bibfnamefont {U.}~\bibnamefont {Schwarz}},\ and\ \bibinfo {author}
  {\bibfnamefont {Y.}~\bibnamefont {Grin}},\ }\bibfield  {title} {\bibinfo
  {title} {“{Excess}” electrons in {LuGe}},\ }\href
  {https://doi.org/10.1002/anie.202014284} {\bibfield  {journal} {\bibinfo
  {journal} {Angewandte Chemie International Edition}\ }\textbf {\bibinfo
  {volume} {60}},\ \bibinfo {pages} {6457} (\bibinfo {year}
  {2021})}\BibitemShut {NoStop}%
\bibitem [{\citenamefont {Landrum}\ and\ \citenamefont
  {Dronskowski}(2000)}]{landrum_orbital_2000}%
  \BibitemOpen
  \bibfield  {author} {\bibinfo {author} {\bibfnamefont {G.~A.}\ \bibnamefont
  {Landrum}}\ and\ \bibinfo {author} {\bibfnamefont {R.}~\bibnamefont
  {Dronskowski}},\ }\bibfield  {title} {\bibinfo {title} {The {Orbital}
  {Origins} of {Magnetism}: {From} {Atoms} to {Molecules} to {Ferromagnetic}
  {Alloys}},\ }\href
  {https://doi.org/10.1002/(SICI)1521-3773(20000502)39:9<1560::AID-ANIE1560>3.0.CO;2-T}
  {\bibfield  {journal} {\bibinfo  {journal} {Angewandte Chemie International
  Edition}\ }\textbf {\bibinfo {volume} {39}},\ \bibinfo {pages} {1560}
  (\bibinfo {year} {2000})}\BibitemShut {NoStop}%
\bibitem [{\citenamefont {Landrum}\ and\ \citenamefont
  {Dronskowski}(1999)}]{landrum_ferromagnetism_1999}%
  \BibitemOpen
  \bibfield  {author} {\bibinfo {author} {\bibfnamefont {G.~A.}\ \bibnamefont
  {Landrum}}\ and\ \bibinfo {author} {\bibfnamefont {R.}~\bibnamefont
  {Dronskowski}},\ }\bibfield  {title} {\bibinfo {title} {Ferromagnetism in
  {Transition} {Metals}: {A} {Chemical} {Bonding} {Approach}},\ }\href
  {https://doi.org/10.1002/(SICI)1521-3773(19990517)38:10<1389::AID-ANIE1389>3.0.CO;2-K}
  {\bibfield  {journal} {\bibinfo  {journal} {Angewandte Chemie International
  Edition}\ }\textbf {\bibinfo {volume} {38}},\ \bibinfo {pages} {1389}
  (\bibinfo {year} {1999})}\BibitemShut {NoStop}%
\end{thebibliography}
%

\end{document}